\documentclass [12pt,eqsecnum,amsfonts,amssymb,aps]{revtex4}

\input epsf

\usepackage[usenames]{color}
\usepackage{graphicx}
\usepackage{bm}
\usepackage{rotating}

\newcommand{\beq}{\begin{equation}}
\newcommand{\eeq}{\end{equation}}
\newcommand{\beqs}{\begin{eqnarray}}
\newcommand{\eeqs}{\end{eqnarray}}

\begin{document}

\baselineskip 6.0mm

\title{Exact Potts/Tutte Polynomials for Hammock Chain Graphs}

\author{Yue Chen and Robert Shrock}

\affiliation{C. N. Yang Institute for Theoretical Physics and \\
  Department of Physics and Astronomy \\
Stony Brook University, Stony Brook, NY 11794}

\begin{abstract}

  We present exact calculations of the $q$-state Potts model partition
  functions and the equivalent Tutte polynomials for chain graphs comprised of
  $m$ repeated hammock subgraphs $H_{e_1,...,e_r}$ connected with line graphs
  of length $e_g$ edges, such that the chains have open or cyclic boundary
  conditions (BC). Here, $H_{e_1,...,e_r}$ is a hammock (series-parallel)
  subgraph with $r$ separate paths along ``ropes'' with respective lengths
  $e_1, ..., e_r$ edges, connecting the two end vertices. We denote the
  resultant chain graph as $G_{\{e_1,...,e_r\},e_g,m;BC}$. We discuss special
  cases, including chromatic, flow, and reliability polynomials. In the case of
  cyclic boundary conditions, the zeros of the Potts partition function in the
  complex $q$ function accumulate, in the limit $m \to \infty$, onto curves
  forming a locus ${\cal B}$, and we study this locus.

\end{abstract}

\maketitle

\newpage
\pagestyle{plain}
\pagenumbering{arabic}

% ===============================================================

\section{Introduction}
\label{intro_section}

The $q$-state Potts model has long been of interest in the study of phase
transitions and critical phenomena \cite{potts,wurev}.  On a lattice, or, more
generally, on a graph $G$, at temperature $T$, the partition function for the 
Potts model is
\beq
Z= \sum_{\{\sigma_i\}}e^{-\beta {\cal H}} \ ,
\label{z}
\eeq
where $\beta = 1/(k_BT)$, $k_B$ is the Boltzmann constant, and the
Hamiltonian is
\beq
{\cal H} = -J\sum_{e_{ij}}\delta_{\sigma_i \sigma_j} \ ,
\label{ham}
\eeq
where $J$ is the spin-spin
interaction constant, $i$ and $j$ denote vertices (= sites) in $G$, $e_{ij}$ is
the edge (= bond) connecting them, and $\sigma_i$ are classical spins taking on
values in the set $\{1,...,q\}$. We use the notation
\beq
K = \beta J \ , \quad  y = e^K \ , \quad v = y -1 \ .
\label{kv}
\eeq
We denote the partition function of the Potts model on a graph $G$ as
$Z(G,q,v)$. Thus, for the Potts ferromagnet (FM, $J > 0$) and antiferromagnet
(AFM, $J < 0$), the physical ranges of $v$ are $v \ge 0$ and $-1 \le v \le 0$,
respectively.  On a graph $G$, we denote the Potts partition function as
$Z(G,q,v)$. For the Potts antiferromagnet (PAF), $J < 0$ so that, as $T \to 0$,
$K \to -\infty$, i.e., $v \to -1$; hence, in this limit, the only contributions
to the PAF partition function are from spin configurations in which adjacent
spins have different values.  The resultant $T=0$ PAF partition function is
therefore precisely the chromatic polynomial $P(G,q)$ of the graph $G$, which
counts the number of ways of assigning $q$ colors to the vertices of $G$,
subject to the condition that no two adjacent vertices have the same color
\cite{birk}-\cite{dkt}. That is, 
\beq
Z(G,q,-1)=P(G,q) \ . 
\label{zp}
\eeq
This is called a proper $q$-coloring of (the vertices
of) $G$. The zeros of $P(G,q)$ in the complex $q$ plane are called chromatic
zeros of $G$. An important feature of the antiferromagnetic Potts model is that
for sufficiently large $q$ on a given graph $G$ with finite maximal vertex
degree, it has a nonzero entropy per site at zero temperature, $S_0=k_B \ln W$,
where $W$ denotes the ground state degeneracy per site
\cite{wurev,baxter70,baxter86,baxter87}. This is important as an
exception to the third law of thermodynamics, that the entropy per site
vanishes at zero temperature. A physical example of this phenomenon is the
residual entropy of ice \cite{pauling}-\cite{lieb}. The Potts partition
function $Z(G,q,v)$ is equivalent to a function of fundamental importance in
mathematical graph theory, the Tutte polynomial, $T(G,x,y)$
\cite{tutte47}-\cite{welsh} (see Eq. (\ref{ztrel}) below). We will therefore
often refer to these together as the Potts/Tutte polynomial.

In Ref. \cite{wa3} with S.-H. Tsai, an exact expression was given (in Eq. (3.7)
of \cite{wa3}) for the partition function of the zero-temperature Potts
antiferromagnet, or equivalently, the chromatic polynomial, of a ``hammock''
graph, denoted $H_{k,r}$, defined as a series-parallel graph containing two end
vertices with $r$ separate ``ropes'' (paths) joining these two end-vertices,
such that on each rope there are $k$ ``knots'' (vertices), including the end
vertices, i.e., $k-1$ edges on each rope.  This was
part of a program of studies \cite{wa3}-\cite{wa2} of families of graphs having
the property that in the limit $r \to \infty$ for fixed $k$ (denoted $L_r$
below), the magnitudes of the chromatic zeros are unbounded, and the continuous
accumulation set of chromatic zeros extends to the origin of the $1/q$
plane. Here we will generalize this family of graphs to encompass the case
where there are different numbers of edges $e_j$, $j=1,...,r$, on the different
ropes and hence denote this hammock graph as $H_{e_1,...,e_r}$. In
Ref. \cite{nec} with S.-H. Tsai, exact results were given for the chromatic
polynomial of open and cyclic chain graphs composed of $m$ repetitions of
$p$-sided polygons connected to each other by line segments of length $e_g$
edges. Each polygon can be described as a hammock graph with $r=2$ ropes, viz.,
$H_{e_1,e_2}$ (where $e_1$ and $e_2$ are, in general, different), such that the
right-hand end-vertex of a given $H_{e_1,e_2}$ subgraph is connected to a
line graph consisting of $e_g$ edges, which then connects to the left-hand
end-vertex of the next $H_{e_1,e_2}$ subgraph in the chain. In \cite{neca}, one
of us generalized \cite{nec} to a calculation of the full Potts/Tutte
polynomial for these graphs.  A motivation for the work in \cite{neca} was to
elucidate how the results for the zero-temperature Potts antiferromagnet
obtained in \cite{nec} generalize to finite-temperature and also to
ferromagnetic as well as antiferromagnetic spin-spin couplings.

\begin{figure}[htbp]
\centering
\leavevmode
\epsfxsize=4.0in
\epsffile{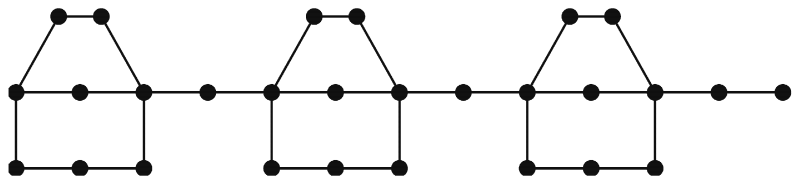}
\vspace{1 cm}
\caption{Illustrations of hammock chain graphs with $r=3$, 
$G_{\{e\}_{r=3},e_g,m;o}$  and $G_{\{e\}_{r=3},e_g,m;c}$, 
where $\{e\}_{r=3} \equiv \{e_1,e_2,e_3\}=\{2,3,4\}$, $e_g=2$, and $m=3$. 
For the cyclic hammock chain graph, the vertex on the right end is identified with the
vertex on the left end of the chain, while for the open hammock chain graph, it is 
a distinct vertex.} 
\label{hamfig}
\end{figure}

In this paper we present a calculation of the Potts/Tutte polynomial for a
further generalization of these families of graphs, namely for a chain graph
comprised of $m$ hammock subgraphs $H_{e_1,...,e_r}$ such that
the right-hand end vertex of a given hammock subgraph is connected to a line
graph of length $e_g$ edges, which then connects to the left-hand end-vertex of
the next hammock subgraph.  We consider both open (free) and cyclic
longitudinal boundary conditions (BC) for the chain, and thus denote a general
chain graph of this type as $G_{\{e_1,...,e_r\},e_g,m;BC}$.  We will use a
compact notation to denote the set of edges in each hammock subgraph as
\beq
\{e\}_r \equiv \{e_1,e_2,...,e_r\} \ ,
\label{ejrset}
\eeq
so that a hammock subgraph is denoted $H_{\{e\}_r}$, and the full chain graph 
comprised of $m$ hammock subgraphs $H_{\{e\}_r}$ connected with
line graphs of length $e_g$ edges, with longitudinal boundary
conditions denoted as $BC=o$ for open or $BC=c$ for cyclic,
is denoted
\beq
G_{\{e\}_r,e_g,m;BC} \equiv G_{\{e_1,...,e_r\},e_g,m;BC} \ , \quad BC = o \ 
{\rm or} \ c \ .
\label{graph}
\eeq
In Fig. \ref{hamfig} we show illustrative examples of the open and cyclic $r=3$
hammock chain graphs with $(\{e_1,e_2,e_3\},e_g,m)=(\{2,3,4\},2,3)$, i.e.,
$G_{\{2,3,4\},2,3;o}$ and $G_{\{2,3,4\},2,3;c}$.  These hammock chain graphs
are recursive in the sense of \cite{bds,bbook}; i.e., $G_{\{e\}_r,e_g,m+1;o}$
can be obtained by attaching another $H_{\{e\}_r}$ subgraph and associated
line graph of length $e_g$ edges to $G_{\{e\}_r,e_g,m;o}$, and similarly,
$G_{\{e\}_r,e_g,m+1;c}$ can be obtained by cutting a $G_{\{e\}_r,e_g,m;c}$
graph at any of the $H_{\{e\}_r}$ end-vertices and inserting and gluing a
$H_{\{e\}_r}$ subgraph with its associated line graph of length $e_g$ edges.
It is clear that the graph $G_{\{e\}_r,e_g,m;BC}$ with $BC=o$ or $BC=c$ is
unchanged if one permutes the values of the individual edges $e_j$ in the set
$\{e\}_r$. Hence, without loss of generality, by convention, when considering
specific sets $\{e\}_r$, we take $e_1 \le e_2 \le ... \le e_r$.  Obviously,
these are planar graphs.

If $r=1$, then $G_{e_1,e_g,m;o}$ and $G_{e_1,e_g,m;c}$ reduce simply
to the line graph with $n=m(e_1+e_g)+1$ vertices and the circuit graph
with $n=m(e_1+e_g)$ vertices, respectively.  Since the Potts/Tutte
polynomials for these are well-known, we will focus on the cases with
$r \ge 2$ here. If $r=2$, then each hammock subgraph is a polygon with
$p$ sides, where
\beq 
r=2: \quad p = e_1+e_2 \ . 
\label{pdef}
\eeq
This was the case studied in Refs. \cite{nec,neca}. Illustrative figures of the
polygon chain graphs were given in Fig. 1 of \cite{nec}.  As noted, the
chromatic polynomial for the case of arbitrary $r$ with $e_i=e_j$ for all $i,j
\in \{1,...,r\}$ was studied in \cite{wa3}.  In addition to being of interest
in its own right, the generalization considered here is 
valuable because it shows how the previous $r=2$ results in
\cite{nec,neca} arise as special cases of general-$r$ properties and how the
results in \cite{wa3} arise as $v=-1$ special cases of general physical values
of $v$, namely $-1 \le v < \infty$. 

Many results are simplest when expressed in terms of Tutte polynomials, so we
will give these first, and then discuss the corresponding Potts model partition
functions.  In general, for a given graph $G$, although the Tutte polynomial
$T(G,x,y)$ is equivalent to the Potts model partition function $Z(G,q,v)$, it
is valuable to include results expressed in terms of the latter as well as the
former, because the variable $q$ plays an important physical role in the
Hamiltonian formulation of Eqs. (\ref{z}) and (\ref{ham}), and only in the
Potts partition function does this variable appear by itself; in $T(G,x,y)$ it
only appears in combination with the temperature-dependent variable $y=v+1$, as
$x=1+q/v$. After giving these results, we then present special cases of
one-variable polynomials, including the chromatic, flow, and reliabiity
polynomials.

% fig. 1

We mention some previous related work by other authors. Read and Tutte
calculated the chromatic polynomial for the case $r=3$, namely
$P(H_{\{e\}_{r=3}},q)$, in \cite{read_tutte}. In Ref. \cite{sokal_roots}, Sokal
calculated the Potts model partition function $Z(H_{\{e\}_r},q,v)$, for general
$r$ and generalized this further to a multivariate case where the $v$
parameters are different for different edges (see also \cite{bhsw}).  To our
knowledge, the Tutte polynomial and corresponding Potts model partition
function have not previously been calculated for the cases considered here,
namely an open or cyclic chain graph for general $r$ and $m$,
$G_{\{e\}_r,e_g,m;BC}$.  We will apply our results to study partition function
zeros and their accumulation locus in the limit $m \to \infty$.  As will be
shown, for fixed $r$, $\{e\}_r$, and $e_g$, in the limit $m \to \infty$, the
zeros of $Z(G_{\{e\}_r,e_g,m;o},q,v)$ are discrete, with no nontrivial
continuous accumulation locus. In contrast, as was already clear from the
results in \cite{nec,neca}, for fixed $r$, $\{e\}_r$, and $e_g$, in the same
limit $m \to \infty$, the zeros of the chromatic polynomial and Potts/Tutte
polynomial for the cyclic hammock chain graph accumulate on a nontrivial
continuous locus.  Hence, in this paper we will focus mainly on the cyclic
chain graphs.

% ===============================================================

\section{Some Background}
\label{background_section}

In this section we give some relevant background on graph theory and the Potts
and Tutte polynomials (see, e.g., \cite{boll,welsh,bbook}).  In general, a
graph $G=(V,E)$ is defined by its set of vertices (sites), $V$, and its set of
edges (bonds), $E$.  We denote the number of vertices in $G$ as $n=n(G)=|V|$
and the number of edges in $G$ as $e(G)=|E|$. The degree $\Delta$ of a vertex
$v_i \in V$ is defined as the number of edges that connect to this vertex.  The
number of connected components of a graph is denoted $k(G)$.  The number of
linearly independent circuits in a graph $G$, called the cyclomatic number of
$G$, is denoted $c(G)$ and satisfies the relation $c(G)=e(G)+k(G)-n(G)$.  A
spanning subgraph of $G$, denoted $G^\prime=(V,E^\prime)$, is a subgraph of
$G=(V,E)$ that has the same vertex set $V$ as $G$ and an edge set $E^\prime
\subseteq E$.

For the open and cyclic hammock chain graphs, the number of vertices and
edges are 
\beq
n(G_{\{e\}_r,e_g,m;o}) =
m\bigg [ \Big (\sum_{j=1}^r e_j \Big ) + e_g - (r-1) \bigg ] + 1 \ , 
\label{nham_open}
\eeq
\beq
n(G_{\{e\}_r,e_g,m;c}) =
m\bigg [ \Big (\sum_{j=1}^r e_j \Big ) + e_g - (r-1) \bigg ]  \ , 
\label{nham_cyclic}
\eeq
and
\beq
e(G_{\{e\}_r,e_g,m;o}) = e(G_{\{e\}_r,e_g,m;c}) =
m\bigg [ \Big (\sum_{j=1}^r e_j \Big ) + e_g \bigg ] \ . 
\label{eham}  
\eeq
The number of linearly independent circuits on the hammock chain graphs 
are thus 
\beq
c(G_{\{e\}_r,e_g,m;o}) = m(r-1) 
\label{c_ham_open}
\eeq
and
\beq
c(G_{\{e\}_r,e_g,m;c}) = m(r-1) + 1 \ . 
\label{c_ham_cyclic}
\eeq

As is evident from Eqs. (\ref{nham_open}) and (\ref{nham_cyclic}), there are 
several different ways to take the limit of infinitely many vertices. These 
include the limits

\begin{enumerate} 

 \item 
$L_{e_g}$: \ $e_g \to \infty$ with $r$, $\{e\}_r$, 
and $m$ fixed; 

\item $L_{e_j}$: \ $e_j \to \infty$ with other $e_\ell \in \{e\}_r$, 
$\ell \ne j$ fixed, and $r$, $e_g$, and $m$ fixed;

\item $L_r$: \ $r \to \infty$ with each $e_j$ in the expanding set 
$\{e\}_r$ finite and $e_g$ and $m$ fixed; 

\item $L_m$: \ $m \to \infty$ with $r$, $\{e\}_r$ and $e_g$ fixed. 

\end{enumerate}
In the limits $L_{e_g}$, $L_{e_j}$, and $L_m$ on $G_{\{e\}_r,e_g,m;BC}$, 
all vertices have bounded degree, whereas in the limit $L_r$, the 
end-vertices of each hammock subgraph in the chain have infinite degree. 

In the context of chromatic polynomials, in \cite{wa3}, Tsai and Shrock applied
their calculations of $P(H_{k,r},q)$ to the $L_r$ limit, while in \cite{nec}
for $r=2$, these authors considered several limits with detailed results for
the $L_m$ limit. For general physical $v$, Ref. \cite{neca} studied the $L_m$
limit, and here also we will apply our new calculations in particular to this
$L_m$ limit, i.e., the limit of an infinitly long hammock chain graph. We will
thus introduce the following compact notation. 
We denote the limit as $m \to \infty$ of $G_{\{e\}_r,e_g,m;BC}$ with 
a given set of values of $r$, $\{e\}_r$, and $e_g$ as
\beq
\lim_{m \to \infty} G_{\{e\}_r,e_g,m;BC} \equiv \{ G_{\{e\}_r,e_g;BC}\} \ .
\label{ginf}
\eeq

We consider cases with both zero and nonzero $e_g$.  Formally, one
could allow one or more of the edges $e_j \in \{e\}_r$ in the hammock
subgraphs to be zero. However, insofar as one is interested in the
application to the Potts spin model, one would take the minimum value
${\rm min}(e_j) \ge 1$ for $e_j \in \{e\}_r$ to maintain the physical
property that spins do not interact with themselves.  Note that if
$e_1=0$ and any other $e_j \in \{e\}_r$ is equal to 1, then the graph
$G_{\{e\}_r,e_g,m;BC}$ contains loop(s), so in the physical context, a
spin would interact with itself. We therefore assume $e_j \ge 1$ for
all $e_j \in \{e\}_r$ here.  The presence of a loop in a graph $G$
also causes the chromatic polynomial of this graph to vanish
identically, since this loop renders it impossible to satisfy the
proper $q$-coloring condition.

A $\Delta$-regular graph $G$ is defined as a graph with the property
that all of its vertices have the same degree,
$\Delta$.  In a $\Delta$-regular graph, one has the
relation $\Delta = 2e(G)/n(G)$.  A hammock chain graph is not, in
general, a $\Delta$-regular graph. However, in the limit of an infinitely
long hammock chain, we may define an effective vertex degree as
\beq
\Delta_{\rm eff}(\{G_{\{e\}_r,e_g}\}) = \lim_{m \to \infty}
\frac{2e(G_{\{e\}_r,e_g,m;BC})}{n(G_{\{e\}_r,e_g,m;BC})} \ , 
\label{Delta_eff_def}
\eeq
where we omit the subscript $BC$ in $\Delta_{\rm
  eff}(\{G_{\{e\}_r,e_g}\})$ because this is independent of the
boundary condition. Note that with cyclic boundary conditions, the ratio
$2e(G_{\{e\}_r,e_g,m;c})/n(G_{\{e\}_r,e_g,m;c})$ is independent of
$m$.  We calculate
\beq
\Delta_{\rm eff}(\{G_{\{e\}_r,e_g}\}) = 
  \frac{2[(\sum_{j=1}^r e_j)+e_g]}{(\sum_{j=1}^r e_j)+e_g-(r-1)} \ .
\label{Delta_eff}
\eeq
In this $m \to \infty$ limit, a fraction $f_2$ of the vertices, namely
the ones in the interior of the ropes on each hammock subgraph and in
interior of the line segments between each hammock subgraph, have
degree 2, while a fraction $f_{he}$ of the vertices, which are hammock
endpoint ($he$) vertices, have degree $r+1$. These fractions are
\beq
f_2 = \frac{(\sum_{j=1}^r e_j) + e_g-(r+1)}
{(\sum_{j=1}^r e_j) + e_g-(r-1)} = 1 - \frac{2}{(\sum_{j=1}^r e_j) + e_g-(r-1)}
\label{f2}
\eeq
and
\beq
f_{he} = \frac{2}{(\sum_{j=1}^r e_j) + e_g-(r-1)} = 1-f_2 \ .
\label{fhe}
\eeq
Thus,
\beq
\Delta_{\rm eff}(\{G_{\{e\}_r,e_g}\}) = 2f_2+(r+1)f_{he} \ . 
\label{Delta_eff_alt}
\eeq
For example, in the $m \to \infty$ limit of the open and cyclic
hammock chain graphs with $r=3$, $\{e_1,e_2,e_3\}=\{2,3,4\}$ and
$e_g=2$ in Fig. \ref{hamfig}, we have $f_2 = 7/9$, $f_{he} = 2/9$ and
$\Delta_{\rm eff}=22/9 = 2.44444..$. We note the limits
\beq
\lim_{(\sum_{j=1}^r e_j) \to \infty} \Delta_{\rm eff}(\{G_{\{e\}_r,e_g}\}) = 2
\quad
  {\rm for \ fixed} \ e_g, \ r 
\label{Delta_ej_inf}
\eeq
and
\beq
\lim_{e_g \to \infty} \Delta_{\rm eff}(\{G_{\{e\}_r,e_g}\}) = 2 \quad
  {\rm for \ fixed} \ \{e\}_r, \ r \ .
\label{Delta_eg_inf}
\eeq
As is evident in Eq. (\ref{Delta_eff}), the value of $\Delta_{\rm
  eff}(\{G_{\{e\}_r,e_g}\})$ is determined by an interplay of the
degree-2 vertices in the interiors of the $r$ ropes and in the line
segments between the hammock subgraphs, on the one hand, and the
degree-$(r+1)$ vertices at the ends of each hammock subgraph.  This is
illustrated, for example, by a family of (open or cyclic) hammock chain
graphs in which all of the edges $e_j \in \{e\}_r$ have the common
value $e_{\rm com}$.  For the $m \to \infty$ limit of a hammock graph
in this family (with either boundary condition),
\beq
e_j = e_{\rm com} \ \Rightarrow \ \Delta_{\rm eff}(\{G_{\{e\}_r,e_g}\}) =
\frac{2(re_{\rm com}+e_g)} {r(e_{\rm com}-1)+ e_g+1} \ . 
\label{Delta_ecom}
\eeq
Now consider the limit $r \to \infty$ with fixed finite $e_g$; 
if $e_{\rm com}=1$, then 
\beq
e_{\rm com}=1 \ \Rightarrow \ \lim_{r \to \infty}
\Delta_{\rm eff}(\{G_{\{e\}_r,e_g}\})  = \infty \ , 
\label{Delta_r_inf_ecom_eq_1}
\eeq
whereas if $e_{\rm com} \ge 2$, then the effective vertex degree remains finite
in this limit:
\beq
e_{\rm com} \ge 2 \ \Rightarrow \ \lim_{r \to \infty}
\Delta_{\rm eff}\{G_{\{e\}_r,e_g}\}) =
\frac{2e_{\rm com}}{e_{\rm com}-1} \ .
\label{Delta_r_inf_ecom_ge_2}
\eeq

The Tutte polynomial of a graph $G$ \cite{tutte47}-\cite{welsh}, 
which we denote as $T(G,x,y)$, is
\beq
T(G,x,y) = \sum_{G' \subseteq G} (x-1)^{k(G')-k(G)}(y-1)^{c(G')} \ . 
\label{t}
\eeq
where $G'$ is a spanning subgraph of $G$.  Since $k(G') - k(G) \ge 0$
and $c(G') \ge 0$, this is, indeed, a polynomial in the given
variables $x$ and $y$. (The first factor can equivalently be written
as $(x-1)^{r(G)-r(G')}$, where $r(G)=n(G)-k(G)$ is the rank of $G$,
since $r(G)-r(G')=k(G')-k(G)$, owing to the property that
$n(G')=n(G)$.)  Although the starting point for the definition of the
Potts model partition function is the Hamiltonian form $Z =
\sum_{\sigma_i} \exp[K \sum_{e_{ij}} \delta_{\sigma_i,\sigma_j}]$, it
can be written as a similar sum over contributions from spanning
subgraphs without explicit reference to the sum over spin
configurations, namely as \cite{fk}
\beq
Z(G,q,v) = \sum_{G' \subseteq G} q^{k(G')}v^{e(G')} \ . 
\label{cluster}
\eeq
Since $k(G') \ge 1$ and $e(G') \ge 0$, this shows that $Z(G,q,v)$ is a
polynomial in $q$ and $v$. Given that $k(G') \ge 1$, $Z(G,q,v)$ always contains
an overall factor of $q$, so one can define a reduced partition function 
$Z_r(G,q,v)$ as
\beq
Z_r(G,q,v) = q^{-1} Z(G,q,v) \ . 
\label{zr}
\eeq
An elementary property of $Z(G,q,v)$ for an arbitrary graph $G$ is that
if $v=0$, then the only nonzero contribution to the sum in Eq. (\ref{cluster})
is from the spanning subgraph with no edges, ${\cal E}_n$, which has
$k(G')=n(G)$ components, so
\beq
v=0 \ \Rightarrow \ Z(G,q,v)=q^{n(G)} \ . 
\label{zv0}
\eeq 
From (\ref{t}) and (\ref{cluster}), it follows that $Z(G,q,v)$ and 
$T(G,x,y)$ are simply related, according to 
\beqs
Z(G,q,v) &=& (x-1)^{k(G)}(y-1)^{n(G)}T(G,x,y) \cr\cr
         &=& (q/v)^{k(G)}v^{n(G)}T(G,x,y) \ , 
\label{ztrel}
\eeqs
where 
\beq
x = 1 + \frac{q}{v} \ , \quad y = v+1  \ . 
\label{xyqv}
\eeq
Note that 
\beq
q=(x-1)(y-1) \ . 
\label{qxy}
\eeq

Given a graph $G$, let us denote $G-e$ as the graph obtained by
deleting the edge $e \in E$ and $G/e$ as the graph obtained by
deleting the edge $e$ and identifying the two vertices that were
connected by this edge of $G$ (known as a contraction of $G$ on $e$.)
An edge that connects a vertex back to itself is called a loop.  An
edge which, if cut, increases the number of disjoint components in
$G$, is called a bridge. A graph consisting of $n$ vertices with no
edges is denoted ${\cal E}_n$, the ``empty'' graph.  The Tutte
polynomial satisfies the properties $T({\cal E}_n,x,y)=1$,
$T(G,x,y)=xT(G-e,x,y)$ if $e$ is a bridge, and $T(G,x,y)=yT(G/e,x,y)$
if $e$ is a loop. If $e$ is neither a bridge nor a loop, then
$T(G,x,y)$ satisfies the deletion-contraction relation (DCR)
\beq
T(G,x,y) = T(G-e,x,y)+T(G/e,x,y) \ . 
\label{tdcr}
\eeq

As mentioned, the special case $v=-1$ corresponds to the $T=0$ Potts
antiferromagnet, and is equivalent to setting $x=1-q$ and $y=0$. Thus,
given that the chromatic polynomial $P(G,q)=Z(G,q,v=-1)$, one has
\beq
Z(G,q,-1) = P(G,q) = (-q)^{k(G)}(-1)^{n(G)}T(G,1-q,0) \ . 
\label{chrompoly}
\eeq
Recall that for the circuit graph with $n$ vertices, $C_n$, 
\beq
T(C_n,x,y) = \frac{x^n -x}{x-1} + y 
\label{tcn}
\eeq
and equivalently
\beq
Z(C_n,q,v) = (q+v)^n + (q-1)v^n \ . 
\label{zcn}
\eeq
Note that $T(C_1,x,y)=y$ and, if $n \ge 2$, then 
$T(C_n,x,y)$ can alternately be expressed as 
$T(C_n,x,y)=(\sum_{j=1}^{n-1} x^j)+y$. 
To save space, we will sometimes use the compact notation 
\beq
c \equiv q-1 = xy-x-y \ . 
\label{c}
\eeq
(The reader should not confuse this $q$-dependent quantity $c$ with the
$G$-dependent cyclotomic number, $c(G)$.) 
It is also useful to define a polynomial of degree $n-2$ in $q$
denoted $D_n$ that appears in various expressions for Tutte and Potts
model polynomials \cite{hs}. Extracting the factor $q(q-1)$ in
$P(C_n,q)$, one can write
\beq
P(C_n,q) =  (q-1)^n + (q-1)(-1)^n = q(q-1)D_n \ ,
\label{pcn}
\eeq
where, for $n \ge 2$,
\beq
D_n = \sum_{j=0}^{n-2}(-1)^j {n-1 \choose j} q^{n-2-j} 
= (q-1)^{n-2} \sum_{j=0}^{r-2} (1-q)^{-j}  \ , 
\label{dk}
\eeq
and ${r \choose s} = r!/[s!(r-s)!]$ is the binomial coefficient. For
$n=1$, $P(C_1,q)=0$ and hence also $D_1 = 0$. Some expressions for
$D_n$ with higher $n$ include $D_2=1$, $D_3=q-2$, $D_4=q^2-3q+3$,
$D_5=(q-2)(q^2-2q+2)$, etc. Since $P(C_n,q)$ is of degree $n$ in $q$,
it follows from the definition (\ref{pcn}) that for $n \ge 2$, $D_n$
is of degree $n-2$ in $q$. Some identities obeyed by $D_n$ are listed
in the Appendix; see also \cite{hs}. 

% ===============================================================

\section{Calculations and Results}
\label{results_section}

Using an iterative application of the deletion-contraction theorem, we have
calculated the Tutte polynomial and equivalent Potts model partition function
for the open and cyclic hammock chain graphs with general $r$, $\{e\}_r$,
$e_g$, and $m$, $G_{\{e\}_r,e_g,m;o}$ and $G_{\{e\}_r,e_g,m;c}$.  For the open
hammock chain graph $G_{\{e\}_r,e_g,m;o}$, we obtain
\beq
T(G_{\{ e \}_r,e_g,m;o},x,y) = (\lambda_{T,0})^m  \ , 
\label{tham_open}
\eeq
where
\beq
\lambda_{T,0} = \frac{x^{e_g}}{q(x-1)^{r-1}} \bigg [ 
\prod_{j=1}^r (x^{e_j}+c) + c \prod_{j=1}^r (x^{e_j}-1) \bigg ] \ . 
\label{tlam0}
\eeq
For the cyclic hammock chain graph $G_{\{e\}_r,e_g,m;c}$ we calculate 
\beq
T(G_{\{e\}_r,e_g,m;c},x,y) =
\frac{1}{x-1}\bigg [ (\lambda_{T,0})^m + c(\lambda_{T,1})^m \bigg ] \ , 
\label{tham_cyclic}
\eeq
where $c$ and $\lambda_{T,0}$ were given in Eqs. (\ref{c}) and (\ref{tlam0}), 
and 
\beq
\lambda_{T,1} = \frac{1}{q(x-1)^{r-1}}\bigg [ 
\prod_{j=1}^r (x^{e_j}+c) - \prod_{j=1}^r (x^{e_j}-1) \bigg ] \ . 
\label{tlam1}
\eeq
These expressions for $\lambda_{T,0}$ and
$\lambda_{T,1}$, and for $T(G_{\{e\}_r,e_g,m;o},x,y)$ and
$T(G_{\{e\}_r,e_g,m;c},x,y)$, are invariant under a permutation of the
$r$ edges in the set $\{e\}_r = \{e_1,e_2,...,e_r\}$. This is a consequence
of the fact that the graphs $G_{\{e\}_r,e_g,m;o}$ and $G_{\{e\}_r,e_g,m;c}$
themselves are invariant under such a permutation.

Since $T(G,x,y)$ is a polynomial in $x$ and $y$ for any graph $G$, it follows
that the expressions in square brackets for $\lambda_{T,0}$ and $\lambda_{T,1}$
each contain a factor of $q(x-1)^{r-1}$ which cancels the prefactor
$1/[q(x-1)^{r-1}]$. The overall prefactor $1/(x-1)$ in Eq. (\ref{tham_cyclic})
is also cancelled. Since $x=1$ $\Rightarrow \ c=-1$, the numerator of
$T(G_{\{e\}_r,e_g,m;c},x,y)$ has the limit 
$(\lambda_{T,0})^m - (\lambda_{T,1})^m$
as $x \to 1$, and hence this cancellation of the prefactor $1/(x-1)$ in
Eq. (\ref{tham_cyclic}) implies that, for arbitrary $\{e\}_r$ and $e_g$,
\beq
\lambda_{T,0;x=1} = \lambda_{T,1;x=1} \ , 
\label{tlam0_eq_tlam1_x1}
\eeq
where the evaluation at $x=1$ is indicated in a subscript. 
We give the general expression for $\lambda_{T,0;x=1}$ below 
in Eq. (\ref{tlam01_x1}). 

One can express $\lambda_{T,0}$ and $\lambda_{T,1}$ equivalently in a form
in which the $r$-fold products in Eqs. (\ref{tlam0}) and (\ref{tlam1}) are
multiplied out.  For this purpose, we introduce some compact notation. 
Certain terms involve $x$ raised to the power 
\beq
\sum_{j=1}^r e_j \equiv (\sum e_j)_r \ , 
\label{sum}
\eeq
where the second term will be taken as shorthand notation for the first.  Other
terms involve sums, each term of which is $x$ raised to a power consisting of
the sum $\sum_{j=1}^r e_j$ with $d$ edges deleted, which will be denoted as
$(\sum e_j)_{r-d}$.  There are ${r \choose d}$ ways to choose the subset
of $d$ edges in $\{e\}_r$ to be deleted, and hence these sums will consist
of ${r \choose d}$ terms.  Expressions for the corresponding Potts model
partition functions involve factors of the form $v^{(\sum e_j)_d} \,
(q+v)^{(\sum e_{j'})_{r-d}}$, where here the notation for the exponent of
$v$ in the factor $v^{(\sum e_j)_d}$ means a sum of the $d$ edges in
$\{e\}_r$ that are deleted in the other factor, $(q+v)^{(\sum
  e_{j'})_{r-d}}$.  We also recall the Heaviside step function, $\theta(x)$,
defined as $\theta(x)=1$ if $x \ge 0$ and $\theta(x)=0$ if $x < 0$.
For the hammock chain graphs $G_{\{e\}_r,e_g,m;BC}$, multiplying out the
$r$-fold products in Eqs. (\ref{tlam0}) and (\ref{tlam1}), we obtain the
following equivalent expressions:
\beqs
\lambda_{T,0} &=& \frac{\theta(r-2)x^{e_g}}{(x-1)^{r-1}} \, \bigg [ 
      x^{(\sum e_j)_r} + 
    \theta(r-3) c D_2\sum_{{r \choose 2} \ {\rm terms}} 
x^{(\sum e_j)_{r-2}}
     \cr\cr
&+& \theta(r-4) c D_3 \sum_{{r \choose 3} \ {\rm terms}} x^{(\sum e_j)_{r-3}} 
+ ... + cD_r \bigg ]
\label{tlam0_alt} 
\eeqs
and
\beqs
\lambda_{T,1} &=& \frac{1}{(x-1)^{r-1}} \bigg [
    \theta(r-2) D_2 \sum_{r \ {\rm terms}}x^{(\sum e_j)_{r-1}}
      + \theta(r-3)D_3 \sum_{{r \choose 2} \ {\rm terms}} x^{(\sum e_j)_{r-2}}
      \cr\cr
 &+& \theta(r-4)D_4 \sum_{{r \choose 3} \ {\rm terms}} x^{(\sum e_j)_{r-3}} 
+ ... + D_{r+1}
      \bigg ] \ , 
\label{tlam1_alt}
\eeqs
In Eqs. (\ref{tlam0_alt}) and (\ref{tlam1_alt}) (and Eqs.
        (\ref{zlam0_alt}) and (\ref{zlam1_alt}) below), we include $D_2$
factors to show the general structure, but note that $D_2=1$.  These
forms for $\lambda_{T,0}$ and $\lambda_{T,1}$ are useful partly
because there are cancellations that occur among certain terms in the
expressions (\ref{tlam0}) and (\ref{tlam1}) involving $r$-fold
products.

For $r=2$, these general results reduce to 
\beq
\lambda_{T,0;r=2} = \frac{x^{e_g}}{x-1} \Big [x^{e_1+e_2} + c \Big ]
\label{tlam0_r2}
\eeq
and
\beq
\lambda_{T,1;r=2} = \frac{1}{x-1} \, [ (x^{e_1}+x^{e_2}) + D_3 ] \ , 
\label{tlam1_r2}
\eeq
in agreement with Eqs. (3.2) and (3.4) in \cite{neca}.
For $r=3$, our general results yield the explicit expressions 
\beq
\lambda_{T,0;r=3} = \frac{x^{e_g}}{(x-1)^2} \Big [
  x^{e_1+e_2+e_3} + c(x^{e_1}+x^{e_2}+x^{e_3}) + cD_3 \Big ]
\label{tlam0_r3}
\eeq
and
\beqs
\lambda_{T,1;r=3} &=& \frac{1}{(x-1)^2} \, \Big [
  (x^{e_1+e_2} + x^{e_1+e_3}+x^{e_2+e_3}) + D_3(x^{e_1}+x^{e_2}+x^{e_3})
  + D_4 \big ] \ , \cr\cr
&& 
\label{tlam1_r3}
\eeqs
while for $r=4$ they yield 
\beqs
\lambda_{T,0;r=4} &=& \frac{x^{e_g}}{(x-1)^3} \Big [
  x^{e_1+e_2+e_3+e_4} \cr\cr
&+& 
c(x^{e_1+e_2}+x^{e_1+e_3}+x^{e_1+e_4}+x^{e_2+e_3}+x^{e_2+e_4}+x^{e_3+e_4})
  \cr\cr
  &+& cD_3 (x^{e_1}+x^{e_2}+x^{e_3}+x^{e_4}) + cD_4 \Big ]
\label{tlam0_r4}
\eeqs
and
\beqs
\lambda_{T,1;r=4} &=& \frac{1}{(x-1)^3} \, \Big [
  (x^{e_1+e_2+e_3} + x^{e_1+e_2+e_4}+x^{e_1+e_3+e_4}+x^{e_2+e_3+e_4})
  \cr\cr
  &+& D_3(x^{e_1+e_2}+x^{e_1+e_3}+x^{e_1+e_4}+x^{e_2+e_3}+x^{e_2+e_4}+
x^{e_3+e_4}) \cr\cr
&+& D_4(x^{e_1}+x^{e_2}+x^{e_3}+x^{e_4}) + D_5 \Big ] \ . 
\label{tlam1_r4}
\eeqs
In a similar manner, one can evaluate our general formulas (\ref{tlam0}) and
(\ref{tlam1}), or equivalently, (\ref{tlam0_alt}) and (\ref{tlam1_alt}), to
obtain explicit expressions for $\lambda_{T,0}$ and $\lambda_{T,1}$ for higher
values of $r$. 

For several applications, including calculations of reliability polynomials and
various graphical quantities such as the number of spanning trees and the
number of connected spanning subgraphs of $G_{\{e\}_r,e_g,m;o}$ and
$G_{\{e\}_r,e_g,m;c}$, one needs to evaluate $T(G_{\{e\}_r,e_g,m;BC},x,y)$ at
$x=1$. Since Eqs. (\ref{tlam0}) and (\ref{tlam1}) have factors $1/(x-1)^{r-1}$,
which are singular at $x=1$, we use an $(r-1)$-fold application of
L'H\^{o}pital's rule for the evaluations of $\lambda_{T,0}$ and $\lambda_{T,1}$
at $x=1$, together with an additional use of L'H\^{o}pital's rule to deal with
the $1/(x-1)$ factor in $T(G_{\{e\}_r,e_g,m;c},x,y)$. For clarity, we sometimes
append the value of $r$ and $x$ in subscripts on $\lambda_{T,0}$ and
$\lambda_{T,1}$.  
We find 
\beq
\lambda_{T,0;x=1} =  \lambda_{T,1;x=1} = \frac{1}{(y-1)} 
\bigg [ \prod_{j=1}^r (e_j + y-1) - \prod_{j=1}^r e_j \bigg ] \ . 
\label{tlam01_x1}
\eeq
As is evident from Eq. (\ref{tlam01_x1}), the second term is always cancelled 
by a term $\prod_{j=1}^r e_j$ arising from the first product in the square
brackets. As examples of the general formula (\ref{tlam01_x1}), if 
$r=2$, then
\beq
\lambda_{T,0;x=1,r=2}= \lambda_{T,1;x=1,r=2} = e_1+e_2 + y-1 \ , 
\label{tlam01_r2_x1}
\eeq
and if $r=3$, 
\beqs
\lambda_{T,0;x=1,r=3} &=& \lambda_{T,1;x=1,r=3} = 
(e_1e_2 + e_1e_3 + e_2e_3) \cr\cr
&+& (y-1)(e_1+e_2+e_3) + (y-1)^2 \ . 
\label{tlam01_r3_x1}
\eeqs

Hence, for the open hammock chain graph, 
\beq
T(G_{\{e\}_r,e_g,m;o},1,y) = (\lambda_{T,0;x=1})^m  \ .
\label{tham_open_x1}
\eeq
For the cyclic hammock chain, given the equality (\ref{tlam0_eq_tlam1_x1}), 
calculating the derivatives $\partial \lambda_{T,0}/{\partial x}$, and 
$\partial \lambda_{T,1}/{\partial x}$ and evaluating them at $x=1$, we have
\beqs
&& T(G_{\{e\}_r,e_g,m;c},1,y) = (\lambda_{T,0})^{m-1} \Bigg [ 
m\bigg ( \frac{\partial \lambda_{T,0}}{\partial x} - 
         \frac{\partial \lambda_{T,1}}{\partial x} \bigg )
+ (y-1)\lambda_{T,0} \Bigg ]\Bigg |_{x=1}  \cr\cr
&=& (\lambda_{T,0;x=1})^{m-1} \bigg [ (me_g+y-1)\lambda_{T,0;x=1}+
m\prod_{j=1}^r e_j \bigg ] \ , 
\label{tham_cyclic_x1}
\eeqs
where in Eqs. (\ref{tham_open_x1}) and (\ref{tham_cyclic_x1}),
$\lambda_{T,0;x=1}$ is given by Eq. (\ref{tlam01_x1}). 
For example, for $r=2$, 
\beq
T(G_{\{e\}_2,e_g,m;o},1,y) = (e_1+e_2+y-1)^m  \ .
\label{tham_open_r2x1}
\eeq
and
\beqs
&& T(G_{\{e\}_{r=2},e_g,m;c},1,y) = 
(e_1+e_2+y-1)^{m-1} \times \cr\cr
&\times& \bigg [ ( me_g + y-1)(e_1+e_2+y-1) + me_1 e_2 \bigg ] \ . 
\label{tham_cyclic_x1_r2}
\eeqs
These $r=2$ results are in agreement with Eqs. (3.1)-(3.4) in \cite{neca},
which used the notation $p=e_1+e_2$. For $r=3$, 
we have $T(G_{\{e\}_{r=3},e_g,m;o},1,y)$ as in Eq. (\ref{tham_open_x1}), and 
\beqs 
&& T(G_{\{e\}_{r=3},e_g,m;c},1,y) = [\lambda_{T,0;x=1,r=3}]^{m-1} \times 
\cr\cr
&\times& \bigg [ 
( me_g + y-1)\lambda_{T,0;x=1,r=3} + me_1 e_2 e_3 \bigg ] \ ,
\label{tham_cyclic_x1_r3}
\eeqs
with $\lambda_{T,0;x=1,r=3}$ given in Eq. (\ref{tlam01_r3_x1}).
Corresponding explicit expressions for $x=1$ can be obtained in a similar
manner for higher values of $r$ from our general formulas for the Tutte
polynomials.

% ======================================================================

\section{Potts Model Partition Functions}
\label{potts_results_section}

Using Eq. (\ref{ztrel}), we calculate the equivalent Potts model partition 
functions
\beq
Z(G_{\{e\}_r,e_g,m;o},q,v) = q(\lambda_{Z,0})^m
\label{zham_open}
\eeq
and
\beq
Z(G_{\{e\}_r,e_g,m;c},q,v) = (\lambda_{Z,0})^m + (q-1)(\lambda_{Z,1})^m \ , 
\label{zham_cyclic}
\eeq
where
\beqs 
\lambda_{Z,0} &=& \frac{(q+v)^{e_g}}{q^r} \bigg [ 
         \prod_{j=1}^r \Big \{ (q+v)^{e_j}+(q-1)v^{e_j} \Big \} 
+ (q-1)\prod_{j=1}^r \Big \{ (q+v)^{e_j}-v^{e_j} \Big \} \bigg ] 
\cr\cr
&&
\label{zlam0}
\eeqs
and
\beqs 
\lambda_{Z,1} &=& \frac{v^{e_g}}{q^r} \bigg [ 
         \prod_{j=1}^r \Big \{ (q+v)^{e_j}+(q-1)v^{e_j} \Big \} 
- \prod_{j=1}^r \Big \{ (q+v)^{e_j}-v^{e_j} \Big \} \bigg ] \ . 
\cr\cr
&&
\label{zlam1}
\eeqs
Note that the $j$'th term in the first product in
Eqs. (\ref{zlam0}) and (\ref{zlam1}) is $Z(C_{e_j},q,v)$, where
$C_n$ is the $n$-vertex circuit graph, so
\beq
\prod_{j=1}^r \Big \{ (q+v)^{e_j}+(q-1)v^{e_j} \Big \} = 
\prod_{j=1}^r Z(C_{e_j},q,v) \ . 
\label{acprod}
\eeq
As a consequence of the 
equality $x=1 \ \Rightarrow \ q=0$, Eq. (\ref{tlam0_eq_tlam1_x1})
implies that, for arbitrary $\{e\}_r$ and $e_g$,
\beq
\lambda_{Z,0} = \lambda_{Z,1} \quad {\rm at} \ q=0 \ . 
\label{zlam0_eq_zlam1_q0}
\eeq
We remark on some special cases. 
For the case of a single hammock graph $H_{\{e\}_r}$, i.e., 
$m=1$ and $e_g=0$, Eq. (\ref{zham_open}) with (\ref{zlam0}) agrees
with (Eq. (2.34) of) \cite{sokal_roots}. For the special case where each
hammock graph is $H_{k,r}$ and $v=-1$, 
Eq. (\ref{zham_open}) with (\ref{zlam0}) agrees with 
(Eq. (3.7) in) \cite{wa3}. For a general hammock chain graph with arbitrary
$m$, our calculations in Eqs. (\ref{zham_open})-(\ref{zlam1}) agree with
the results in Refs. \cite{nec,neca} for the $r=2$ case considered in those
studies. 

  As in the case of the Tutte polynomials, one can express these results in a
form with the $r$-fold products in Eqs. (\ref{zlam0}) and (\ref{zlam1})
multiplied out.  Using the same notation as before, we have 
\beqs
\lambda_{Z,0} &=& \frac{(q+v)^{e_g}}{q^{r-1}} \bigg [ (q+v)^{(\sum e_j)_r} +
\theta(r-3) a \sum_{{r \choose 2} \ {\rm terms}}
v^{(\sum e_j)_2}(q+v)^{(\sum e_{j'})_{r-2}} \cr\cr
&+&
\theta(r-4) a D_3 \sum_{{r \choose 3} \ {\rm terms}}
v^{(\sum e_j)_3}(q+v)^{(\sum e_{j'})_{r-3}} + ... +
a D_r v^{(\sum e_j)_r} \bigg ]
\label{zlam0_alt}
\eeqs
and 
\beqs
\lambda_{Z,1} &=& \frac{v^{e_g}}{q^{r-1}}  \bigg [ 
  \theta(r-2)D_2
  \sum_{r \ {\rm terms}} v^{(\sum e_j)_1} (q+v)^{(\sum e_{j'})_{r-1}} 
\cr\cr
&+& \theta(r-3) D_3 \sum_{{r \choose 2} \ {\rm terms}} v^{(\sum e_j)_2}
(q+v)^{(\sum e_{j'})_{r-2}} \cr\cr
&+& \theta(r-4) D_4 \sum_{{r \choose 3} \ {\rm terms}} v^{(\sum e_j)_3}
(q+v)^{(\sum e_{j'})_{r-3}} + ... + D_{r+1}v^{(\sum e_j)_r} \bigg ] \ ,
\cr\cr
&& 
\label{zlam1_alt}
\eeqs
where in terms of the form $v^{(\sum e_j)_d} (q+v)^{(\sum
  e_{j'})_{r-d}}$, the edges in the sums $(\sum e_j)_d$ and 
$(\sum e_{j'})_{r-d}$ in the
first and second exponents are orthogonal subsets of the full set of edges
$\{e\}_r$.

We display the explicit evaluations of these general formulas for the cases
$r=2$ and $r=3$.  For $r=2$, 
\beq
\lambda_{Z,0} = q^{-1} (q+v)^{e_g} 
\Big [(q+v)^{e_1+e_2} + (q-1)v^{e_1+e_2} \Big ]
\label{zlam0_r2}
\eeq
\beq
\lambda_{Z,1} = q^{-1} v^{e_g} 
\bigg [ v^{e_1}(q+v)^{e_2} + v^{e_2}(q+v)^{e_1} +(q-2)v^{e_1+e_2} \bigg ] \ , 
\label{zlam1_r2}
\eeq
in agreement with Eqs. (4.3) and (4.4) in \cite{neca}. For $r=3$, 
\beqs
\lambda_{Z,0} &=& q^{-2} (q+v)^{e_g} \bigg [ (q+v)^{e_1+e_2+e_3} 
\cr\cr
&+& (q-1) \Big \{ v^{e_1+e_2}(q+v)^{e_3}+v^{e_1+e_3}(q+v)^{e_2} 
+ v^{e_2+e_3}(q+v)^{e_1} \Big \} 
\cr\cr
&+& (q-1)(q-2)v^{e_1+e_2+e_3} \bigg ] 
\label{zlam0_r3}
\eeqs
and
\beqs
\lambda_{Z,1} &=& q^{-2} v^{e_g} \bigg [ \Big \{
v^{e_1}(q+v)^{e_2+e_3} + v^{e_2}(q+v)^{e_1+e_3} + v^{e_3}(q+v)^{e_1+e_2} 
\Big \} 
\cr\cr
&+& (q-2)\Big \{ v^{e_1+e_2}(q+v)^{e_3} + v^{e_1+e_3}(q+v)^{e_2} + 
v^{e_2+e_3}(q+v)^{e_1}  \Big \} \cr\cr
&+& (q^2-3q+3)v^{e_1+e_2+e_3} \bigg ] \ . 
\label{zlam1_r3}
\eeqs
In a similar manner, one can use our general formulas (\ref{zlam0}) and
(\ref{zlam1}), or equivalently, (\ref{zlam0_alt}) and (\ref{zlam1_alt}) to
obtain explicit expressions for $\lambda_{Z,0}$ and $\lambda_{Z,1}$ for higher
values of $r$.

The fact that $Z(G,q,v)$ is a polynomial in $q$ (as well as $v$) for any graph
$G$ implies that the expressions in square brackets in Eqs. (\ref{zlam0_alt})
and (\ref{zlam1_alt}) each contain a factor $q^{r-1}$ that cancels the
prefactor $q^{-(r-1)}$ in these expressions.  By taking the limit $q \to 0$,
one sees that this implies the identities
\beq
1+(q-1)\sum_{j=2}^r {r \choose j} D_j = q^{r-1}
\label{sn_sum_identity}
\eeq
and
\beq
\sum_{j=1}^r {r \choose j} D_{1+j} = q^{r-1} \ .
\label{dn_sum_identity}
\eeq
If $q=1$, then an elementary result for any graph $G$ is that 
\beq
Z(G,q=1,v)= e^{Ke(G)} = (v+1)^{e(G)} \ .
\label{zq1}
\eeq

We remark on the limits of infinite and zero temperature.  The
infinite-temperature limit is $\beta \to 0$, whence also $K = \beta J
\to 0$ and $v \to 0$. In this limit, for an arbitrary graph,
$Z(G,q,v=0)=q^{n(G)}$, as in Eq. (\ref{zv0}). This is evident from the
cluster representation and has the physical significance that in this
limit the Boltzmann factor $e^{-\beta {\cal H}} = 1$, so the partition
function just enumerates the total set of possible spin
configurations, which is $q^{n(G)}$.  It is easy to check 
that the above results satisfy Eq. (\ref{zv0}), since 
\beq
\lambda_{Z,0} = q^{(\sum_{j=1}^r e_j)+e_g-(r-1)} \quad {\rm at} \ v=0
\label{zlam0_v0}
\eeq
and
\beq
\lambda_{Z,1} = 0 \quad {\rm at} \ v=0 \ , 
\label{zlam1_v0}
\eeq
so for the open and cyclic hammock chain graphs, 
\beqs
Z(G_{\{e\}_r,e_g,m;o},q,v=0) &=& q(\lambda_{Z,0})^m = 
q^{m[(\sum_{j=1}^r e_j)+e_g-(r-1)]+1} \cr\cr
&=& q^{n(G_{\{e\}_r,e_g,m;o})}
\label{z_open_v0}
\eeqs
and
\beqs
Z(G_{\{e\}_r,e_g,m;c},q,v=0) &=& (\lambda_{Z,0})^m = 
q^{m[(\sum_{j=1}^r e_j)+e_g-(r-1)]} \cr\cr
&=& q^{n(G_{\{e\}_r,e_g,m;c})} \ .
\label{z_cyclic_v0}
\eeqs

One may also consider the limit of zero temperature. 
If $J > 0$, i.e., the spin-spin interaction is ferromagnetic, then as 
$T \to 0$, i.e., $K=\beta J \to +\infty$ (and thus also $v \to \infty$), 
the spins are all aligned in 
one of the $q$ states of the Potts model, so for an arbitrary graph $G$, 
\beq
\lim_{v \to \infty} Z(G,q,v) = q\exp[K e(G)] \ ,
\label{zvinf}
\eeq
where  $e(G)$ denotes the number of edges of $G$. Again, in addition to the
physical explanation, this is evident from the cluster 
representation (\ref{cluster}).  One checks that our 
calculations for the open and cyclic hammock chain graphs satisfy this general 
result as follows. Here the terms 
$\lambda_{Z,0}$ and $\lambda_{Z,1}$ have the same limiting form: 
\beq
\lambda_{Z,0} \to v^{(\sum_{j=1}^r e_j) + e_g} \to 
\exp\Big [K \Big ((\sum_{j=1}^r e_j) + e_g \Big ) 
\Big ] \quad {\rm as} \ K \to \infty 
\label{zlam0_kinf}
\eeq
and
\beq
\lambda_{Z,1} \to v^{(\sum_{j=1}^r e_j) + e_g} \to 
\exp\Big [K \Big ((\sum_{j=1}^r e_j) + e_g \Big ) 
\Big ] \quad {\rm as} \ K \to \infty \ . 
\label{zlam1_kinf}
\eeq
Hence, using Eq. (\ref{eham}), one sees that for both open and cyclic 
boundary conditions, 
\beq
Z(G_{\{e\}_r,e_g,m;BC},q,v) \to q \exp[K e(G_{\{e\}_r,e_g,m;BC})] \quad 
{\rm as} \ K \to \infty \ , 
\label{zfm_zerotemp}
\eeq
in agreement with the general result (\ref{zvinf}).  If $T\to 0$ with 
$J < 0$, i.e., $K \to -\infty$ and $v \to -1$, one has the identity
$Z(G,q,v=-1)=P(G,q)$, as noted above. 

% ==================================================

\section{Chromatic Polynomials}
\label{chrompoly_section}

In this section we discuss a particularly interesting special case of
the Potts/Tutte polynomials, namely chromatic polynomials (for some 
background, see, e.g., \cite{read_tutte}-\cite{dkt}, \cite{boll,bbook} ). 
Given the relation (\ref{zp}), we make the substitution 
$v=-1$ in our general results for $Z(G_{\{e\}_r,e_g,m;o},q,v)$ and 
$Z(G_{\{e\}_r,e_g,m;c},q,v)$ to obtain the corresponding
chromatic polynomials 
$P(G_{\{e\}_r,e_g,m;o},q)$ and 
$P(G_{\{e\}_r,e_g,m;c},q)$. We have
\beq
P(G_{\{e\}_r,e_g,m;o},q) = q(\lambda_{P,0})^m
\label{p_ham_open}
\eeq
and
\beq
P(G_{\{e\}_r,e_g,m;c},q) = (\lambda_{P,0})^m + (q-1)(\lambda_{P,1})^m \ , 
\label{p_ham_cyclic}
\eeq
where 
\beqs
\lambda_{P,0} &=& (\lambda_{Z,0})|_{v=-1} \cr\cr
&=& \frac{(q-1)^{e_g}}{q^r} \bigg [ 
\Big \{ \prod_{j=1}^r P(C_{e_j},q) \Big \} + 
\frac{1}{(q-1)^{r-1}} \Big \{ \prod_{j=1}^r P(C_{e_j+1},q) \Big \} \bigg ]
\cr\cr
&&
\label{plam0}
\eeqs
and
\beqs
\lambda_{P,1} &=& (\lambda_{Z,1})|_{v=-1}  \cr\cr
&=& \frac{(-1)^{e_g}}{q^r} \bigg [ 
\Big \{ \prod_{j=1}^r P(C_{e_j},q) \Big \} - 
\frac{1}{(q-1)^r} \Big \{ \prod_{j=1}^r P(C_{e_j+1},q) \Big \} \bigg ] \ . 
\cr\cr
&&
\label{plam1}
\eeqs
We will synonymously use the short-form notation $\lambda_{P,0}$ and
$\lambda_{P,1}$ as in Eqs. (\ref{plam0}) and (\ref{plam1}) and, where necessary
for clarity, the long-form notation $\lambda_{P,0;\{e\}_r,e_g}$ and
$\lambda_{P,1;\{e\}_r,e_g}$.  The poles at $q=0$ and $q=1$ in various
prefactors in Eqs. (\ref{plam0}) and (\ref{plam1}) are cancelled by factors
arising from the polynomials $P(C_{e_j},q)$ and $P(C_{e_j+1},q)$ in the
$r$-fold products. This cancellation is rendered explicit by the use of the
factorization in Eq. (\ref{pcn}) with (\ref{dk}) to express $\lambda_{P,0}$ and
$\lambda_{P,1}$ in terms of the $D_{e_j}$ and $D_{e_j+1}$ polynomials, yielding
\beqs
\lambda_{P,0} &=& (q-1)^{e_g} \bigg [ 
      (q-1)^r \Big \{ \prod_{j=1}^r D_{e_j} \Big \} + 
(q-1)\Big \{ \prod_{j=1}^r D_{e_j+1} \Big \} \bigg ] \cr\cr
&&
\label{plam0dk}
\eeqs
and
\beqs
\lambda_{P,1} &=& (-1)^{e_g} \bigg [ 
      (q-1)^r \Big \{ \prod_{j=1}^r D_{e_j} \Big \} - 
\Big \{ \prod_{j=1}^r D_{e_j+1} \Big \} \bigg ] \ . \cr\cr
&&
\label{plam1dk}
\eeqs

The special case $m=1$ and $e_g=0$ corresponds to a single hammock graph,
$H_{\{e\}_r}$. In this special case, one has
\beqs
P(H_{\{e\}_r},q) &=& \frac{1}{q^{r-1}} \, \bigg [ 
\Big \{\prod_{j=1}^r P(C_{e_j},q)\Big \} + 
\frac{1}{(q-1)^{r-1}} \, \Big \{\prod_{j=1}^r P(C_{e_j+1},q)\Big \} \bigg ] 
\cr\cr
&=& q(q-1)\bigg [ (q-1)^{r-1}\Big \{ \prod_{j=1}^r D_{e_j}\Big \} + 
\Big \{\prod_{j=1}^r D_{e_j+1}\Big \} \bigg ] \ . 
\label{pham}
\eeqs

In general, the minimum value of $q$ for which one can perform a proper
$q$-coloring of a graph $G$ is the chromatic number, $\chi(G)$. For the hammock
graphs, $\chi(G_{\{e\}_r,e_g,m,BC})$ is 2 if all circuits on
$G_{\{e\}_r,e_g,m,BC}$ are of even length, and 3 otherwise.  We recall that the
zeros of a chromatic polynomial $P(G,q)$ are called the chromatic zeros of the
graph $G$.  These always include $q=0$ and for a graph with at least one edge,
also $q=1$.  Thus, $P(G_{\{e\}_r,e_g,m,BC},q)$ always contains a factor
$q(q-1)$. If $G_{\{e\}_r,e_g,m,BC}$ contains a circuit of odd length, then
$P(G_{\{e\}_r,e_g,m,BC},q)$ also contains a factor $(q-2)$ and hence vanishes
at $q=2$. It is straightforward to derive this property from the analytic
results given above for $\lambda_{P,0}$ and $\lambda_{P,1}$.  Let us consider
first the case of an open hammock chain and observe that at $q=2$, $D_k=1$ if
$k$ is even and $D_k=0$ if $k$ is odd (cf. Eq. (\ref{dn_q2})).  If all $e_j \in
\{e\}_r$ are even, then $\Big [\prod_{j=1}^r D_{e_j}\Big ]_{q=2}=1$, while
$\Big [\prod_{j=1}^r D_{e_j+1}\Big ]_{q=2}=0$, so, from Eq. (\ref{p_ham_open}),
it follows that $P(G_{\{e\}_r,e_g,m;0},q)=2(\lambda_{P,0})^m$ has the following
values at $q=2$:
\beq
P(G_{\{e\}_r,e_g,m;o},2)= 
\cases{ 2 & if $e_j \in \{e\}_r$ are all even or all odd \cr
        0 & if $\{e\}_r$ contains both even and odd edges} \ . 
\label{pham_open_q2}
\eeq
Next, let us consider the cyclic hammock chain.  Here we observe that at $q=2$,
$P(G_{\{e\}_r,e_g,m;c},2) = (-1)^{e_g} \Big [ (\lambda_{P,0})^m -
(\lambda_{P,1})^m \Big ]$.  Now, with $q=2$, if all $e_j \in \{e\}_r$ are even,
then $\lambda_{P,1}=(-1)^{e_g}$, while if all $e_j \in \{e\}_r$ are odd, then
$\lambda_{P,1}=-(-1)^{e_g}$, and if the set $\{e\}_r$ contains both even and
odd edges, then $\lambda_{P,1}=0$. So for this cyclic hammock chain, 
the chromatic polynomial $P(G_{\{e\}_r,e_g,m;c},q)$ evaluated at $q=2$,
has the following values: 
\beqs
&& {\rm all} \ e_j \ {\rm even} \ \Rightarrow \ P(G_{\{e\}_r,e_g,m;c},2) = 
1 + (-1)^{me_g} = \cases{ 2 & if $me_g$ is even \cr
                          0 & if $me_g$ is odd } \cr\cr
&& 
\label{pham_cyclic_q2_ej_even}
\eeqs
while 
\beqs
&& {\rm all} \ e_j \ {\rm odd} \ \Rightarrow \ P(G_{\{e\}_r,e_g,m;c},2) = 
1 + (-1)^{m(e_g+1)} = \cases{ 2 & if $m(e_g+1)$ is even \cr
                              0 & if $m(e_g+1)$ is odd } \ . \cr\cr
&&
\label{pham_cyclic_q2_ej_odd}
\eeqs
Finally, 
\beq
P(G_{\{e\}_r,e_g,m;c},2) = 0 \ {\rm if} \ \{e\}_r \ {\rm contains \ \ both \ \ 
even \ \ and \ \  odd} \ \ e_j \ . 
\label{pham_cyclic_q2_ejmixed}
\eeq
These analytic results are in accord with the general statement above that
$P(G_{\{e\}_r,e_g,m;o},2)$ and $P(G_{\{e\}_r,e_g,m;c},2)$ vanish whenever the
respective graph contains a circuit of odd length.  For an arbitrary graph $G$,
the property that $P(G,2)=2$ is equivalent to the property that $G$ is
bipartite.

As noted in the Introduction, a feature of particular interest here is that 
for sufficiently large $q$ on a given graph $G$ with finite maximal vertex
degree, the zero-temperature Potts
antiferromagnet has a nonzero entropy per vertex, $S_0=k_B \ln W$, 
or equivalently, a configurational degeneracy per vertex $W$ that is greater
than unity. Here, for a finite graph $G$, $W$ is given by 
\beq
W(G,q) = [P(G,q)]^{1/n(G)} \ , 
\label{wfinite}
\eeq
and thus, in the $n \to \infty$ limit, 
\beq
W(\{G\},q) = \lim_{n(G) \to \infty} P(G,q)^{\frac{1}{n(G)}} \ . 
\label{w}
\eeq
As discussed in \cite{w,a}, for certain values of $q$, denoted $q_s$,
one must take account of the noncommutativity
\beq
\lim_{n(G) \to \infty} \lim_{q \to q_s} [P(G,q)]^{1/n(G)} \ne
\lim_{q \to q_s}\lim_{n(G) \to \infty} [P(G,q)]^{1/n(G)} \ .
\label{wnoncom}
\eeq
The special values of $q_s$ here include $q \in \{0,1\}$, since 
$P(G_{\{e\}_r,e_g,m;BC},q)$ vanishes at these values and also the value 
$q=2$ if $G_{\{e\}_r,e_g,m;BC}$ contains a circuit with an odd number of
edges. 

% =======================================================

\section{Chromatic Zeros, Their Accumulation Locus in the $L_m$ Limit, 
and Ground State Degeneracy $W$}
\label{pzeros_section}

\subsection{General} 

We next discuss the zeros of the chromatic polynomials of the open and cyclic
hammock chain graphs and their continuous accumulation locus in the limit of
infinite chain length, $m \to \infty$, with fixed $r$, $\{e\}_r$, and $e_g$,
denoted $L_m$ limit in Section \ref{background_section}.  For an arbitrary
graph $G$, the coefficients of powers of $q$ in $P(G,q)$ are real (actually,
integers), and hence the set of zeros of $P(G,q)$ in the $q$ plane is invariant
under complex conjugation.  The chromatic zeros of the open hammock chain,
$G_{\{e\}_r,e_g,m;o}$ are discrete and do not form a continuous locus in the
$q$ plane in the $L_m$ limit. In contrast, in this $L_m$ limit, almost all of
the chromatic zeros for the cyclic hammock chain graph $G_{\{e\}_r,e_g,m;c}$
merge to form a continuous locus, denoted ${\cal B}$, consisting of one or more
closed curves in the $q$ plane.  As in \cite{w} and subsequent works, here the
symbol ${\cal B}$ stands for ``boundary''.  Henceforth, in our discussion of
${\cal B}$, we implicitly restrict to the case of cyclic boundary conditions.
Properties of chromatic zeros in the case of a single hammock graph
$H_{\{e\}_r}$ were discussed in \cite{wa3,nec,sokal_roots,bhsw}; in particular,
it was shown in \cite{wa3} that in the $L_r$ limit, these zeros have unbounded
magnitude, they do merge to form a nontrivial continuous locus comprised of
curves, and this locus passes through the origin of the $1/q$ plane.  Here we
restrict to the $L_m$ limit. As a preface to the discussion below, it is useful
to recall that for an arbitrary graph, a chromatic polynomial has the following
zero-free regions on the real axis: (i) $(-\infty,0)$, (ii) (0,1)
\cite{woodall}, and (iii) $(1,\frac{32}{27})$ \cite{jackson,thomassen,dkt}.

The location of the continuum accumulation set of chromatic zeros in the $L_m$
limit is determined by an application of a theorem due to Beraha, Kahane, and
Weiss (BKW) \cite{bkw,bkw_chrom}.  For values of $q$ where $\lambda_{P,0}$ and
$\lambda_{P,1}$ are both nonzero, the locus ${\cal B}$ occurs where
\beq
|\lambda_{P,0}|=|\lambda_{P,1}| \ . 
\label{lameq}
\eeq
This is easily understood, since, as is evident in
Eq. (\ref{p_ham_cyclic}), $P(G_{\{e\}_r,e_g,m;c},q)$ is a sum of
$m$'th powers of $\lambda_{P,0}$ and $\lambda_{P,1}$, so in order to
have a cancellation between these terms yielding a zero in this
chromatic polynomial, it is necessary that they should have the same
magnitude. The situation where $\lambda_{P,0}=\lambda_{P,1}=0$ at a
given $q$ requires a more detailed analysis, as was discussed in
\cite{nec} and is elaborated upon further below.  For moderately large
values of $m$, most chromatic zeros lie close to (or on) the
asymptotic locus ${\cal B}$. For a given limit $\{G\}$, the maximal
point at which ${\cal B}$ intersects the real axis is denoted $q_c =
q_c( \{G\})$.  In passing, we note that an isolated chromatic zero not
on ${\cal B}$ for $\{G_{\{e\}_r,e_g;c}\}$ is the zero at $q=1$. For an
arbitrary graph, since the set of zeros of $P(G,q)$ is invariant under
complex conjugation $q \to q^*$, it follows that the continuous
accumulation locus ${\cal B}$ is also invariant under complex
conjugation: ${\cal B}^* = {\cal B}$. The locus ${\cal B}$ is an
algebraic curve, in the terminology of algebraic geometry.  We recall
that a multiple point (MP) on an algebraic curve is defined as a point
where several branches of the curve intersect
\cite{hartshorne,shafarevich}. If there are $n_i$ branches intersecting
at a multiple point, and if these have different tangents at the
intersection point, then the multiple point is said to have index
$n_i$. In this case, $2n_i$ curves emanate out from the multiple point
(or, equivalently, go into it), forming the $n_i$ branches.  As one
crosses a curve on ${\cal B}$ (away from possible multiple points),
there is a switching between dominant terms $\lambda$; for the cyclic
hammock chain graphs there are just two such $\lambda$ terms
contributing to the chromatic polynomial (\ref{p_ham_cyclic}), namely
$\lambda_{P,0}$ and $\lambda_{P,1}$ in Eqs. (\ref{plam0}) and
(\ref{plam1}). Correspondingly, there is nonanalytic change in the $W$
function, as discussed below.

An important question concerning the Potts antiferromagnet on the $n \to
\infty$ limit of a given family of graphs is the form of this locus ${\cal B}$
and, in particular, the value of $q_c$.  On a graph with vertices of bounded
degree, for sufficiently large (real) $q$, $W(\{ G \},q) \sim q$.  As $q$
decreases through (real) positive values, $W(\{ G \},q)$ first changes its
analytic form as $q$ decreases through $q_c$, crossing the boundary ${\cal B}$.
As in earlier works such as \cite{w,nec}, we denote the region in the complex
$q$ plane that is analytically connected with the line segment $q > q_c$ as
region $R_1$.  In the evaluation of Eq. (\ref{w}), there are actually $n(G)$
different $1/n(G)$'th roots of $P(G,q)$. To ensure that $W(\{G\},q)$ is real
and nonnegative in physical applications, one picks the canonical real positive
root for real positive $q > q_c$.  In contrast, in regions not analytically
connected with $R_1$, there is no canonical choice of root to take in
Eq. (\ref{w}), so one can only determine the magnitude $|W(\{ G \},q)|$
\cite{w}. From our general calculations, for the $L_m$ limit of a cyclic
hammock chain and for $q \in R_1$, where $\lambda_{P,0}$ is dominant, we have
\beq W(\{G_{\{e\}_r,e_g;c},q) =
[\lambda_{P,0;\{e\}_r,e_g}]^{\frac{1}{[(\sum_{j=1}^r e_j)+e_g+(r-1)]}} \quad
{\rm for} \ q \in R_1 \ .
\label{w_cyclic_region1}
\eeq
As noted above, when one crosses a generic point on a curve on the boundary
locus ${\cal B}$ away from any possible multiple points, there is a switch in
the dominant $\lambda$ in the chromatic polynomial (\ref{p_ham_cyclic}). Thus,
crossing ${\cal B}$ from the region $R_1$ away from a multiple point, and
entering another region, denoted generically as $R_x$, the dominant
$\lambda$ in this other region is $\lambda_{P,1}$, so 
\beq
W(\{G_{\{e\}_r,e_g;c}\},q) = 
[\lambda_{P,1;\{e\}_r,e_g}]^{\frac{1}{[(\sum_{j=1}^r e_j)+e_g+(r-1)]}} 
\quad {\rm for} \ q \in R_x \ . 
\label{w_cyclic_regionx}
\eeq
In both Eqs. (\ref{w_cyclic_region1}) and (\ref{w_cyclic_regionx}), we take the
limit $n(G) \to \infty$ first and then $q \to q_s$, where $q_s$ denotes the
special points mentioned in connection with the noncommutativity
(\ref{wnoncom}). 

If one crosses the locus ${\cal B}$ at a multiple point and enters one
of the loop regions connected to this multiple point (e.g. at any of
the three multiple points on the locus ${\cal B}$ for the $r=2$ family
$(\{e_1,e_2\},e_g)=(\{4,4\},0)$ in Fig. \ref{zeros_r2_440}, the
dominant $\lambda$ in the loop region is still $\lambda_{P,0}$, as is
true for the two loop regions connected to the multiple point at $q=2$
on the locus ${\cal B}$ for the $r=3$ family $(\{e_1,e_2,e_3\},e_g)=
(\{2,2,2\},0)$ in Fig. \ref{zeros_r3_2220}. This property was noted
for $r=2$ in \cite{nec}, and our results with higher $r$ again exhibit
the same property.  Thus, as will be discussed below, when one
increases $e_g$ from 0 to nonzero values and these loop regions
connected to multiple points on ${\cal B}$ separate to form bubble
regions, the dominant $\lambda$ in these bubble regions continues to
be $\lambda_{P,1}$. Let us denote the number of regions bounded by
components of ${\cal B}$, by $N_{\rm reg.}$ and the number of disjoint
components on ${\cal B}$ as $N_{\rm comp.}$. For families where the
respective loci ${\cal B}$ do not contain any multiple points, $N_{\rm
  comp.}$ and $N_{\rm reg.}$ are simply related by \cite{nec}
\beq
N_{\rm reg.} = N_{\rm comp.}+1  \quad 
{\rm if \ no \ multiple \ point \ on} \ {\cal B} \ . 
\label{region_component_rel}
\eeq
As was discussed in \cite{nec,wa3}, for a given family of graphs,
e.g., the $r=2$ cyclic hammock graphs with loci ${\cal B}$ not
containing any multiple point(s), there is an upper bound on $N_{\rm
  comp.}$ from the Harnack theorem in algebraic geometry
\cite{shafarevich,wilson_hilbert16}.  However, as shown in \cite{nec},
it is not very restrictive; for the simplest $r=2$ case,
$(\{e_1,e_2\},e_g)=(\{2,2\},1)$, this bound is $N_{\rm comp.} \le 37$,
while the actual number is $N_{\rm comp.}=2$.  As $e_g$ increases in
this $(\{2,2\},e_g)$ family, the Harnack upper bound increases above
37, but $N_{\rm comp.}$ remains equal to 2 and hence is even less
restrictive.  This assessment in \cite{nec} also applies to families
of cyclic hammock graphs with higher $r$. Parenthetically, we note that
the Harnack upper bound is known to be sharp; i.e., there exist curves
(different from those considered in this study) that saturate it
\cite{wilson_hilbert16}.

With regard to the boundary locus ${\cal B}$, first, if $r=1$, then
$G_{e_1,e_g,m;c}$ is just the circuit graph with $n=m(e_1+e_g)$ vertices, so
${\cal B}$ is the unit circle $|q-1|=1$ in the complex $q$ plane and $q_c=2$.
Henceforth, we restrict to the cases where $r \ge 2$.  
In Ref. \cite{nec} a number of general results were proved for ${\cal B}$ for
the $r=2$ case considered there. These included the following:

\begin{itemize} 

\item

(B1) ${\cal B}$ is compact.

\item

(B2) ${\cal B}$ passes through $q=0$ as the most leftward crossing on the 
real $q$ axis.

\item

(B3) ${\cal B}$ encloses regions in the $q$ plane.

\item

(B4) (with the ordering $e_1 \le e_2$) if $e_1 = 1$, then 
${\cal B}$ is the unit circle $|q-1|=1$ independent
of the values of $e_2$ and $e_g$; thus, in this case, $q_c=2$.  The
chromatic zeros of $G_{\{e_1,e_2\},e_g,m}$ (aside from the discrete zero at 
$q=1$) lie exactly on this locus for general $e_2$, $e_g$, and $m$. 

\item

(B5) if $p=e_1+e_2$ is even, then $q_c=2$.

\item

(B6) if $p$ is odd and $e_1 \ge 2$, then $q_c < 2$ and for fixed $(e_1,e_2)$,
$q_c$ increases monotonically as $e_g$ increases, approaching 2 from below
as $e_g \to \infty$. 

\end{itemize}

We find that properties (B1), (B2), and (B3) also hold for general $r$
and are proved in the same way as for the $r=2$ case.  Each region is
bounded by a component of the total boundary ${\cal B}$.  Concerning 
the property (B2), we find that as $q$ increases from negative real
values through $q=0$ to small positive values, the dominant $\lambda$
switches from $\lambda_{P,0}$ to $\lambda_{P,1}$. As in \cite{nec}, we
denote the region that is contiguous with region $R_1$ at $q=0$ as
$R_2$. This region $R_2$ contains the discrete zero at $q=1$. 

There is also a generalization of property (B4). As a step in the proof of
(B4), Ref.  \cite{nec} showed (with the given ordering $e_1 \le e_2$) that if
$e_1=1$, then the chromatic polynomial for the cyclic polygon chain reduces to
\beq
P(G_{\{e\}_{r=2},e_g,m;c},q) = (D_{e_1+e_2})^m P(C_{n_c},q) 
= q(q-1)(D_{1+e_2})^m D_{n_c} \ , 
\label{p_cyclic_r2_reduction}
\eeq
where 
\beq
n_c = m(e_1+e_g)=m(1+e_g) \ . 
\label{nc}
\eeq
The subscript $c$ indicates that in this $e_1=1$ case, there is a global
circuit formed as one traverses a route along the edge $e_1$ of a given $r=2$
hammock (polygon), then the $e_g$ edges, then the $e_1$ edge of the next
polygon, and so forth, returning to the starting vertex after a length of $n_c$
vertices, equal to the same number of edges.  This is the origin of the product
factor $P(C_{n_c},q)$ in the reduction formula
(\ref{p_cyclic_r2_reduction}). The result then follows, since ${\cal B}$ is
determined by the behavior of the zeros of $P(C_{n_c},q)$, and is the
unit circle $|q-1|=1$ in the complex $q$ plane.  Aside from the discrete zero
at $q=1$, the zeros of $P(C_{n_{\rm c}},q)$ lie exactly on this locus.

Here we generalize property (B4) to arbitrary $r$. The case $r=1$ has
already been discussed above and for that case ${\cal B}$ is the unit
circle $|q-1|=1$ for arbitrary $e_1$ and $e_g$. We thus focus on the
cases $r \ge 2$. Again using our standard ordering of edge values $e_1
\le e_2 \le ...\le e_r$, we find that if $e_1=1$, then
$P(G_{\{e\})r,e_g,m;c},q)$ reduces as
\beqs
&& P(G_{\{e\}_r,e_g,m;c},q) = \bigg [ \prod_{j=2}^r D_{1+e_j} \bigg ]^m 
\, P(C_{n_c},q) 
= q(q-1)\bigg [ \prod_{j=2}^r D_{1+e_j} \bigg ]^m \, D_{n_c} \ ,
\cr\cr
&& 
\label{p_cyclic_r_reduction}
\eeqs
where each factor $D_{1+e_j}$ can equivalently be written
$D_{e_1+e_j}$.  Thus, the same result follows for ${\cal B}$ for
general $r$ as for $r=2$ (and $r=1$), namely that in this case, ${\cal
  B}$ is the unit circle $|q-1|=1$ in the complex $q$ plane.

% --------------------------------------------------------------------

\subsection{Determination of $q_c$ }

The locus ${\cal B}$ crosses the real $q$ axis at a left-most point, $q=0$, and
a right-most point, $q_c$. In this section we determine $q_c$. This involves
generalizing the properties (B5) and (B6) presented for the $r=2$ case in
Ref. \cite{nec}.  We find that, depending on the edge parameters $\{e\}_r$ and
$e_g$, $q_c$ is at most 2, so one part of the analysis consists of determining
the conditions under which it is equal to 2. Since if ${\min}(e_j)=1$, then
${\cal B}$ is fully determined to be the circle $|q-1|=1$, and hence $q_c=2$,
the only cases that we need to consider are those with $e_j \ge 2$ for all
$j=1,...,r$.

It is instructive first to review the situation for $r=2$ studied in \cite{nec}
(which used the notation $a_1 = \lambda_{P,0}$ and $a_2 = \lambda_{P,1}$);
here, $\lambda_{P,0}$ and $\lambda_{P,1}$ are :
\beq
\lambda_{P,0;r=2} = (q-1)^{e_g+1}D_{e_1+e_2}
\label{plam0_r2}
\eeq
and
\beq
\lambda_{P,1;r=2} = (-1)^{e_1+e_2+e_g} \, q^{-1} \,
\Big [ (1-q)^{e_1}+(1-q)^{e_2}+ q-2 \Big ] \ . 
\label{plam1_r2}
\eeq
In this $r=2$ case, if $q=2$ (as indicated in the subscript), then  
\beq
\lambda_{P,0;\{e\}_{r=2},e_g,q=2} = \cases{ 1 &if $e_1+e_2$ is even \cr
                                            0 &if $e_1+e_2$ is odd \cr}
\label{plam0_r2_q2}
\eeq
and 
\beq
\lambda_{P,1;\{e\}_{r=2},e_g,q=2} = 
\cases{ (-1)^{e_g}  &if $e_1$ and $e_2$ are both even \cr
       -(-1)^{e_g}  &if $e_1$ and $e_2$ are both odd  \cr
                  0 &if $e_1+e_2$ is odd \cr} \ . 
\label{plam1_r2_q2}
\eeq
In analyzing whether a special (sp) point denoted $q_{sp}$ is on ${\cal B}$,
the criterion $|\lambda_{P,0}|=|\lambda_{P,1}|$ suffices if these are
both nonzero at $q_{sp}$, but if they are both zero at this point, then
one must examine the Taylor series expansions of $\lambda_{P,0}$ and
$\lambda_{P,1}$ around this point. If the respective leading terms in
these expansions for $\lambda_{P,0}$ and $\lambda_{P,1}$ occur at the
same order $O((q-q_{sp})^t)$ and have coefficients of $(q-q_{sp})^t$ that
are equal in magnitude, then $q_{sp}$ is on the locus ${\cal B}$. If they
occur at different orders, or at the same order with
coefficients of different magnitudes, then as $m \to \infty$ and $q
\to 2$, the contribution of the smaller term in the set 
$\{ (\lambda_{P,0})^m, \ (\lambda_{P,1})^m \}$ becomes negligibly
small compared with the contribution from the larger one, precluding
any possibility of a cancellation that could lead to a zero in the
chromatic polynomial and hence implying that the point $q=2$ is not on
${\cal B}$, which is the continuous accumulation set of these zeros.
As indicated in Eqs. (\ref{plam0_r2_q2}
) and (\ref{plam1_r2_q2}), in
the $r=2$ case considered in \cite{nec}, if one of the set $\{e_1,e_2
\}$ is odd and the other is even, then $\lambda_{P,0}=\lambda_{P,1}=0$
at $q=2$. The Taylor series expansions for this case are \cite{nec}
\beq
\lambda_{P,0;\{e\}_{r=2},e_g} = \frac{1}{2}(e_1+e_2-1)(q-2) + ... \quad
       {\rm as} \ q \to 0 
\label{alpha_plam0_r2_q2}
\eeq
and
\beq
\lambda_{P,1;\{e\}_{r=2},e_g} = 
\frac{(-1)^{e_g+1}}{2}(1+e_{\rm even} - e_{\rm odd})(q-2) + ... \ , 
\label{alpha_plam1_r2_q2}
\eeq
where, for clarity, we indicate the parameters as subscripts; 
$e_{\rm even}$ and $e_{\rm odd}$ refer to whichever of the set 
$\{e_1,e_2\}$ are even and odd; and $+ ...$ indicates higher-order terms in
the respective Taylor series expansions. As discussed in \cite{nec}, since
\beq
\frac{|1+e_{\rm even}-e_{\rm odd}|}{e_1+e_2-1} < 1 
\label{alpha_ineq_r2_q2}
\eeq
for the relevant case where ${\rm min}(e_j) \ge 2$, it follows that
for this situation in which $e_1+e_2$ is odd, the point $q=2$ is not
on ${\cal B}$. Since $|\lambda_{P,0}| > |\lambda_{P,1}|$ for (real) $q
> 2$, it also follows that $q_c < 2$ in this case; the actual value of
$q_c$ depends on $e_1$, $e_2$, and $e_g$. Illustrative plots showing
${\cal B}$ for $e_1=2$, $e_2=3$, and $0 \le e_g \le 3$ were included
as Figs. 3(a,b) and 4(a,b) in \cite{nec}.  For $e_1=2$, $e_2=3$, and
$e_g=0, \ 1, \ 2$, $q_c$ has the respective values (given to the indicated
floating-point precision) 1.453398, \ 1.569840, \ 1.636883, which are
the unique real solutions to the respective equations
\beq
q_c((\{2,3\},0)): \quad q^3-3q^2+5q-4=0 \ ,
\label{eq_qc_r2_230}
\eeq
\beq
q_c((\{2,3\},1)): \quad q^3-4q^2+7q-5=0 \ ,
\label{eq_qc_r2_231}
\eeq
and
\beq
q_c((\{2,3\},2)): \quad q^5-5q^4+11q^3-13q^2+9q-4=0 \ .
\label{eq_qc_r2_232}
\eeq
The property that (in this case of odd
$e_1+e_2$) $q_c$ increases toward 2 from below as $e_g \to \infty$
follows because $\lambda_{P,0}$ contains a factor $(q-1)^{e_g+1}$,
while $|\lambda_{P,1}|$ is independent of $e_g$, so in order for
$|\lambda_{P,0}|$ to be equal to $|\lambda_{P,1}|$ as must be true at
$q_c$, it is necessary that $q \nearrow 2$ as $e_g \to \infty$.

We now generalize this analysis to arbitrary $r$.  For this purpose, we first
observe that at $q=2$, $\lambda_{P,0}$ and $\lambda_{P,1}$ take the respective
forms 
\beqs 
\lambda_{P,0;q=2} &=& \frac{1}{2^r}\bigg [ 
         \prod_{j=1}^r \Big \{ 1 + (-1)^{e_j} \Big \} 
+ \prod_{j=1}^r \Big \{ 1 - (-1)^{e_j} \Big \} \bigg ] 
\cr\cr
&&
\label{plam0_q2}
\eeqs
and
\beqs 
\lambda_{P,1;q=2} &=& \frac{(-1)^{e_g}}{2^r} \bigg [ 
         \prod_{j=1}^r \Big \{ 1 + (-1)^{e_j} \Big \} 
- \prod_{j=1}^r \Big \{ 1 - (-1)^{e_j} \Big \} \bigg ] \ . 
\cr\cr
&&
\label{plam1_q2}
\eeqs
If all $e_j \in \{e\}_r$ are even, then the first $r$-fold product in
Eqs. (\ref{plam0_q2}) and (\ref{plam1_q2}) is $2^r$, while the second
product is zero.  If all $e_j \in \{e\}_r$ are odd, then the first
$r$-fold product in Eqs. (\ref{plam0_q2}) and (\ref{plam1_q2}) is zero
while the second is $2^r$, Therefore if all $e_j \in \{e\}_r$ are
even, then $\lambda_{P,0;q=2}=1$ and $\lambda_{P,1;q=2}=(-1)^{e_j}$,
while if all $e_j \in \{e\}_r$ are odd, then $\lambda_{P,0;q=2}=1$ and
$\lambda_{P,1;q=2}=-(-1)^{e_g}$.  This shows that if
(i) all $e_j \in \{e\}_r$ are even or (ii) all $e_j \in \{e\}_r$ are
odd, then $|\lambda_{P,0;q=2}| = |\lambda_{P,1;q=2}| \ne 0$, so the
locus ${\cal B}$ contains the point $q=2$, and $q_c=2$.

If the set of edge values $\{e\}_r$ contains both even and odd
members, then both the first and second $r$-fold products in
Eqs. (\ref{plam0_q2}) and (\ref{plam1_q2}) vanish, so
$\lambda_{P,0;q=2}=\lambda_{P,1;q=2}=0$, and it is necessary to analyze
the Taylor series expansions of these terms around the point $q=2$ to
determine if the point $q=2$ is on the locus ${\cal B}$. To do this,
it is convenient to use a different convention for the ordering of the
edge values $e_j \in \{e\}_r$ than the one used in the rest of this
paper, namely with $e_1 \le e_2 \le ... \le e_r$; instead of that
convention, here, if there are $\ell$ odd edges and $r-\ell$ even edges in
the set $\{e\}_r$, we order them so that the first $\ell$ edges are odd,
and the remaining $r-\ell$ edges are even.  We define
\beq
\epsilon \equiv q-2 \ .
\label{eps}
\eeq
Substituting $q=2+\epsilon$ into the general expressions (\ref{plam0})
and (\ref{plam1}) and expanding in powers of $\epsilon$, we obtain the
leading-order results
\beq
\lambda_{P,0} = \frac{1}{2^r}\bigg [ 2^{r-\ell}
  \Big \{ \prod_{j=1}^\ell (e_j-1) \Big \} \epsilon^\ell +
2^\ell \Big ( \prod_{j=\ell+1}^r e_j \Big ) \epsilon^{r-\ell} \bigg ] + ... 
\label{plam0_taylor}
\eeq
and
\beq
\lambda_{P,1} = \frac{(-1)^{e_g}}{2^r}\bigg [ 2^{r-\ell}
  \Big \{ \prod_{j=1}^\ell (e_j-1) \Big \} \epsilon^\ell -
2^\ell \Big ( \prod_{j=\ell+1}^r e_j \Big ) \epsilon^{r-\ell} \bigg ] + ... 
\label{plam1_taylor}
\eeq
where, as before, the $+ ...$ indicates higher-order terms in the small
quantity $\epsilon$. 
If $\ell < r-\ell$, i.e., $\ell < r/2$, then the first term in
the square brackets in Eqs. (\ref{plam0_taylor}) dominates over the second
term, and similarly in Eq. (\ref{plam1_taylor}). We then have, for the
leading-order behavior as $q \to 2$, 
\beq
\lambda_{P,0} =
2^{-\ell} \Big \{ \prod_{j=1}^\ell (e_j-1) \Big \} \epsilon^\ell + ...
\label{plam0_taylor_dominant_smaller_ell}
\eeq
and
\beq
\lambda_{P,1} =
2^{-\ell} (-1)^{e_g} \Big \{ \prod_{j=1}^\ell (e_j-1) \Big \} \epsilon^\ell
+ ... 
\label{plam1_taylor_dominant_smaller_ell}
\eeq
Since the leading contributions in the respective Taylor series
expansions of $\lambda_{P,0}$ and $\lambda_{P,1}$ around $q=2$ are
equal in magnitude, the point $q=2$ is on the locus ${\cal B}$ and
$q_c=2$. If $\ell > r-\ell$, i.e., $\ell > r/2$, then the second term in
the square brackets in Eqs. (\ref{plam0_taylor}) dominates over the first
term, and similarly in Eq. (\ref{plam1_taylor}). Hence, the 
leading-order behavior as $q \to 2$ is
\beq
\lambda_{P,0} =
2^{-(r-\ell)} \Big ( \prod_{j=\ell+1}^r e_j \Big ) \epsilon^{r-\ell} + ...
\label{plam0_taylor_dominant_larger_ell}
\eeq
and
\beq
\lambda_{P,1} =
2^{-(r-\ell)} (-1)^{e_g+1}
\Big ( \prod_{j=\ell+1}^r e_j \Big ) \epsilon^{r-\ell} + ...
\label{plam1_taylor_dominant_larger_ell}
\eeq
Again, the leading contributions in the respective Taylor series
expansions of $\lambda_{P,0}$ and $\lambda_{P,1}$ around $q=2$ are
equal in magnitude, so the point $q=2$ is on the locus ${\cal B}$ and $q_c=2$.

Finally, if $\ell = r-\ell$, i.e., $r=2\ell$, then both terms in the
square brackets in Eqs. (\ref{plam0_taylor}) for $\lambda_{P,0}$ are
of equal order in the small quantity $\epsilon$, as are both terms in
Eq. (\ref{plam1_taylor}) for $\lambda_{P,1}$. Hence, in this case, the
leading-order behavior as $q \to 2$ is
\beq
\lambda_{P,0} =
2^{-\ell} \epsilon^\ell \bigg [ \prod_{j=1}^\ell (e_j-1)
  + \prod_{j=\ell+1}^{2\ell} e_j \bigg ] + ...
\label{plam0_taylor_ellhalf}
\eeq
and
\beq
\lambda_{P,1} =
2^{-\ell} (-1)^{e_g}\epsilon^\ell \bigg [ \prod_{j=1}^\ell (e_j-1)
  - \prod_{j=\ell+1}^{2\ell} e_j \bigg ] + ...
\label{plam1_taylor_ellhalf}
\eeq
Thus, in this case with $r=2\ell$, the magnitudes of $\lambda_{P,0}$
and $\lambda_{P,1}$ can be expressed concisely as $\lambda_{P,0} =
|\lambda_{P,0}| = 2^{-\ell} \epsilon^\ell [ A + B]$ and
$|\lambda_{P,1}|=2^{-\ell} \epsilon^\ell |A-B|$, where $A \equiv
\prod_{j=1}^\ell (e_j-1)$ and $B \equiv \prod_{j=\ell+1}^{2\ell} e_j$.
Since we have excluded zero values for edge(s) $e_j \in \{e\}_r$
(which would lead to spins interacting with themselves in the physical
Potts model context), it follows that $A+B > |A-B|$, so
$|\lambda_{P,0}| \ne |\lambda_{P,1}|$.  Hence, if $r=2\ell$, i.e., $r$
is even and half of the edges in $\{e\}_r$ are even while the other
half are odd, then the point $q=2$ is not on ${\cal B}$. Furthermore,
$\lambda_{P,0} > |\lambda_{P,1}|$ at $q=2$, which is the same
inequality that holds in the region $R_1$ extending beyond the point
$q=2$ to large positive $q$. Hence, $q_c < 2$ for this case. The
actual value of $q_c$ in this case depends on the values of the edges
$\{e\}_r$ and $e_g$, as was discussed above for $r=2$. As examples
for higher $r$, we take $r=4$ and $\{e_1,e_2,e_3,e_4\}=\{2,2,3,3\}$.
For $e_g=0$, $q_c=1.586793$, while for $e_g=1$, $q_c=1.669201$, which
are, respectively, the unique real solutions to the equations
\beq
q_c((\{2,2,3,3\},0)): \quad q^5-6q^4+19q^3-36q^2+37q-16=0
\label{eq_qc_r4_22330}
\eeq
and
\beq
q_c((\{2,2,3,3\},1)): \quad q^5-7q^4+23q^3-42q^2+41q-17=0 \ .
\label{eq_qc_r4_22331}
\eeq
For the same reason given in the
discussion for $r=2$, as $e_g \to \infty$ for this case of $r=2\ell$
with $\ell$ even and $\ell$ odd edges in $\{e\}_r$, $q_c$ increases
monotonically, approaching 2 from below in this limit. 
A corollary of this general-$r$ analysis is that the point $q=2$ is on
the locus ${\cal B}$ and $q_c=2$ for the $m \to \infty$ limit of all
cyclic hammock chain graphs $G_{\{e\}_r,e_g,m;c}$ with odd $r$, while
$q=2$ is not on ${\cal B}$ and $q_c < 2$ if $r$ is even.  This
corollary follows since the equation $\ell=r-\ell$, i.e., $\ell=r/2$
has no integral solution if $r$ is odd.

As an explicit example of these general-$r$ results, we consider the case
$r=3$. Our general results above yield the following expressions for
$\lambda_{P,0}$ and $\lambda_{P,1}$ at $q=2$ (with the values of $r$ and $q$
again indicated with subscripts):
\beq
\lambda_{P,0;\{e\}_{r=3},e_g;q=2} = \frac{1}{4}\bigg [ 
1 + \Big \{(-1)^{e_1+e_2} + (-1)^{e_1+e_3} + (-1)^{e_2+e_3} \Big \} \bigg ]
\label{plam0_r3_q2}
\eeq
and
\beqs
&& \lambda_{P,1;\{e\}_{r=3},e_g;q=2} = \frac{(-1)^{e_g}}{4}\bigg [ 
(-1)^{e_1+e_2+e_3} + \Big \{ (-1)^{e_1} + (-1)^{e_2} + (-1)^{e_3} \Big \} 
\bigg ] \ . \cr\cr
&& 
\label{plam1_r3_q2}
\eeqs
Hence, 

\begin{itemize}

\item (i) If all of the $e_j$, $1 \le j \le 3$, are even, then 
$\lambda_{P,0;\{e\}_{r=3},e_g;q=2} = 1$ and 
$\lambda_{P,1;\{e\}_{r=3},e_g;q=2} = (-1)^{e_g}$. 

\item (ii) If two of the $e_j$ are even and the third is odd, then 
$\lambda_{P,0;\{e\}_{r=3},e_g;q=2} = \lambda_{P,1;\{e\}_{r=3},e_g;q=2} = 0$.

\item (iii) If one of the $e_j$ is even and two are odd, then 
$\lambda_{P,0;\{e\}_{r=3},e_g;q=2} = \lambda_{P,1;\{e\}_{r=3},e_g;q=2} = 0$.

\item (iv) If all of the $e_j$ are odd, then 
$\lambda_{P,0;\{e\}_{r=3},e_g;q=2} = 1$ and 
$\lambda_{P,1;\{e\}_{r=3},e_g;q=2} = -(-1)^{e_g}$.

\end{itemize}

Thus, in accord with our general-$r$ analysis above, if all of the
edge values in $\{e\}_{r=3} \equiv \{e_1,e_2,e_3\}$) are even, or all
of these edges are odd, then $|\lambda_{P,0;\{e\}_{r=3},e_g;q=2}| =
|\lambda_{P,1;\{e\}_{r=3},e_g;q=2}| \ne 0$, so the point $q=2$ is on
the locus ${\cal B}$ and $q_c=2$. The remaining two cases in which
the $\{e_1,e_2,e_3\}$ set contains both even and odd edges are analyzed
via a Taylor series expansion.
Since that $\lambda_{Z,0}$ and $\lambda_{Z,1}$ are symmetric
functions of the $r$ edge values in the set $\{e\}_r$, 
we will use the notation $e_{\rm odd}$ to denote the single odd edge value in
case (ii) and $e_{\rm even}$ to denote the single even edge value in case
(iii).  We have 
\beq
{\rm case} \ (ii): \quad  \lambda_{P,0;\{e_j \}_{r=3},e_g} = 
\frac{1}{2}(e_{\rm odd}-1)(q-2) + ... 
\label{alpha_plam0_r3_q2_even_even_odd}
\eeq
\beq
{\rm case} \ (ii): \quad 
\lambda_{P,1;\{e_j \}_{r=3},e_g} =
\frac{1}{2} (-1)^{e_g}(e_{\rm odd}-1)(q-2) + ...  
\label{alpha_plam1_r3_q2_even_even_odd}
\eeq
and 
\beq
{\rm case} \ (iii): \quad  \lambda_{P,0;\{e_j \}_{r=3},e_g} = 
\frac{1}{2}e_{\rm even}(q-2) + ... 
\label{alpha_plam0_r3_q2_even_odd_odd}
\eeq
\beq
{\rm case} \ (iii): \quad \lambda_{P,1;\{e_j \}_{r=3},e_g} = 
\frac{1}{2}(-1)^{e_g+1}e_{\rm even}(q-2) + ... 
\label{alpha_plam1_r3_q2_even_odd_odd}
\eeq
in agreement with our general-$r$ results in Eqs.
(\ref{plam0_taylor_dominant_smaller_ell})-(\ref{plam1_taylor_dominant_larger_ell}).
Hence, in both cases (ii) and (iii) the leading terms in the Taylor
series expansions of $\lambda_{P,0}$ and $\lambda_{P,1}$ are equal in
magnitude, and therefore in all four case (i)-(iv) for the family of
cyclic hammock graphs with $r=3$, the point $q=2$ is on the locus
${\cal B}$ and $q_c=2$, in agreement with our general-$r$ analysis
above.

% ============================================================

\section{Loci ${\cal B}$ for Chromatic Zeros} 

\begin{figure}[htbp]
  \begin{center}
    \includegraphics[height=7cm,width=7cm]{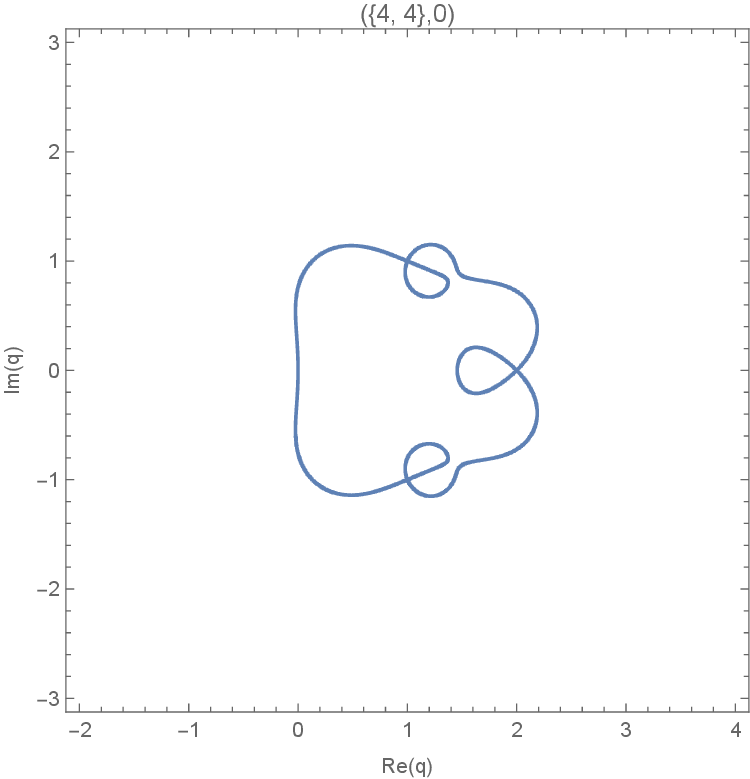}
  \end{center}
\caption{Locus ${\cal B}$ for $r=2$ , $(\{e_1,e_2\},e_g)=(\{4,4\},0)$.}
\label{zeros_r2_440}
\end{figure}

\begin{figure}[htbp]
  \begin{center}
    \includegraphics[height=7cm,width=7cm]{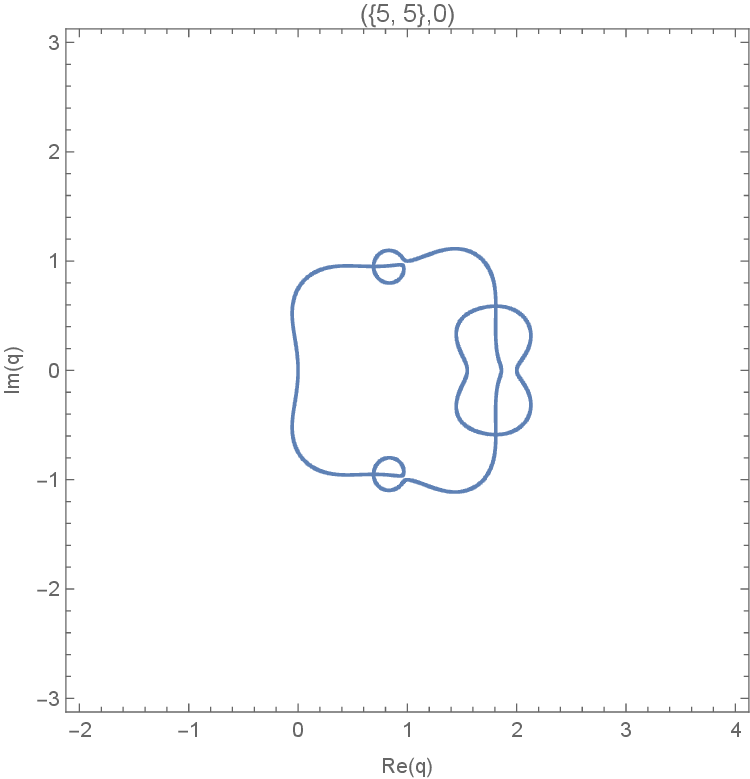}
  \end{center}
\caption{Locus ${\cal B}$ for $r=2$, $(\{e_1,e_2\},e_g)=(\{5,5\},0)$.}
\label{zeros_r2_550}
\end{figure}

\begin{figure}[htbp]
  \begin{center}
    \includegraphics[height=7cm,width=7cm]{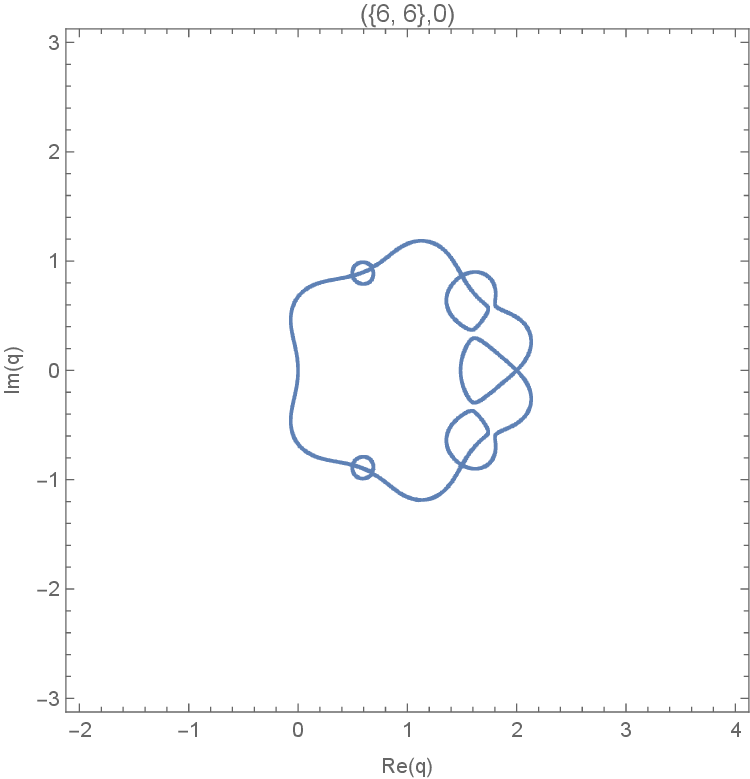}
  \end{center}
\caption{Locus ${\cal B}$ for $r=2$, $(\{e_1,e_2\},e_g)=(\{6,6\},0)$.}
\label{zeros_r2_660}
\end{figure}

\begin{figure}[htbp]
  \begin{center}
    \includegraphics[height=7cm,width=7cm]{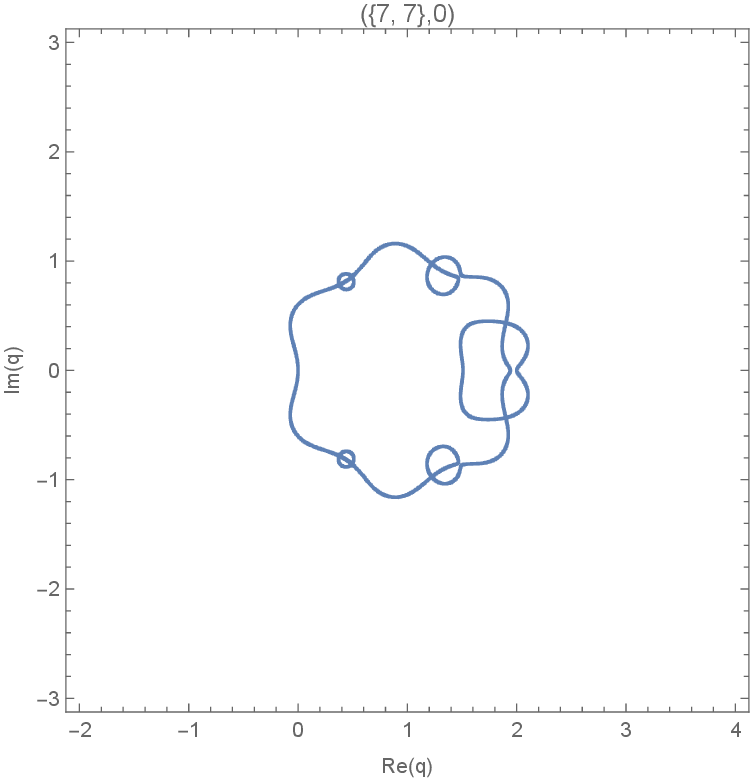}
  \end{center}
\caption{Locus ${\cal B}$ for $r=2$, $(\{e_1,e_2\},e_g)=(\{7,7\},0)$.}
\label{zeros_r2_770}
\end{figure}

\begin{figure}[htbp]
  \begin{center}
    \includegraphics[height=7cm,width=7cm]{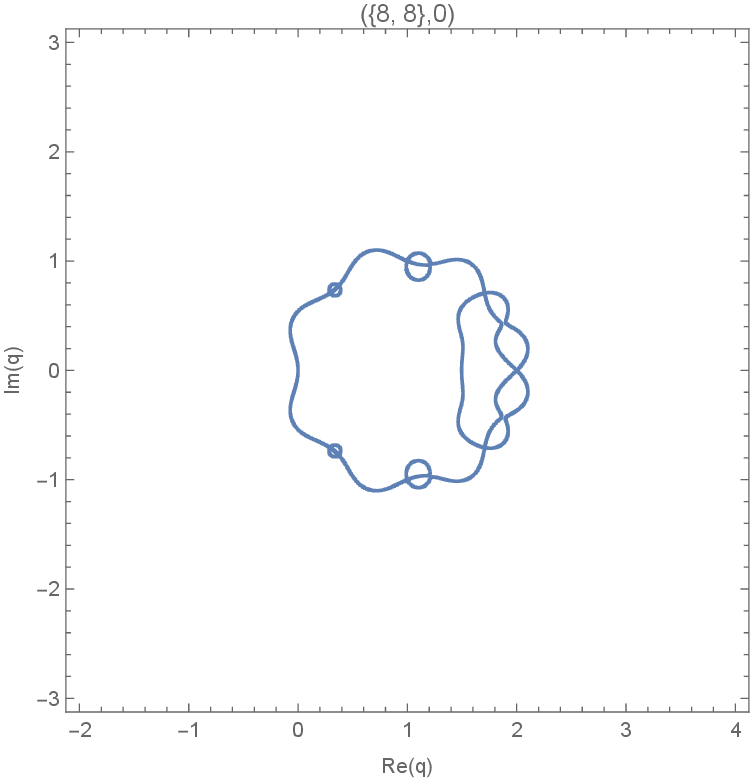}
  \end{center}
\caption{Locus ${\cal B}$ for $r=2$, $(\{e_1,e_2\},e_g)=(\{8,8\},0)$.}
\label{zeros_r2_880}
\end{figure}

\begin{figure}[htbp]
  \begin{center}
    \includegraphics[height=7cm,width=7cm]{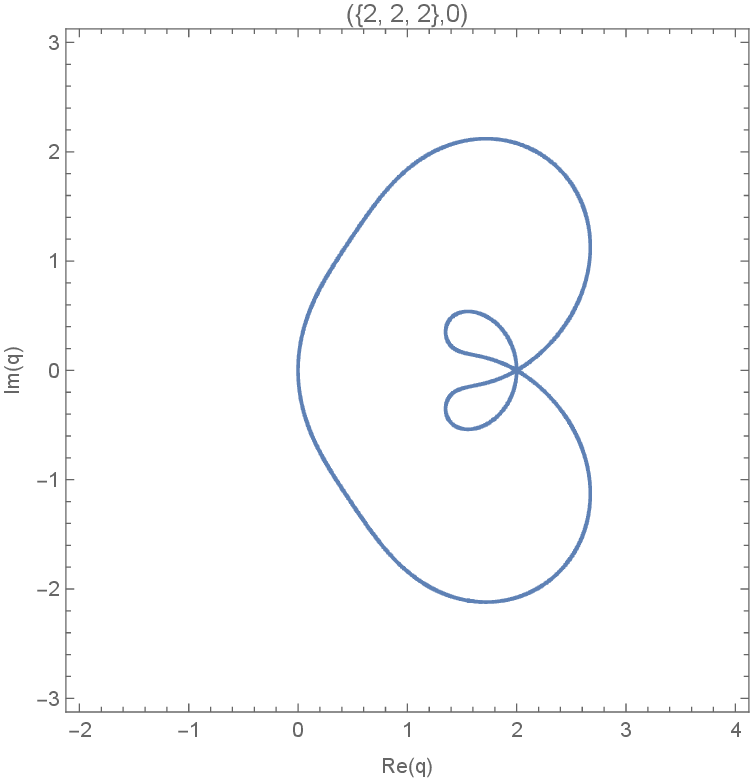}
  \end{center}
\caption{Locus ${\cal B}$ for $r=3$, $(\{e_1,e_2,e_3\},e_g)=(\{2,2,2\},0)$.}
\label{zeros_r3_2220}
\end{figure}

\begin{figure}[htbp]
  \begin{center}
    \includegraphics[height=7cm,width=7cm]{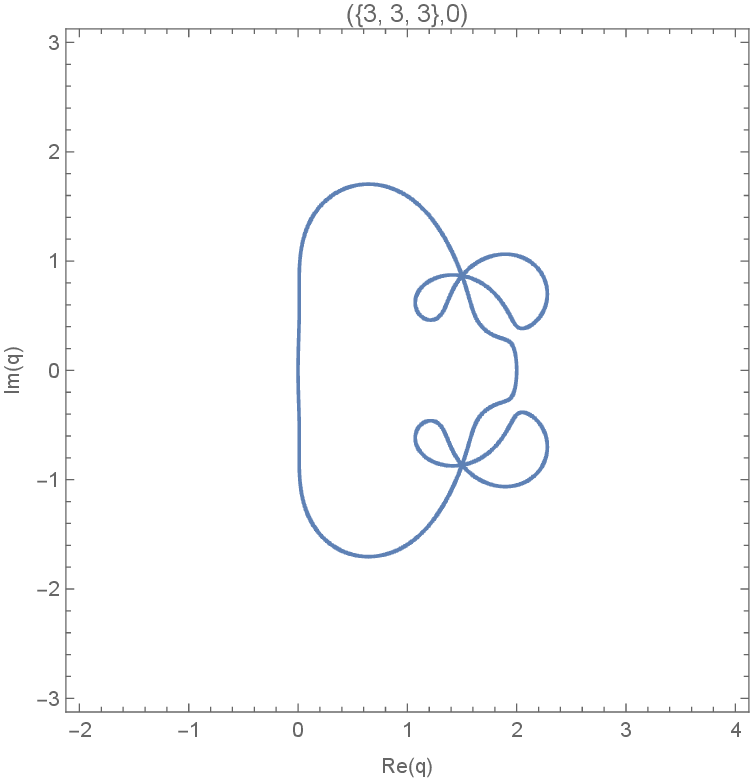}
  \end{center}
\caption{Locus ${\cal B}$ for $r=3$, $(\{e_1,e_2,e_3\},e_g)=(\{3,3,3\},0)$.}
\label{zeros_r3_3330}
\end{figure}

\begin{figure}[htbp]
  \begin{center}
    \includegraphics[height=7cm,width=7cm]{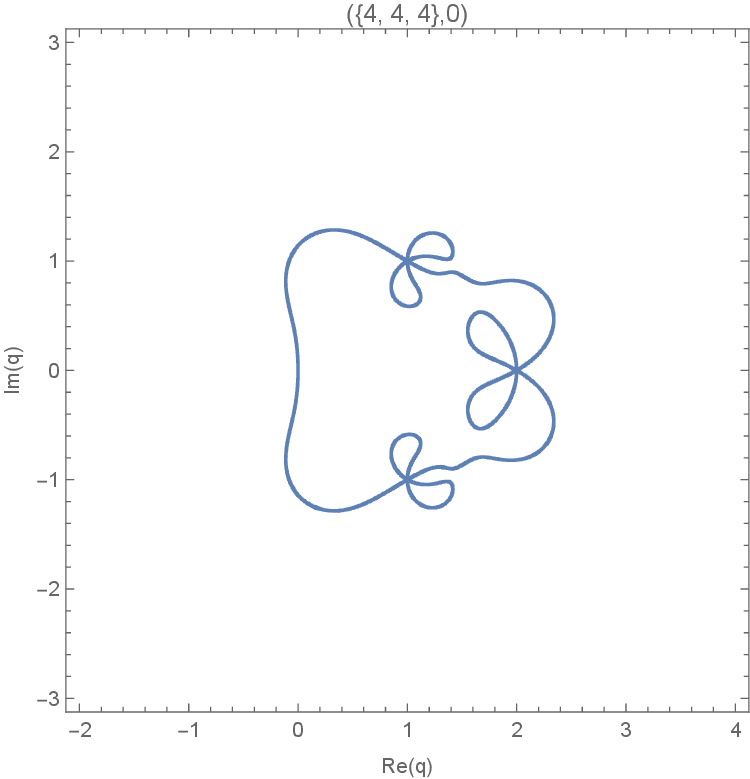}
  \end{center}
\caption{Locus ${\cal B}$ for $r=3$, $(\{e_1,e_2,e_3\},e_g)=(\{4,4,4\},0)$.}
\label{zeros_r3_4440}
\end{figure}

\begin{figure}[htbp]
  \begin{center}
    \includegraphics[height=7cm,width=7cm]{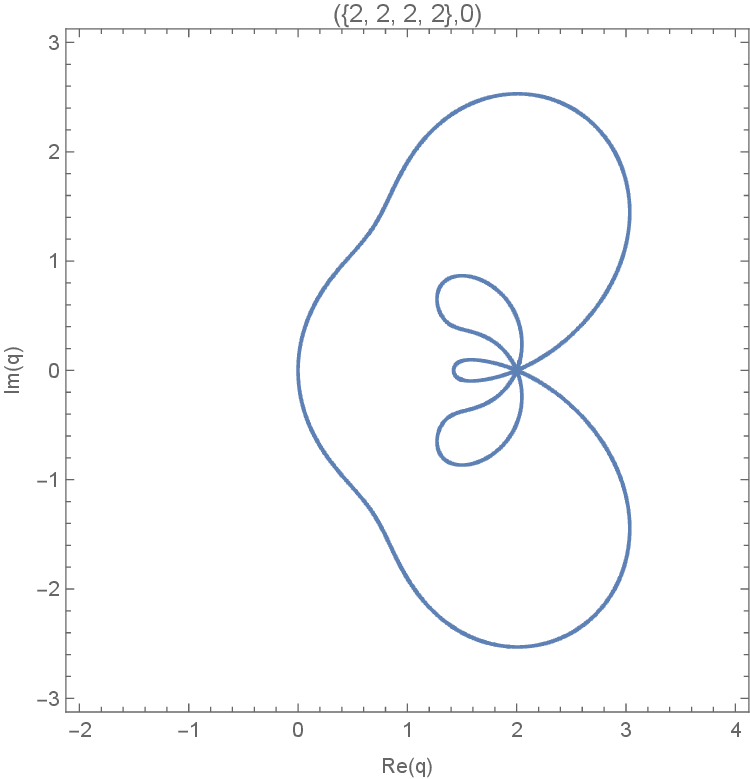}
  \end{center}
\caption{Locus ${\cal B}$ for $r=4$, $(\{e_1,e_2,e_3,e_4\},e_g)
=(\{2,2,2,2\},0)$.}
\label{zeros_r4_22220}
\end{figure}

\begin{figure}[htbp]
  \begin{center}
    \includegraphics[height=7cm,width=7cm]{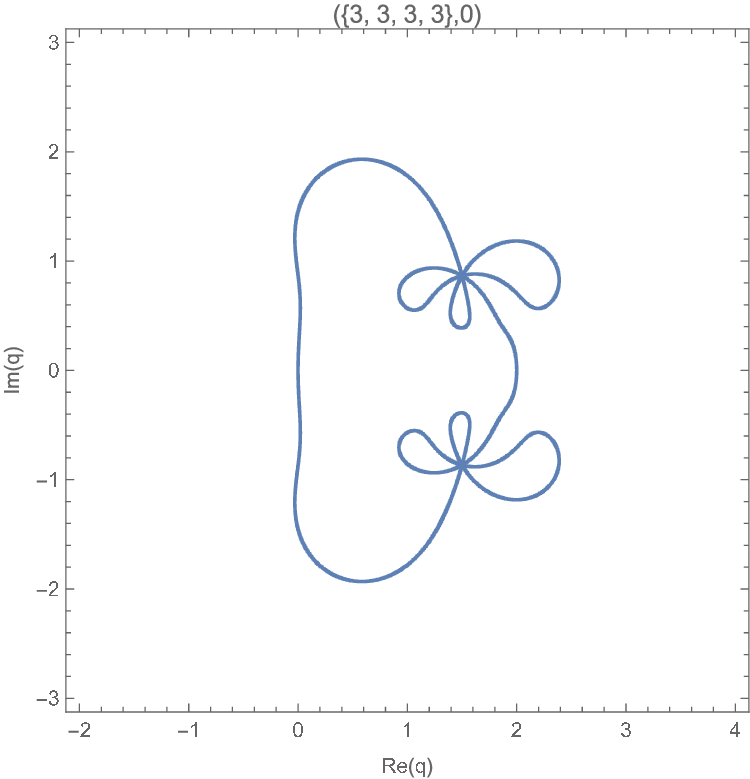}
  \end{center}
\caption{Locus ${\cal B}$ for $r=4$, $(\{e_1,e_2,e_3,e_4\},e_g)
=(\{3,3,3,3\},0)$.}
\label{zeros_r4_33330}
\end{figure}

\begin{figure}[htbp]
  \begin{center}
    \includegraphics[height=7cm,width=7cm]{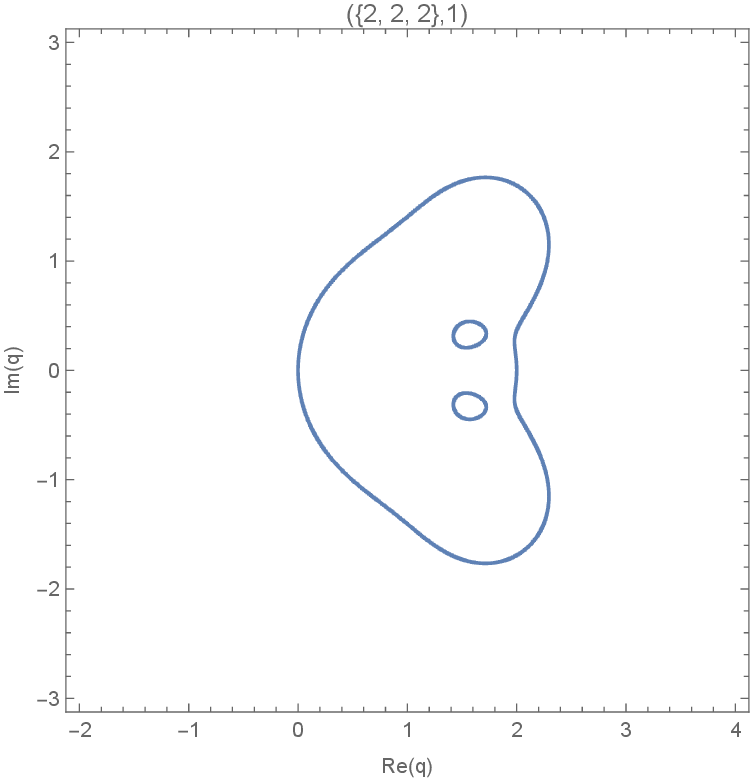}
  \end{center}
\caption{${\cal B}$ for $r=3$, $(\{e_1,e_2,e_3\},e_g)=(\{2,2,2\},1)$.}
\label{zeros_r3_2221}
\end{figure}

\begin{figure}[htbp]
  \begin{center}
    \includegraphics[height=7cm,width=7cm]{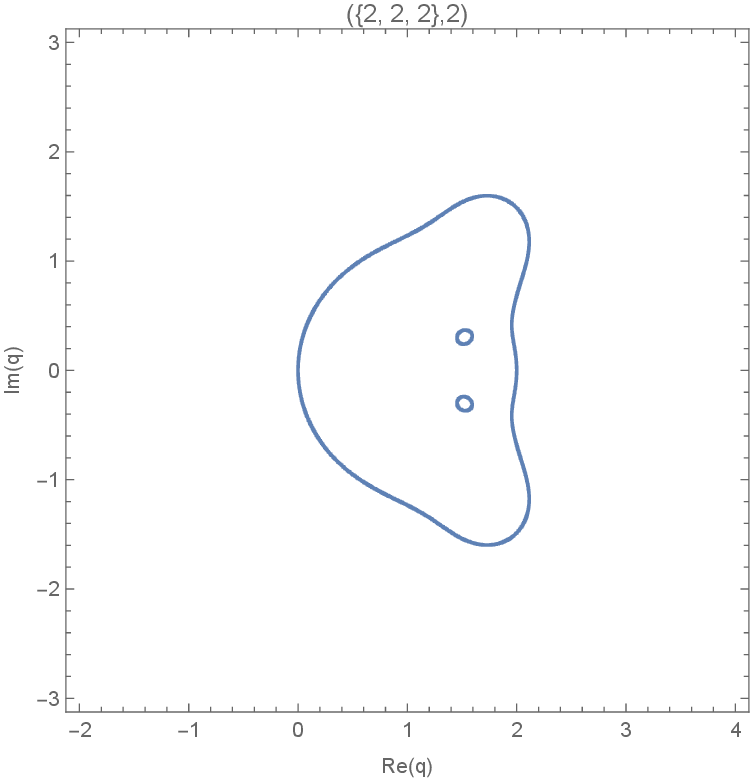}
  \end{center}
\caption{${\cal B}$ for $r=3$, $(\{e_1,e_2,e_3\},e_g)=(\{2,2,2\},2)$.}
\label{zeros_r3_2222}
\end{figure}

\begin{figure}[htbp]
  \begin{center}
    \includegraphics[height=7cm,width=7cm]{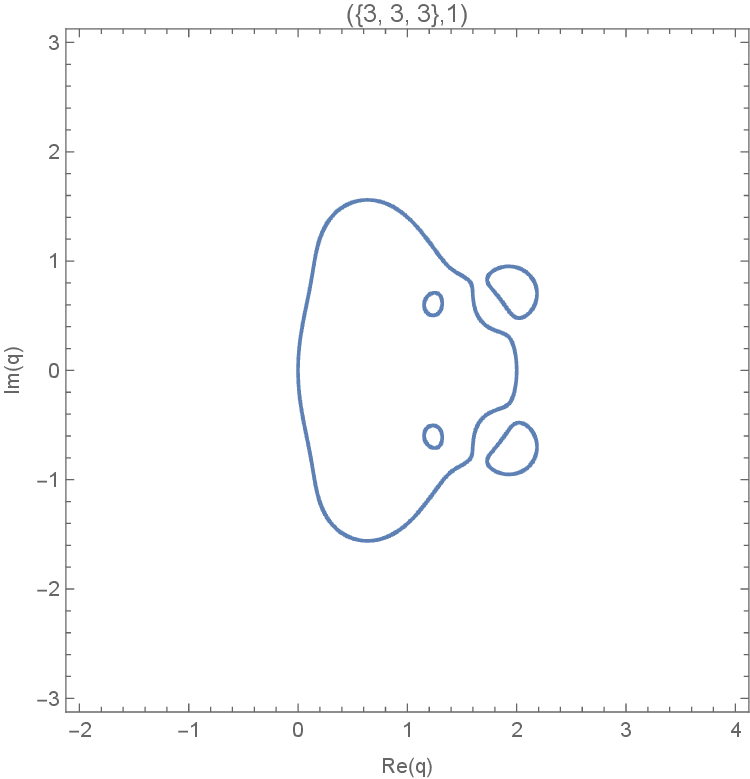}
  \end{center}
\caption{${\cal B}$ for $r=3$, $(\{e_1,e_2,e_3\},e_g)=(\{3,3,3\},1)$.}
\label{zeros_r3_3331}
\end{figure}

\begin{figure}[htbp]
  \begin{center}
    \includegraphics[height=7cm,width=7cm]{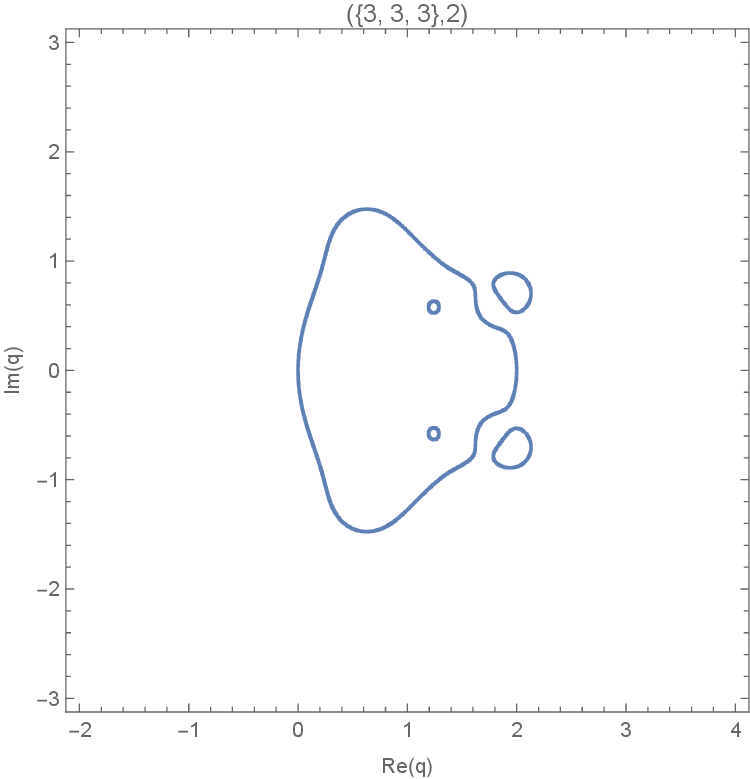}
  \end{center}
\caption{${\cal B}$ for $r=3$, $(\{e_1,e_2,e_3\},e_g)=(\{3,3,3\},2)$.}
\label{zeros_r3_3332}
\end{figure}

\begin{figure}[htbp]
  \begin{center}
    \includegraphics[height=7cm,width=7cm]{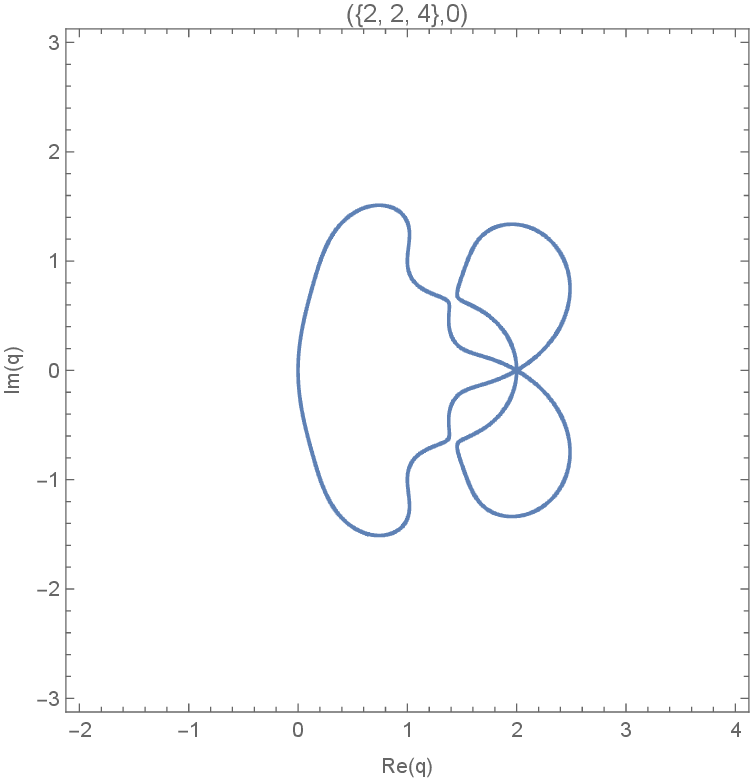}
  \end{center}
\caption{${\cal B}$ for $r=3$, $(\{e_1,e_2,e_3\},e_g)=(\{2,2,4\},0)$.}
\label{zeros_r3_2240}
\end{figure}

\begin{figure}[htbp]
  \begin{center}
    \includegraphics[height=7cm,width=7cm]{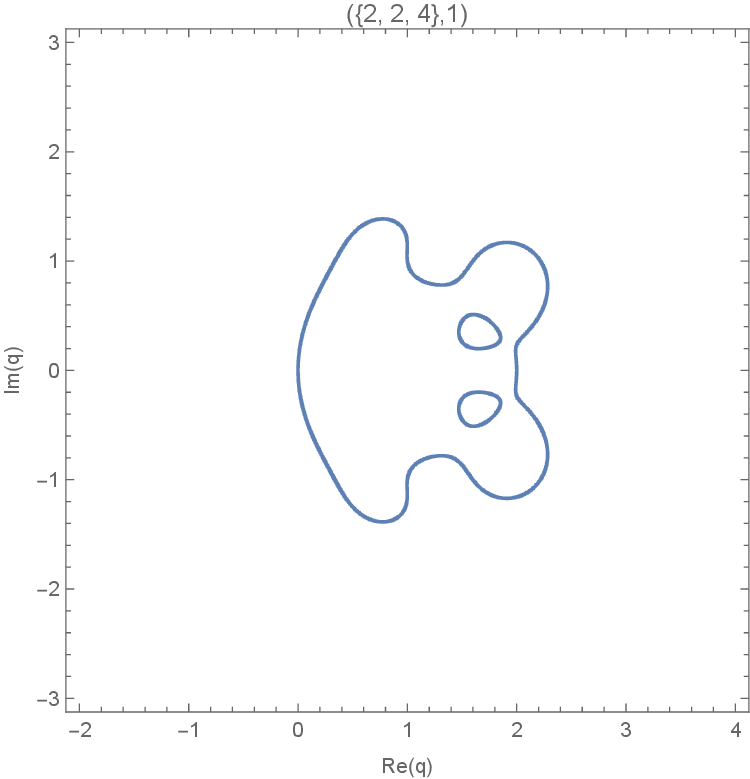}
  \end{center}
\caption{${\cal B}$ for $r=3$, $(\{e_1,e_2,e_3\},e_g)=(\{2,2,4\},1)$.}
\label{zeros_r3_2241}
\end{figure}

\begin{figure}[htbp]
  \begin{center}
    \includegraphics[height=7cm,width=7cm]{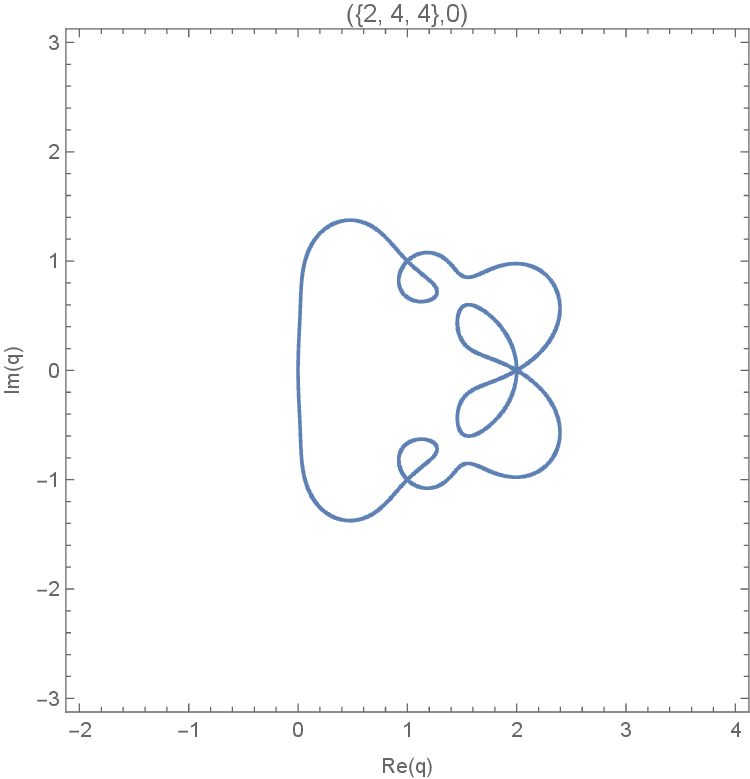}
  \end{center}
\caption{${\cal B}$ for $r=3$, $(\{e_1,e_2,e_3\},e_g)=(\{2,4,4\},0)$.}
\label{zeros_r3_2440}
\end{figure}

\begin{figure}[htbp]
  \begin{center}
    \includegraphics[height=7cm,width=7cm]{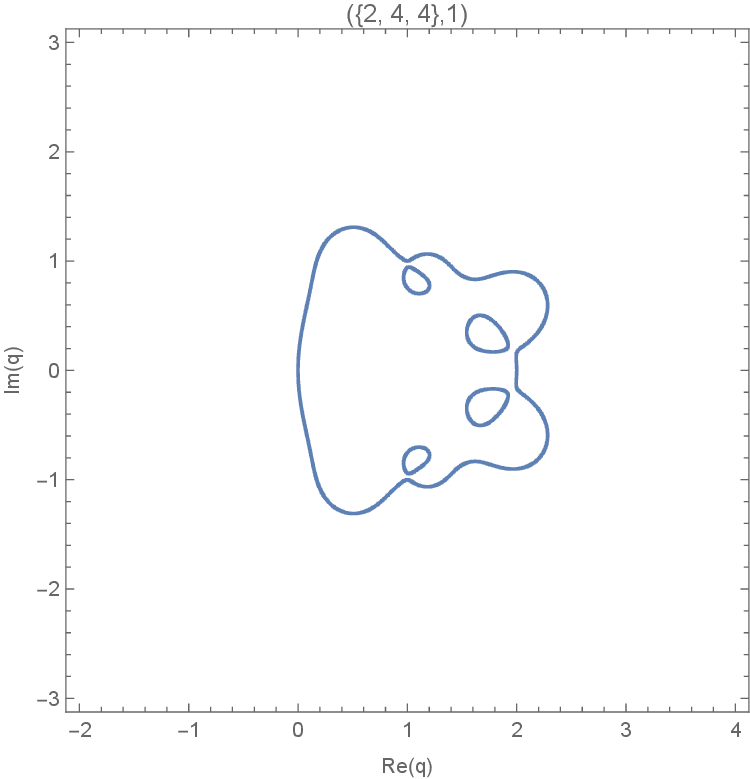}
  \end{center}
\caption{${\cal B}$ for $r=3$, $(\{e_1,e_2,e_3\},e_g)=(\{2,4,4\},1)$.}
\label{zeros_r3_2441}
\end{figure}

\begin{figure}[htbp]
  \begin{center}
    \includegraphics[height=7cm,width=7cm]{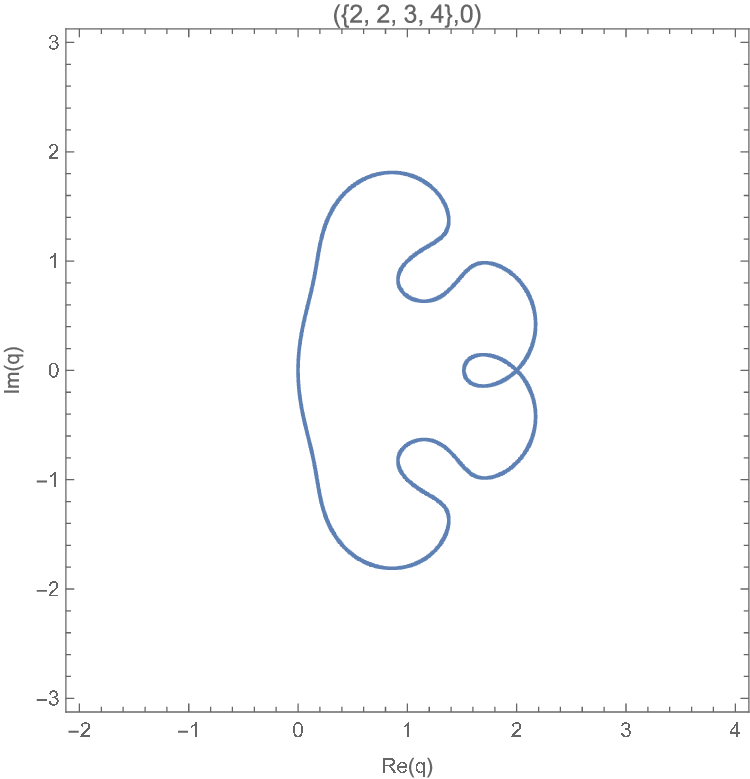}
  \end{center}
\caption{${\cal B}$ for $r=4$, $(\{e_1,e_2,e_3,e_4\},e_g)=(\{2,2,3,4\},0)$.}
\label{zeros_r4_22340}
\end{figure}

\begin{figure}[htbp]
  \begin{center}
    \includegraphics[height=7cm,width=7cm]{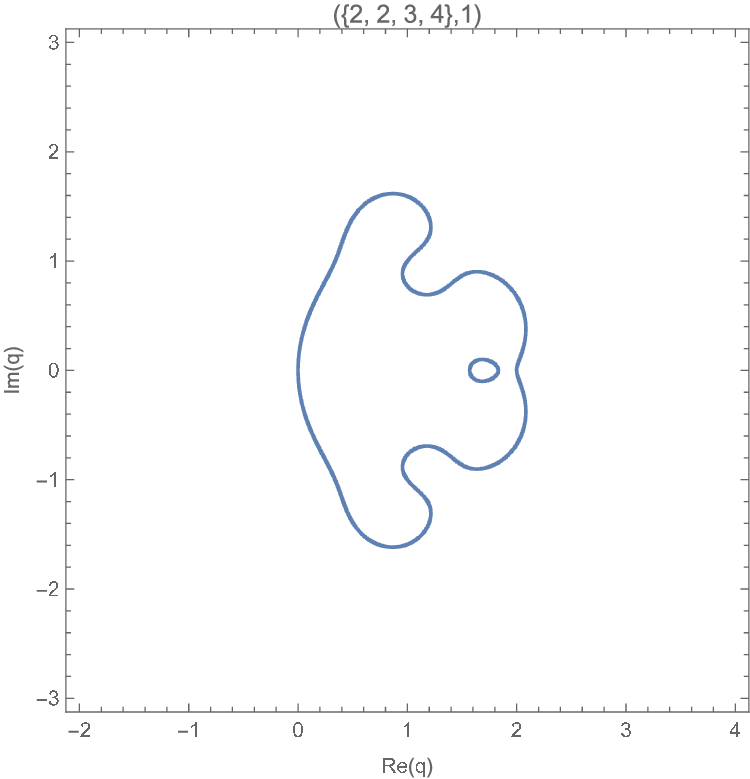}
  \end{center}
\caption{${\cal B}$ for $r=4$, $(\{e_1,e_2,e_3,e_4\},e_g)=(\{2,2,3,4\},1)$.}
\label{zeros_r4_22341}
\end{figure}

\begin{figure}[htbp]
  \begin{center}
    \includegraphics[height=7cm,width=7cm]{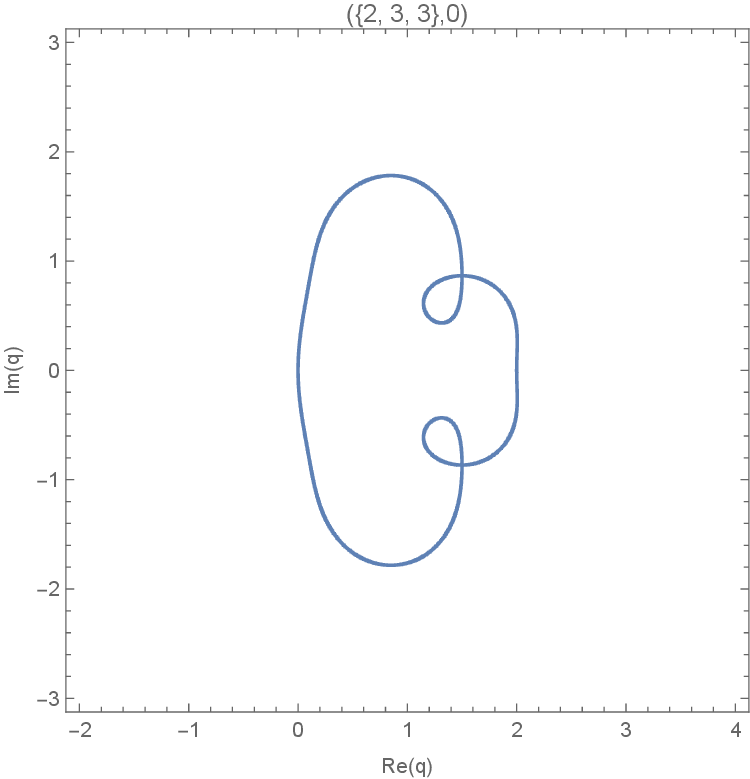}
  \end{center}
\caption{${\cal B}$ for $r=3$, $(\{e_1,e_2,e_3\},e_g)=(\{2,3,3\},0)$.}
\label{zeros_r3_2330}
\end{figure}

\begin{figure}[htbp]
  \begin{center}
    \includegraphics[height=7cm,width=7cm]{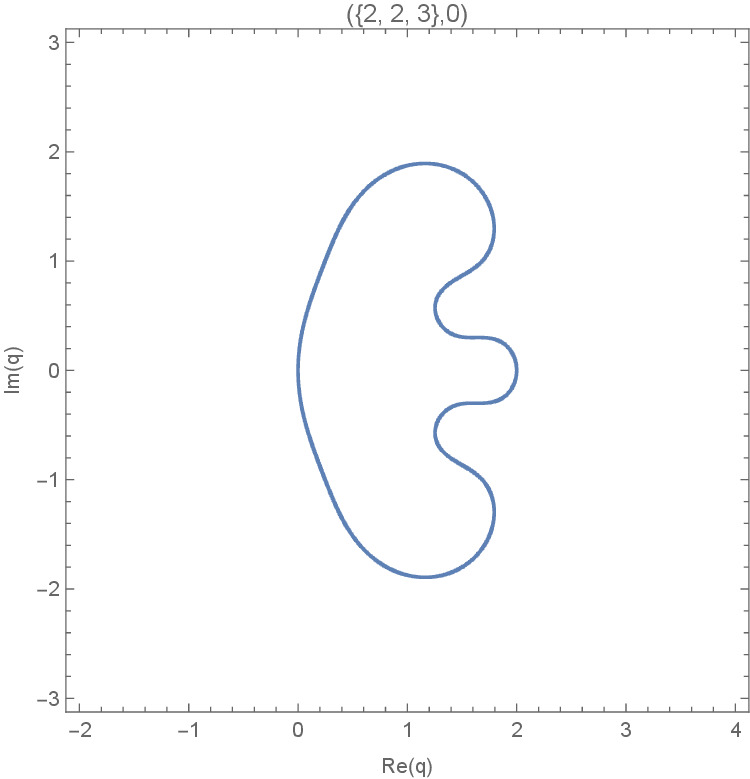}
  \end{center}
\caption{${\cal B}$ for $r=3$, $(\{e_1,e_2,e_3\},e_g)=(\{2,2,3\},0)$.}
\label{zeros_r3_2230}
\end{figure}

\begin{figure}[htbp]
  \begin{center}
    \includegraphics[height=7cm,width=7cm]{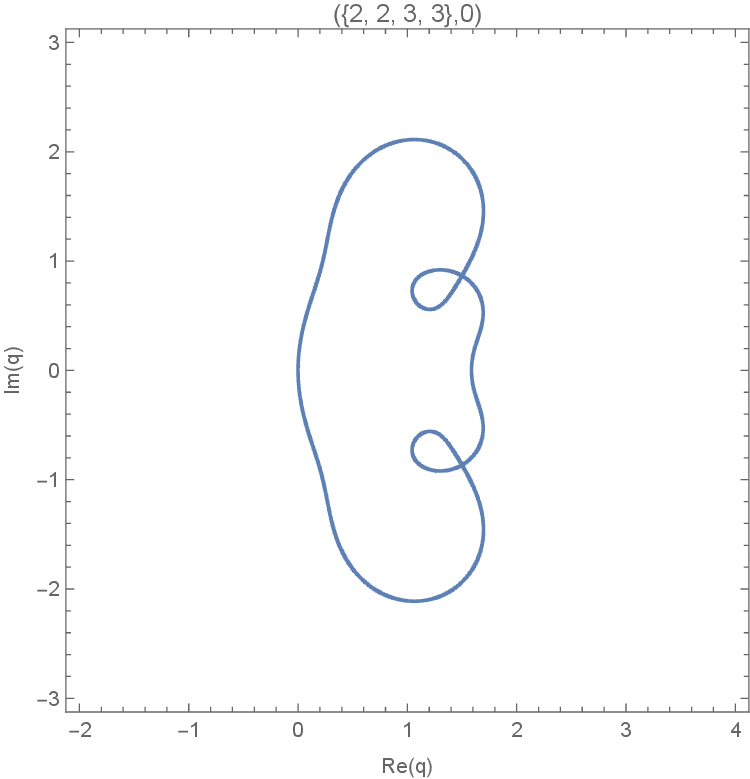}
  \end{center}
\caption{${\cal B}$ for $r=4$, $(\{e_1,e_2,e_3,e_4\},e_g)=(\{2,2,3,3\},0)$.}
\label{zeros_r4_22330}
\end{figure}

We have calculated the continuous accumulation sets of chromatic zeros
${\cal B}$ for the $L_m$ limits of a large number of different cyclic
hammock chain graphs $G_{\{e\}_r,e_g,m;c}$ with various $r \ge 2$ and
edge sets $(\{e\}_r, e_g)$.  This limit is denoted
$\{G_{\{e\}_r,e_g;c}\}$, i.e., with outer brackets, as defined in
(\ref{ginf}), and, with the cyclic boundary condition and $L_m$ limit
understood implicitly, these will often be denoted simply as
$(\{e\}_r,e_g)=(\{e_1,...,e_r\},e_g)$.  In
Figs. \ref{zeros_r2_440}-\ref{zeros_r4_22330} we show some
illustrative plots of these loci for some different choices of
$(\{e\}_r,e_g)$. Below we will discuss how these results change when
$v$ is changed from its value $v=-1$ for the $T=0$ Potts
antiferromagnet, but here we restrict to this $v=-1$ case. Since it
was demonstrated in earlier works such as \cite{baxter87,w,nec},
\cite{strip}-\cite{tt} (for further references, see, e.g.,
\cite{jemrev}), that when ${\cal B}$ is compact, the zeros of
$Z(G_m,q,v)$ in the $q$ plane for strips of various families of graphs
lie close to (or on) the asymptotic accumulation locus ${\cal B}$ for
the $n(G) \to \infty$ limit, it will suffice for our present purposes
to display just these loci ${\cal B}$.

% --------------- locus figures ------------------------

% fig. 2
% B for ({4,4},0)

% fig. 3
% B for ({5,5},0)

% fig. 4
% B for ({6,6},0)

% fig. 5
% B for ({7,7},0)

% fig. 6
% B for ({8,8},0)

% fig. 6 
% B for (2,2,2,0)
%

% fig. 7
% B for (3,3,3,0)
%

% fig. 8 
% B for (4,4,4,0)
%

% fig. 9 
% B for (2,2,2,2,0)
%

% fig. 10
% B for (3,3,3,3,0)
%

% fig. 11 
% B for (2,2,2,1)
%

% fig. 12
% B for (2,2,2,2)
%

% fig. 13 
% B for (3,3,3,1)
%

% fig. 14
% B for (3,3,3,2)
%

% fig. 15
% B for (2,2,4,0)

% fig. 16
% B for (2,2,4,1)

% fig. 17
% B for (2,4,4,0)

%\clearpage

% fig. 18 
% B for (2,4,4,1)

% fig. 19 
% B for (2,2,3,4,0)

% fig. 20 
% B for (2,2,3,4,1)

% fig. 21
% B for (2,3,3,0)

% fig. 22
% B for (2,2,3,0)

% fig. 23
% B for (2,2,3,3,0)

% ========== finite temperature graphs ==============

In the case of the $L_m$ limit of cyclic hammock graphs with $r=2$, i.e.,
cyclic polygon chain graphs, it was shown in \cite{nec} that if $e_g=0$, and
$e_1$ and $e_2$ are both even, then the locus ${\cal B}$ has a multiple point
(MP) of index 2 at $q=2$.  Ref. \cite{nec} also determined the locations
of the multiple points for an interesting special $r=2$ class of cyclic hammock
graphs, namely those for which $e_g=0$ and each of the two ropes have the same
number of edges, i.e., $e_1=e_2\equiv e_{\rm com}$:
\beq
q_\ell = 1 + e^{i\theta_\ell}
\label{qjmp_ecom}
\eeq
where
\beq
\theta_\ell = \pm \frac{2 \ell \pi}{e_{\rm com}}, \quad \ell = 0, 1, ..., 
\bigg (\frac{e_{\rm com}-2}{2} \bigg ) 
\quad {\rm for} \quad e_{\rm com} \ \ {\rm even}
\label{qjmp_ecom_even}
\eeq
\beq
\theta_\ell = \pm \frac{(2\ell+1)\pi}{e_{\rm com}}, \quad \ell = 0, 1, ...,
\bigg ( \frac{e_{\rm com}-3}{2} \bigg ) 
\quad {\rm for} \quad e_{\rm com} \ \ {\rm odd} \ . 
\label{qjmp_ecom_odd}
\eeq

From our calculations of loci ${\cal B}$ for cyclic hammock chain graphs with a
variety of $r \ge 2$ and edge sets $(\{e\}_r, \ e_g)$, we find several general
features. We begin by describing properties of the loci for the
class of cyclic chain graphs in which all of the edges on the $r$ ropes have
the same common value, $e_{\rm com}$, i.e, $e_j = e_{\rm com}$ for all $e_j \in
\{e\}_r$. This class is denoted as the EJC ($e_j$-common-value) class.  We find
the following general properties of ${\cal B}$ for this EJC class.

\begin{enumerate}

% 1 
\item ${\cal B}$ passes through $q=2$ and this point is the maximal point at
  which it intersects the real $q$ axis, so $q_c=2$. 

% 2
\item In this EJC class, a necessary and sufficient condition for ${\cal B}$ to
  have one or more multiple points is that $e_g=0$. Unless otherwise specified,
  all of the properties (iii)-(viii) below assume that this condition $e_g=0$
  is satisfied, in addition to the EJC condition.

% 3
\item If $e_g=0$, then ${\cal B}$ has $N_{MP}=e_{\rm com}-1$ multiple
  points. The locations found for these $N_{MP}$ multiple points in
  \cite{nec} for the case $r=2$ considered there generalize to $r \ge
  3$; i.e., they are again given by
  Eqs. (\ref{qjmp_ecom})-(\ref{qjmp_ecom_odd}).  A corollary is that,
  provided $e_g=0$, one can describe the $N_{MP}$ multiple points as
  follows.  If $e_{\rm com}$ is even, so $N_{MP}$ is odd, then one
  (and only one) of the MPs is located on the real axis at $q=2$,
  while the remaining $N_{MP}=e_{\rm com}-2$ MPs are located at the
  points (\ref{qjmp_ecom_even}) in the complex $q$ plane forming
  $(e_{\rm com}-2)/2$ complex-conjugate pairs. If $e_{\rm com}$ is
  odd, so $N_{MP}$ is even, then there is no MP on the real axis and
  the $e_{\rm com}-1$ MPs occur at the points (\ref{qjmp_ecom_odd})
  comprising $(e_{\rm com}-1)/2$ complex-conjugate pairs.

% 4
\item At each multiple point, $r$ branches intersect, i.e., there are $2r$
  curves emanating out from the MP, with each pair having equal
  tangents at the MP forming one of these $r$ branches. At each MP, the angle 
  between two adjacent branches on ${\cal B}$ is $\phi = \pi/r$.

% 5
\item At each multiple point, two of the $2r$ curves emanating from
  this MP continue outward to connect to other parts of the outer
  envelope of ${\cal B}$, while the remaining $2r-2$ curves form $r-1$
  loops connected (a) to this multiple point or (b) to another
  multiple point. Examples of case (a) are shown, e.g., in Figs,
  \ref{zeros_r2_440}, \ref{zeros_r2_660}, and
  \ref{zeros_r3_2220}-\ref{zeros_r4_33330}, while examples of
  case (b) are Figs. \ref{zeros_r2_550}, \ref{zeros_r2_770}, and
  \ref{zeros_r2_880}. For case (a), especially if there are several
  loops, the corresponding regions can have the appearance of petals
  of a (generically asymmetric) flower.

% 6  
\item
  As a special case of the general property that at each point on
  ${\cal B}$, $|\lambda_{P,0}|=|\lambda_{P,1}|$, it follows that at
  each of the points on the real axis where ${\cal B}$ crosses this
  axis, $|\lambda_{P,0}|=|\lambda_{P,1}|$.  When one crosses from a
  region where $\lambda_{P,0}$ is dominant to a region where
  $\lambda_{P,1}$ is dominant, there is a non-analytic change in $W(\{
  G_{\{e\}_r,e_g} \},q)$.  As in the earlier papers
  (e.g. \cite{w,nec}), the region in the complex plane including the
  real intervals $q > q_c$ and $q < 0$ is denoted region $R_1$, while
  the interior region including the real interval in the neighborhood
  of, and to the right of, the origin, is denoted $R_2$, and so forth,
  with higher-number subscripts for any other regions, including the
  notation $R_s$ and $R_s^*$ for complex-conjugate regions.  At each
  point on ${\cal B}$, both $\lambda_{P,0}$ and $\lambda_{P,1}$ have
  the same magnitude.

% 7  
\item If $e_{\rm com}$ is even, so $N_{MP}$ is odd, then one of the
  MPs occurs at $q=2$. There are then two subcases to consider: if (a)
  $r$ is even, then one of the curves on ${\cal B}$ connecting to the
  MP at $q=2$ is self-conjugate and crosses the real axis at an
  interior (int) point denoted $q_{\rm int}$, while (b) if $r$ is odd,
  then the $r-1$ loops form $(r-1)/2$ complex-conjugate pairs that do
  not intersect the real axis. Examples of subcase (a) are the loci
  ${\cal B}$ for $e_{\rm com}=4, \ 6, 8$ and $r=2$ shown in Figs.
  \ref{zeros_r2_440}, \ref{zeros_r2_660}, and \ref{zeros_r2_880}, and
  for $e_{\rm com}=2$ and $r=4$ in Fig. \ref{zeros_r4_22220}.
  Examples of subcase (b) are the loci ${\cal B}$ for $e_{\rm com}=2,
  \ 4$ and $r=3$ shown in Figs. \ref{zeros_r3_2220} and
  \ref{zeros_r3_4440}.  We will discuss values of $q_{\rm int}$
  further below. In subcase (a), the dominant $\lambda$ terms on the
  real axis are as follows: $\lambda_{P,0}$ is dominant if $q > 2$ or
  $q < 0$ or $q_{\rm int} < q < 2$, while $\lambda_{P,1}$ is dominant
  if $0 < q < q_{\rm int}$.  In subcase (b), $\lambda_{P,0}$ is
  dominant if $q > 2$ or $q < 0$, while $\lambda_{P,1}$ is dominant if
  $0 < q < 2$.

% 8
\item If $e_{\rm com}$ is odd, so $N_{MP}$ is even, then all of these
  multiple points occur away from the real axis, forming a
  complex-conjugate set. There are again two subcases to consider,
  namely: (a$^\prime$) where curves connecting to complex-conjugate MPs
  on ${\cal B}$ cross the real $q$ axis; and (b$^\prime$) where
  curve(s) connecting to the MPs with ${\rm Im}(q) \ne 0$ form loops
  that do not intersect the real axis. Examples of subcase (a$^\prime$)
  are the loci ${\cal B}$ for $(\{5,5\},0)$ in Fig. \ref{zeros_r2_550}
    and $(\{7,7\},0)$ in Fig. \ref{zeros_r2_770}, while examples of
    subcase (b$^\prime$) are the loci ${\cal B}$ for $(\{3,3,3\},0)$ in
    Fig.  \ref{zeros_r3_3330} and for $(\{3,3,3,3\},0)$ in
    Fig. \ref{zeros_r4_33330}.  The situation for subcase (a$^\prime$)
    is illustrated by loci involving two points at which these curves
    cross the real axis in the interior of the outer envelope, i.e.,
    with $q < 2$; we denote these two interior crossing points as
    $q_{{\rm int},1}$ and $q_{{\rm int},2}$, ordered as
    $q_{{\rm int},1} < q_{{\rm int},2}$. For subcase (a$^\prime$), in
    addition to the intervals $q > 2$ and $q < 0$, $\lambda_{P,0}$ is
    also dominant in the real interval $q_{\rm int,1} < q < q_{\rm
      int,2}$, while $\lambda_{P,1}$ is dominant in the intervals $0 <
    q < q_{{\rm int},1}$ and $q_{{\rm int},2} < q < 2$, together with
    the regions in the complex $q$ plane that can be reached from
    these respective real intervals without crossing any part of
    ${\cal B}$. For subcase (b$^\prime$), $\lambda_{P,0}$ is dominant in
    the intervals $q > 2$ and $q < 0$, while $\lambda_{P,1}$ is dominant
    in the interval $0 < q < 2$, together with the regions in the $q$ plane
    that can be reached from these respective real intervals without
    crossing ${\cal B}$. 

% 9
\item For a given $r$, as $e_{\rm com}$ increases, one observes that
  parts of ${\cal B}$, especially near to some MPs, exhibit avoided
  near-crossings with close-by parts of ${\cal B}$. Hence, if one
  views a diagram with a given locus at low resolution and
  magnification, there can be apparent intersections, while at higher
  resolution and magnification, these are seen to be places where the
  curves nearly intersect, but do not actually do so.  These often
  form approximate barred rings, such that where the bar would
  intersect the ring at both ends, it only does so at the left hand,
  while the form of ${\cal B}$ on the right side of the ring is an
  avoided near-crossing.  An easily visible example of this is evident
  in Fig. \ref{zeros_r2_440}, and this can also be seen, e.g., in
  Fig. \ref{zeros_r2_550}, where the avoided near-crossings involve
  smaller distances between parts of ${\cal B}$. Similar comments on
  avoided near-crossings having the approximate appearance of barred
  rings apply to other figures.  In Fig. \ref{zeros_r2_880} there are
  four complex-conjugate avoided near-crossings that have the
  appearence, at low magnification and resolution, of barred rings,
  together with a complex-conjugate pair of additional avoided
  near-crossings on the right-hand side of the locus ${\cal B}$ that
  are not of the near-barred ring type. We describe the member of this
  pair with ${\rm Im}(q) > 0$ as follows. Starting from the multiple
  point at $q=2$, two curves emanate upward at angles $\pi/4$ and
  $3\pi/4$.  These curves then bend around and approach each other,
  but do not actually intersect in an avoided near-crossing at $q
  \simeq 1.9 + 0.4i$, after which they do cross at a multiple
  point at $q \simeq 1.8 + 0.7i$. 

% 10
\item Here, within the EJC class, we describe how the locus ${\cal B}$
  changes as $e_g$ is increases from 0 to nonzero values. As this
  increase of $e_g$ is carried out, all of the multiple points that were
  present for $e_g=0$ disappear and are replaced by certain sets of
  ``bubble'' regions. Thus, in these cases, the locus ${\cal B}$ gains
  new disconnected components. For a given value of $e_{\rm com}$,
  each of these bubble regions contract monotonically as $e_g$
  increases above 1 and approach points in the limit $e_g \to \infty$.
  As with the avoided near-crossings that can occur on ${\cal B}$, if
  one views plots of ${\cal B}$ with small nonzero values of $e_g$ at
  moderate resolution or magnification, it can be difficult to see the
  detachment, since the inner part of the boundary of the bubble
  region is very close to the position of the former multiple point,
  but the fact that they are separated becomes evident if one uses
  higher magnification. As $e_g$ increases further, the separation of
  the bubble regions can easily be seen.

% 11
\item 
  For both zero and nonzero $e_g$, as $e_{\rm com}$ increases beyond 4, the
  outer envelope of the locus ${\cal B}$ approaches (aside from the presence of
  the $e_{\rm com}-1$ multiple points and associated avoided near-crossings)
  the form of an undular deformation of the unit circle $|q-1|=1$ on the upper,
  lower, and left-hand part of the locus. The angular wavelength interval in
  the even-$e_{\rm com}$ case is $2\pi/e_{\rm com}$, while for the odd-$e_{\rm
    com}$ case, it is $2\pi/(e_{\rm com}-1)$. 

\end{enumerate}

In Ref. \cite{nec}, several plots of ${\cal B}$ (and associated chromatic
zeros) for the $r=2$ class of cyclic hammock chain graphs were presented. These
included several figures for the $r=2$ EJC class, namely $(\{2,2\},e_g)$ and
$(\{3,3\},e_g)$ with $0 \le e_g \le 3$. We note that the locus ${\cal B}$ 
for the family $(\{2,2\},0)$ has a self-conjugate loop extending to the left 
of $q=2$ and crossing the real axis at 
\beqs
(\{2,2\},0): \quad q_{\rm int} &=& 
\frac{4}{3} + \frac{1}{3}(1+3\sqrt{57})^{1/3}
- \frac{8}{3}(1+3\sqrt{57})^{-1/3} \cr\cr
&=& 1.361103,
\label{220_qint}
\eeqs
This value is listed, together with values of $q_{\rm int}$ for higher even
values of $e_{\rm com}$ extending up to 10 in this $r=2$ EJC class in Table
\ref{qint_r2_ejc_egzero_table}

\begin{table}[htbp]
\caption{\footnotesize{ Values of $q_{\rm int}$ on ${\cal B}$ for members of
the $r=2$, $e_g=0$ EJC class of cyclic hammock graphs with 
$e_{\rm com}$ taking even values from 2 to 10.}}
\begin{center}
\begin{tabular}{|c|c|} \hline\hline
  $e_{\rm com}$ & $q_{\rm int}$ \\ \hline
2  & 1.361103   \\
4  & 1.455877   \\
6  & 1.486632   \\
8  & 1.496312   \\
10 & 1.4990415  \\
\hline\hline
\end{tabular}
\end{center}
\label{qint_r2_ejc_egzero_table}
\end{table}

We proceed to discuss some features of the loci ${\cal B}$ further
for various cases $(\{e\}_r,e_g)$. 
In Figs. \ref{zeros_r2_440}-\ref{zeros_r2_880}, we show a series of
loci ${\cal B}$ for the $r=2$ EJC class with higher values of $e_{\rm
  com}$, namely $4 \le e_{\rm com} \le 8$.  For plots with $e_g=0$,
since $r=2$, it follows that $2r=4$ curves emanate from each multiple
point, forming $r=2$ branches, with angular separation $\phi=\pi/r =
\pi/2$ between each curve at a given MP.  In Fig. \ref{zeros_r2_440}
showing ${\cal B}$ for the case $(\{4,4\},0)$ there are $e_{\rm
  com}=3$ multiple points, and, in agreement with the general formulas
(\ref{qjmp_ecom})-(\ref{qjmp_ecom_odd}) from \cite{nec}, these three
MPs are located at $q=2$ and $q=1 \pm i$. In this
Fig. \ref{zeros_r2_440}, the MP at $q=1-i$ is part of an $e$-shaped
portion of ${\cal B}$ involving an avoided near-crossing at the
right-hand side of the $e$, and the complex-conjugate MP at $q=1+i$
has the form of a reflection of this about the horizontal axis to
produce an inverted $e$.  The MP at $q=2$ in this
Fig. \ref{zeros_r2_440} has a loop extending to the left and crossing
the real axis at $q=1.455877$, as listed in Table
\ref{qint_r2_ejc_egzero_table}.  In Fig. \ref{zeros_r2_550} showing
${\cal B}$ for the case $(\{5,5\},0)$, there are
$e_{\rm com}-1=4$ MPs, comprised of two complex-conjugate pairs. From
Eq. (\ref{qjmp_ecom_odd}), it follows that these are located at
\beqs
\{q_{MP,\ell=1}, \ q^*_{MP,\ell=1} \} &=& 
1 + e^{\pm i\pi/5} = \frac{5 + \sqrt{5}}{4} \pm 
\frac{i\sqrt{2(5 - \sqrt{5})}}{4} \cr\cr
&=& 1.809017 \pm 0.587785i
\label{550_q1}
\eeqs
and
\beqs
\{q_{MP,\ell=2}, \ q^*_{MP,\ell=2} \}
  &=& 1 + e^{\pm 3i\pi/5} \cr\cr
  &=& 0.690983 \pm 0.9587785i
\label{550q2}
\eeqs
The locus ${\cal B}$ in this Fig. \ref{zeros_r2_550} crosses the real
axis at the maximal point $q_c=2$, and (aside from the ever-present
crossing at $q=0$) at two interior points to the left of $q_c$, namely
\beq
(\{5,5\},0): \quad q_{\rm int,1} = 1.548348 \ , \quad 
                  q_{\rm int,2} = 1.860499 
\label{550_qint12}
\eeq
These values are listed, together with values of $q_{\rm int,1}$ and $q_{\rm
  int,2}$ for higher odd values of $e_{\rm com}$ extending up to 11 in the
$r=2$, $e_g=0$ EJC class in Table \ref{qint12_r2_ejc_egzero_table}.
In accordance with our general discussion above, the term
$\lambda_{P,0}$ is dominant on the real intervals $q > 2$, $q < 0$, and $q_{\rm
  int,1} < q < q_{\rm int,2}$, together with the regions in the complex plane
that can be reached from these real intervals without crossing any part of
${\cal B}$. The term $\lambda_{P,1}$ is dominant in the real intervals $0 < q <
q_{\rm int,1}$ and $q_{\rm int,2} < q < 2$, together with the regions in the
complex plane that can be reached from these real intervals without crossing
any part of ${\cal B}$. In addition, $\lambda_{P,0}$ is dominant in the inner
part of the ``e''-shaped region associated with the multiple point
$q_{MP,\ell=1}^*$ and the corresponding region associated inverted ``e''-shaped
region associated with the MP $q_{MP,\ell=1}$.

\begin{table}[htbp]
\caption{\footnotesize{ Values of $q_{\rm int,1}$ and $q_{\rm int,2}$ on 
${\cal B}$ for members of
the $r=2$, $e_g=0$ EJC class of cyclic hammock graphs with 
$e_{\rm com}$ taking odd values from 3 to 11.
The dash for $e_{\rm com}=3$ means that there are no interior crossing points
on ${\cal B}$ in this case.}}
\begin{center}
\begin{tabular}{|c|c|c|} \hline\hline
  $e_{\rm com}$ & $q_{{\rm int},1}$ & $q_{{\rm int},2}$ \\ \hline
3  & $-$        & $-$      \\
5  & 1.548348   & 1.860499 \\
7  & 1.508787   & 1.942621  \\
9  & 1.502023   & 1.979817  \\
11 & 1.500493   & 1.996826 \\
\hline\hline
\end{tabular}
\end{center}
\label{qint12_r2_ejc_egzero_table}
\end{table}

As is evident from these figures and from Table
\ref{qint_r2_ejc_egzero_table}, in the $r=2$, $e_g=0$ EJC class with
even $e_{\rm com}$, as $e_{\rm com}$ increases from 2, the inner
crossing point $q_{\rm int}$ increases monotonically, approaching 3/2
from below as $e_{\rm com} \to \infty$. As is also evident from the
figures and from Table \ref{qint12_r2_ejc_egzero_table}, in the $r=2$,
$e_g=0$ EJC class with odd $e_{\rm com}$, as $e_{\rm com}$ increases
from 3, the smaller interior crossing point $q_{\rm int,1}$ decreases
monotonically, approaching 3/2 from above as $e_{\rm com} \to \infty$,
while the larger interior crossing point $q_{\rm int,2}$ increases
monotonically, approaching the right-most crossing point at $q_c=2$
from below as $e_{\rm com} \to \infty$.  Similar analyses of the
values of interior crossing points on ${\cal B}$ can be carried out
for $r \ge 3$.

In Figs. \ref{zeros_r3_2220}-\ref{zeros_r4_33330} we show the respective loci
${\cal B}$ for some cyclic hammock graphs in the EJC $e_g=0$ class with higher
rope values $r=3$ and $r=4$ in each hammock subgraph. These exhibit the general
properties that we have discussed above. Our calculations of ${\cal B}$ for
higher $r$ generalize the finding in \cite{nec} concerning the disappearance of
multiple points in loci for families with $e_g=0$ and their replacement by
bubble regions as $e_g$ is increased from zero to nonzero values. We illustrate
this with Figs. \ref{zeros_r3_2221}-\ref{zeros_r3_3332}. As $r$ increases,
more complicated types of behavior occur as one increases $e_g$ from zero to
nonzero values. For example, Fig. \ref{zeros_r4_22220} showed ${\cal B}$ for
the $r=4$ family $(\{2,2,2,2\},0)$, which has a multiple point of index 4 at
$q=2$ with three loop regions in the interior, relative to the outer
envelope. When one increases $e_g$ from 0 to 1, the locus ${\cal B}$ for the
family $(\{2,2,2,2\},1)$ exhibits a single bubble region in the interior, and
for the family $(\{2,2,2,2\},2)$, the upper and lower right-hand bulb-like
parts of the outer curve on ${\cal B}$ have constricted to produce new bubble
regions in these areas. Thus, with this family, one observes the appearance of
a complex-conjugate pair of bubble regions that did not originate immediately
from the disappearance of a multiple point.

We have also calculated loci ${\cal B}$ for higher $r \ge 3$ with edge sets
$\{e\}_r$ containing some unequal $e_j \in \{e\}_r$. The family $r=2$ with
$(\{e_1,e_2\},e_g)=(\{2,4\},0)$ was shown in \cite{nec} to have a locus ${\cal
  B}$ with a multiple point of index 2 at $q=2$ that disappears and is replaced
by the appearance of a new bubble region for $(\{2,4\},e_g)$ with $e_g > 0$,
with contraction of the bubble region as $e_g$ is increased through nonzero
values. We find a variety of behavior for higher $r$. For example,
Figs. \ref{zeros_r3_2240} and \ref{zeros_r3_2241} show the respective loci
${\cal B}$ for the $r=3$ families $(\{2,2,4\},e_g)$ with $e_g=0$ and $e_g=1$,
respectively.  The locus ${\cal B}$ for the $(\{2,2,4\},0)$ family exhibits a
multiple point of index 3 at $q=2$, while for $(\{2,2,4\},1)$, this MP has
disappeared and has been replaced with two bubble regions in the interior of
the outer envelope curve. Fig. \ref{zeros_r3_2440} shows the locus ${\cal B}$
for the $r=3$ family $(\{2,4,4\},0)$, which exhibits a MP of index 3 at $q=2$,
together with a complex-conjugate pair of MPs of index 2 at $q=1\pm i$. When
one increases $e_g$ from 0 to 1, these MPs are replaced by four bubble regions,
as shown in Fig. \ref{zeros_r3_2441}. Fig. \ref{zeros_r4_22340} shows ${\cal
  B}$ for the $r=4$ family $(\{2,2,3,4\},0)$, exhibiting a single multiple
point of index 2 at $q=2$.  This disappears and is replaced by an interior
bubble region when $e_g$ is increased to nonzero values, as shown in Fig.
\ref{zeros_r4_22341} for the family $(\{2,2,3,4\},1)$.

We include two other illustrative examples of loci ${\cal B}$ for
higher $r$ with some unequal $e_j$ edge values in
Figs. \ref{zeros_r3_2330} and \ref{zeros_r3_2230}.
Fig. \ref{zeros_r3_2330} shows ${\cal B}$ for the $(\{2,3,3\},0)$
family, with index-2 multiple points at $q=1+e^{\pm \pi i/3}$, which
disappear and are replaced by bubble regions for the families
$(\{2,3,3\},e_g)$ with $e_g \ge 1$.  Fig. \ref{zeros_r3_2230} depicts
${\cal B}$ for the $e_g=0$ member of the $r=3$ families
$(\{2,2,3\},e_g)$ with a similar property of a smooth locus ${\cal B}$
with no multiple points. Both of these families have $q_c = 2$.
Ref. \cite{nec} also presented results for the $r=2$ families
$(\{2,3\},e_g)$ with $0 \le e_g \le 3$ for which the respective loci
${\cal B}$ have no multiple points and $q_c$ values less than 2. In
our discussion above, we have generalized the determination of $q_c$
to $r \ge 3$. As an illustration of a higher-$r$ case with $q_c < 2$,
and specifically of the analysis in Eqs. (\ref{plam0_taylor_ellhalf})
and (\ref{plam1_taylor_ellhalf}), we show ${\cal B}$ for the $r=4$
family $(\{2,2,3,3\},0)$, which has $q_c < 2$ (and two
complex-conjugate) multiple points at $q=1+e^{\pm i\pi/3}$).

% ======================================================================

\section{Zeros of the Potts Model Partition Function}
\label{zzeros_section}

In the previous section, we have discussed the continuous accumulation loci of
the chromatic zeros for the $L_m$ limit of cyclic hammock graphs. In the
thermodynamic context, these chromatic zeros are the zeros of the partition
function of the $T=0$ (i.e., $v=-1$) Potts antiferromagnet on these cyclic
hammock graphs.  More generally, it is of interest to investigate the zeros of
the partition functions $Z(G_{\{e\}_r,e_g,m;o},q,v)$ and
$Z(G_{\{e\}_r,e_g,m;c},q,v)$ in the complex $q$ plane for fixed $v$ and in the
complex $v$ plane for fixed $q$, especially in the $L_m$ limit. Some results on
this were given in \cite{neca}, and we extend these here.  Since
$Z(G_{\{e\}_r,e_g,m;o},q,v)$ is a polynomial in $q$ and $v$, one may study the
partition function zeros in the $q$ plane for fixed $v$ or the zeros in the $v$
plane for fixed $q$.  As in previous works such as \cite{a}, we denote the respective continuous accumulation sets of
these zeros in the $L_m$ limit as ${\cal B}_q(v)$ and ${\cal B}_v(q)$. These
are slices through the algebraic variety defined as the continuous accumulation
set of the partition function zeros in the ${\mathbb C}^2 \approx {\mathbb
  R}^4$ space of the variables $(q,v)$ with $q, \ v \in {\mathbb C}$ in the
$L_m$ limit.  

Given Eq. (\ref{zham_open}), the zeros of $Z(G_{\{e\}_r,e_g,m;o},q,v)$ in the
$q$ plane, besides the one at $q=0$, are the set of zeros of $\lambda_{Z,0}$,
which are discrete (i.e., isolated). The locations of these zeros are
independent of $m$, although each of them has multiplicity $m$. In contrast,
$Z(G_{\{e\}_r,e_g,m;c},q,v)$ has $q$-plane zeros resulting from cancellations
between $(\lambda_{Z,0})^m$ and $(\lambda_{Z,1})^m$.  In the $L_m$ limit,
almost all of these zeros merge to form the continuous locus ${\cal B}_q(v)$.
Hence, ${\cal B}_q(v)$ for the $L_m$ of cyclic hammock chain graphs is
nontrivial (for $v \ne 0$), while ${\cal B}_q(v)$ for the $L_m$ limit of open
hammock chain graphs is trivial. As with our analysis of chromatic zeros in
previous sections, we therefore generally restrict here to considering cyclic
hammock chain graphs. For an arbitrary graph $G$, $Z(G,q,v)$ obeys the property
(\ref{zv0}), that $Z(G,q,v)=q^{n(G)}$ if $v=0$, so all of the zeros of
$Z(G,q,v=0)$ occur at $q=0$ (with multiplicity $n(G)$), and ${\cal
  B}_q(v=0)=\emptyset$. Hence, in the rest of our discussion in this section,
unless otherwise specified, we implicitly assume $v \ne 0$, i.e., in the
physical context, the temperature is not infinite. 
Since the loci ${\cal B}_q(v)$ and ${\cal B}_v(q)$ are formed by zeros of
$Z(G_{\{e\}_r,e_g,m;c},q,v)$ in the $L_m$ limit, which, in turn, 
arise because of the
above-mentioned cancellation, one examines the solutions to the equation 
$|\lambda_{Z,0}|=|\lambda_{Z,1}|$ to study these loci. 

The locus ${\cal B}_q(v)$ will be of primary interest here, especially since in
the limit $v \to -1$, this becomes the continuous accumulation set of chromatic
zeros.  The other locus, ${\cal B}_v$ (the accumulation set of Fisher zeros) is
of somewhat less interest, since the $L_m$ limit of these hammock chain graphs
is essentially one-dimensional, and hence, for physical values of $q$, the
corresponding Potts model with either sign of $J$ does not exhibit a phase
transition at any nonzero temperature.

We first recall some elementary properties of ${\cal B}_q(v)$,
generalizing our discussion above of ${\cal B}(v)$ for $v=-1$. As is
evident from Eq. (\ref{cluster}), for an arbitrary graph $G$, if $v
\in {\mathbb R}$, then $Z(G,q,v)$ is a polynomial in $q$ with real
coefficients and hence, for a given $v$, the action of complex
conjugation, $q \to q^*$, induces an automorphism of the resultant set
of zeros of $Z(G,q,v)$ in the complex $q$ plane. Given that $v$ is in
the real ranges listed above for the physical Potts antiferromagnet or
ferromagnet, it follows in the specific context of the $L_m$ limit of
hammock chain graphs that ${\cal B}_q(v)$ maps into itself under the
complex conjugation $q \to q^*$. Similarly, if $q \in {\mathbb R}$,
then $Z(G,q,v)$ is a polynomial in $v$ with real coefficients and
hence, for a fixed physical (and hence real, nonnegative) $q$, the
action of complex conjugation, $v \to v^*$, induces an automorphism of
the set of zeros of $Z(G,q,v)$ in the complex $v$ plane.  Again, it
follows that for a fixed physical value of $q$, the continuous
accumulation locus ${\cal B}_v(q)$ is invariant under the complex
conjugation $v \to v^*$.  For general antiferromagnetic $v$, the
maximal point at which ${\cal B}_q(v)$ crosses the real-$q$ axis is
denoted $q_c=q_c(v)$. Since we focus on ${\cal B}_q(v)$ rather than
${\cal B}_v(q)$ below, unless otherwise indicated, we use the
simplified notation ${\cal B}_q(v) = {\cal B}_q$.

\begin{figure}[htbp]
  \begin{center}
    \includegraphics[height=7cm,width=7cm]{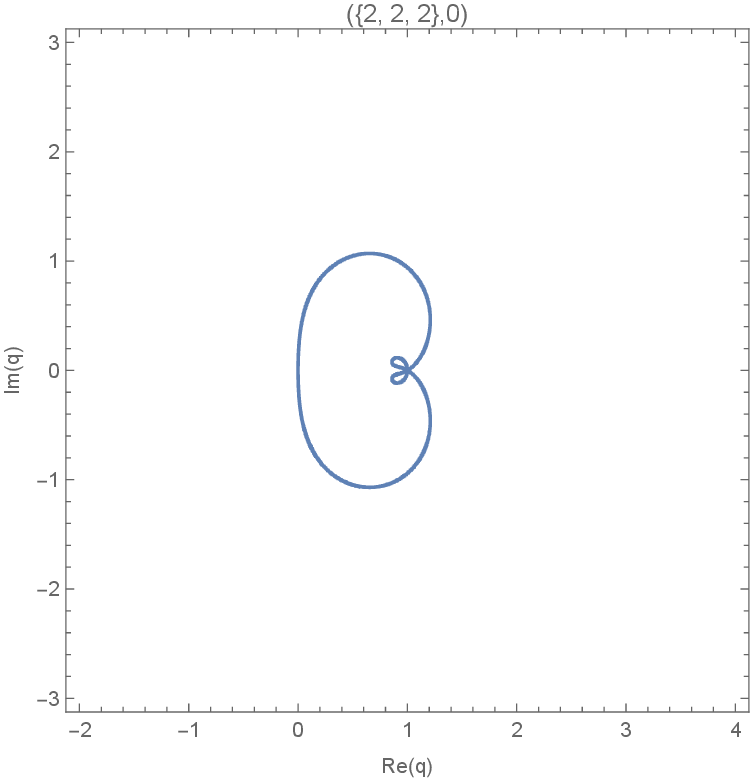}
  \end{center}
\caption{${\cal B}$ for $\{ G_{\{e\}_{r=3},e_g;c} \}$ with 
$(\{e_1,e_2,e_3\},e_g)=(2,2,2,0)$ and $v=-0.5$.}
\label{zeros_r3_2220_0p5}
\end{figure}

\begin{figure}[htbp]
  \begin{center}
    \includegraphics[height=7cm,width=7cm]{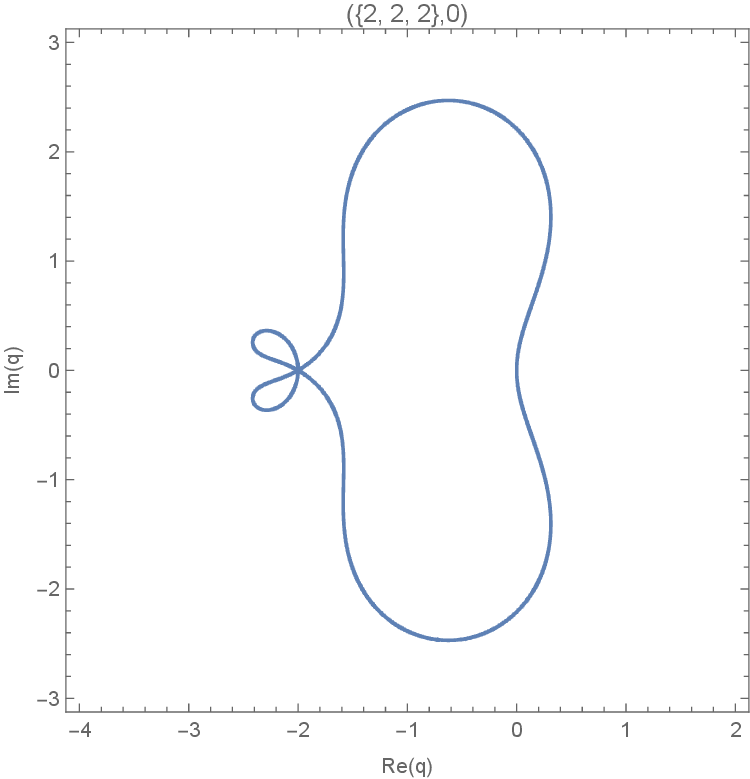}
  \end{center}
\caption{${\cal B}$ for $\{ G_{\{e\}_{r=3},e_g;c} \}$ with 
$(\{e_1,e_2,e_3\},e_g)=(\{2,2,2\},0)$ and $v=1$.}
\label{zeros_r3_2220_1}
\end{figure}

In Figs. \ref{zeros_r3_2220_0p5} and \ref{zeros_r3_2220_1} we show
${\cal B}_q$ for illustrative finite-temperature values of $v$, namely
$v=-0.5$, for the Potts antiferromagnet and $v=1$ for the Potts
ferromagnet.  Comparing the locus in Fig. \ref{zeros_r3_2220_0p5} with
the zero-temperature (i.e., $v=-1$) locus in Fig. \ref{zeros_r3_2220},
one sees that ${\cal B}_q$ contracts both vertically and horizontally,
while continuing to pass through $q=0$. This contraction is a general
feature observed before for other families of graphs
(e.g. \cite{a,s3a}) and continues as $v \to 0^-$, ending as ${\cal B}$
degenerates to a point at $q=0$ for $v=0$, in accordance with
Eq. (\ref{zv0}). The locus ${\cal B}_q$ in Fig. \ref{zeros_r3_2220_1}
for the illustrative finite-temperature ferromagnetic value $v=1$ exhibits
the feature that ${\cal B}$ does not cross the positive-$q$ axis, and this
is a general property for the ferromagnetic range $v \ge 0$.

% fig. 29  
% B for (2,2,2,0) v=-0.5 
%

% fig. 30
% B for (2,2,2,0) v=1 
%

% ------------------------------------------------------------------

There are several properties of ${\cal B}_q$ for the cyclic hammock
chain graphs that hold for general physical $v$.  These include the
following \cite{neca} : (i) ${\cal B}$ is compact; (ii) ${\cal B}$
passes through the point $q=0$; and (iii) ${\cal B}$ encloses regions
in the complex $q$ plane.  These are the generalizations of properties
(B1)-(B3) proved in \cite{nec} for $r=2$ and the special case $v=-1$
and are proved in a similar manner for general $v$. As is clear from
our discussion of ${\cal B}_q$ for the zero-temperature
antiferromagnetic value $v=-1$, the extension to the analysis of
${\cal B}(v)$ for general finite-temperature antiferromagnetic and
ferromagnetic values of $v$ has an interesting wealth of properties.
Further analysis of these properties of ${\cal B}_q(v)$ will be given
in a sequel article.

% ======================================================================

\section{Free Energy}
\label{free_energy_section}

In the context of statistical physics, it is of interest to calculate the Gibbs
free energy per vertex, ${\cal G}$, of the $q$-state Potts model on these
hammock chain graphs.  It is convenient to define a reduced, dimensionless free
energy $f$ according to ${\cal G} = -k_B T f$, where, as above, $T$ is the
temperature. We take the limit $m \to \infty$ in evaluating $f$ and
recall the notation $\{ G_{\{e\}_r,e_g,BC}\}$.

For a strip (chain) graph $G$ and a set of special values of $q$, 
denoted $\{q_x\}$, one can encounter the noncommutativity \cite{a}
\beq
\lim_{q \to q_x} \lim_{n(G) \to \infty} Z(G,q,v)^{1/n} \ne
\lim_{n(G) \to \infty} \lim_{q \to q_x} Z(G,q,v)^{1/n} \ . 
\label{fnoncomm}
\eeq
The corresponding noncommutativity of limits in the case of $P(G,q)$
had also been noted in \cite{w}.  For the graphs of interest here, the
set $\{q_x\}$ includes $\{0,1\}$. When studying the $m \to \infty$
limit of $[Z(G_{\{e\}_r,e_g,m;BC},q,v)]^{1/n}$ here, we choose the
first order of limits, i.e., we take the limit $n \to \infty$ first
and then the limit $q \to q_x$.  It should also be recalled that if $J
> 0$ (so $v \ge 0$), then the cluster representation (\ref{cluster})
defines a Gibbs measure for nonnegative real, as well as nonnegative
integral $q$, as is clear since each term $q^{k(G')}v^{e(G')}$ in
(\ref{cluster}) is nonnegative.  This is also true in the
antiferromagnetic case $J<0$ if $q$ is a positive integer, since one
can revert back to the Hamiltonian form $Z(G,q,v)=\exp[K\sum_{e_{ij}}
  \delta_{\sigma_i \sigma_j}]$, which is positive-definite.  However,
if $J < 0$, then for positive but nonintegral $q$, the cluster
representation (\ref{cluster}) can yield a negative value for
$Z(G,q,v)$ and hence not satisfy the condition for a Gibbs measure. We
will consider the limit of an infinite-length chain graph, i.e., $m
\to \infty$ for the calculation of the free energy per vertex.  As a
quasi-one-dimensional system, this corresponds to a physical 1D
thermodynamic limit. In this limit, with physical values of input
parameters, the dimensionless free energy per vertex for the Potts
model, denoted $f$, is
\beqs
f(\{G_{\{e\}_r,e_g}\},q,v) & = & \lim_{m \to \infty} \ \frac{1}{n}
\ln \Big [ Z(G_{\{e\},e_g,m;BC}) \Big ] \cr\cr
& = & \frac{1}{[(\sum_{j=1}^r e_j) + e_g-(r-1)]} \, \ln (\lambda_{Z,0}) \ , 
\label{fham}
\eeqs
where $\lambda_{Z,0}$ was given in Eq. (\ref{zlam0}).  In Eq. (\ref{fham}) we
have dropped the subscript $BC$ in $\{G_{\{e\}_r,e_g;BC}\}$ since $f$ is
independent of boundary conditions here, as is a necessary
condition of the thermodynamic limit. We recall that the
independence of boundary conditions implies that the dominant term among
$(\lambda_{Z,0})^m$ and $(\lambda_{Z,1})^m$, must be $(\lambda_{Z,0})^m$, since
this is the only term that contributes to both $Z(G_{\{e_j\},e_g,m;o},q,v)$ and
to $Z(G_{\{e_j\},e_g,m;c},q,v)$. 

From the dimensionless free energy per vertex, $f(\{G_{\{e\}_r,e_g}\},q,v)$,
one can then calculate the internal energy $U$ and specific heat $C$ per
vertex.  Since the expressions are somewhat lengthy, we do not include them
here.  A salient difference between the hammock graphs for $r=2$ and for $r \ge
3$ is that in the $r=2$ case, as is evident from Eq. (\ref{zlam0_r2}), the
dependence of $\lambda_{Z,0;r=2}$, and hence also the dimensionless free energy
$f$, on the edges $e_1$ and $e_2$ in the set $\{e\}_{r=2}$ is only through
their sum, $p=e_1+e_2$, the number of sides of each polygon in the chain.  As
is clear from Eq. (\ref{zlam0}), for hammock graphs with $r \ge 3$, this
simplification no longer holds, i.e., the dependence of $\lambda_{Z,0;r \ge 3}$
on the $r$ edges $e_j \in \{e\}_r$ is not expressible simply as a function of
their sum.

% ======================================================================

% ====================================================================

\section{Flow Polynomials}
\label{flowpoly_section}

Another important one-variable special case of the Tutte polynomial is
the flow polynomial of a graph $G$, denoted $F(G,q)$
\cite{ford_fulkerson}.  To define this polynomial, one considers a
connected graph and chooses an orientation for each edge. Next, one
assigns an element of a finite abelian group ${\mathbb Z}_q$ as a
discretized flow along a given oriented edge and imposes two
conditions: (i) there is flow conservation mod $q$ at each vertex,
i.e., no sources or sinks, and (ii) the flow along each edge must be
nonzero, since a zero flow on an edge would be equivalent to the
absence of this edge in the graph. This defines a nowhere-zero flow on
the graph.  Henceforth, for brevity we will refer to nowhere-zero
flows on a graph simply as flows on the graph.  Just as in solving
Kirchhoff's equations in electrical circuit theory, the enumeration of
these flows is independent of the convention that one takes for the
orientation of each edge \cite{boll}. The flow polynomial thus depends
only on the structure of the graph $G$ itself and is given in terms of
the Tutte polynomial by the relation
\beq
F(G,q) = (-1)^{c(G)}T(G,x=0,y=1-q) \ ,
\label{ftrel}
\eeq
where, as above, $c(G)$ denotes the number of linearly independent
circuits on $G$. The minimum (integer) number $q$ such that a graph
$G$ has a (nowhere-zero) $q$-flow is denoted $\phi(G)$. If a
(connected) graph $G$ contains any bridge, then the flow polynomial
vanishes identically.  The open hammock chain graph
$G_{\{e\}_r,e_g,m;o}$ contains at least one bridge if (i) $e_g \ge 1$
or (ii) $r=1$ and $e_1 \ge 1$. Hence,
\beq
F(G_{\{e\}_r,e_g,m;o},q) = 0 \quad {\rm if} \ e_g \ge 1 
\label{flow_open_bridge1}
\eeq
and
\beq
F(G_{\{e\}_r,e_g,m;o},q) = 0 \quad {\rm if} \ r =1 \ {\rm and} \ e_1 \ge 1
\label{flow_open_bridge2}
\eeq
From our general results for $T(G_{\{e\}_r,e_g,m,BC},x,y)$, we
calculate the flow polynomials
\beq
F(G_{\{e\}_r,e_g,m;o},q) = \delta_{e_g,0} \, [(q-1)D_r]^m 
\label{flow_ham_open}
\eeq
and 
\beq
F(G_{\{e\}_r,e_g,m;c},q) = \delta_{e_g,0} \, [(q-1)D_r]^m 
+ (q-1)[D_{r+1}]^m \ , 
\label{flow_ham_cyclic}
\eeq
where $\delta_{ij}$ is the Kronecker delta function.  For $r=2$, these
general-$r$ results agree with Eqs. (8.3) and (8.4)) in \cite{neca}.
A notable property of these results is that $F(G_{\{e\}_r,e_g,m;o},q)$
and $F(G_{\{e\}_r,e_g,m;c},q)$ depend on $r$, but are independent of
the values of the edges $e_j \in \{e\}_r$.  This property is easily
understandable, since if a given flow is allowed, it does not depend
on how many edges there are in any of the ``ropes'' making up a given
hammock subgraph.

We comment on some properties of our general results
(\ref{flow_ham_open}) and (\ref{flow_ham_cyclic}), beginning with the
open hammock chain graphs. Let us assume that $e_g=0$
and $r \ge 2$ so that there are no bridges in
$G_{\{e\}_r,e_g,m;o}$. From Eq. (\ref{flow_ham_open}) we see that (i)
$G_{\{e\}_r,e_g=0,m;o}$ has a 2-flow (i.e.,
$F(G_{\{e\}_r,e_g=0,m;o},q=2) > 0$) only if $r$ is even; (ii) there is
precisely one 2-flow on $G_{\{e\}_r,e_g=0,m;o}$ if $r$ is even, i.e.,
$F(G_{\{e\}_r,e_g=0,m;o},q=2)=1$ for even $r$; and (iii) for even $r$,
$\phi(G_{\{e\}_r,e_g=0,m;o})=2$.  We proceed to prove these
properties.  Since $D_r$ vanishes at $q=2$ if $r$ is odd, it follows
that $F(G_{\{e\}_r,e_g,m;o},q)=0$ at $q=2$ if $r$ is odd.  If $q=2$,
i.e., the abelian group used for the flows is ${\mathbb Z}_2$, then
$1=-1$ mod 2. Now (with $e_g=0$ so flows are possible on this open
hammock chain graph) if $q=2$, then there is only one nonzero current
of flow on a given oriented edge $\vec e_j$, namely a flow of 1 unit
in the direction of the arrow. If and only if $r$ is even, then the
unit flows on $r/2$ ropes entering the end vertex of each hammock
subgraph on the open chain can flow out again as unit flows on the
orthogonal set of $r/2$ ropes, but this is not possible if $r$ is odd.
Equivalently, we can regard this flow as unit flows on all $r$ ropes
going into the end vertex and use the fact that $r \times 1 = 0$ mod 2
if and only if $r$ is even. This proves property (i). Clearly, there
is only one such flow configuration, which proves (ii).  The existence
of this flow proves property (iii).

For sufficiently large fixed value of $q$,
$F(G_{\{e\}_r,e_g=0,m;o},q)$ is a monotonically increasing function of
$r$.  This is a consequence of the property that $D_r$ is a polynomial
of degree $r-2$, going like $q^{r-2}$ for large $q$.  If $r$ is odd,
then there is a small interval of $q$ extending from $q=2$ to a
slightly larger value of $q$ for which, if $r$ is odd and $s$ is even,
then $D_r < D_s$ even though $r > s$, reflecting the zero at $q=2$ in
$D_r$ for odd $s$. For example, let us compare $D_4=q^2-3q+3$ and
$D_5=(q-2)(q^2-2q+q)$ in the real-$q$ interval slightly above $q=2$.
In the interval $2 \le q \le 2.54369$ (where the upper end of this
interval is determined as the unique real root of the cubic equation
$q^3-5q^2+9q-7=0$), $D_4 > D_5$.  However, for all $q > 2.54369$, $D_5
> D_4$. Similarly, $D_7 < D_6$ for $2 \le q \le 2.38809$, but $D_7 >
D_6$ for $q > 2.38809$.  Hence, if one restricts $q$ to integral
values $q \ge 3$, then $F(G_{\{e\}_r,e_g=0,m;o},q)$ is a monotonically
increasing function of $r$.

We also remark on properties of $F(G_{\{e\}_r,e_g=0,m;c},q)$ for the
cyclic hammock chain graphs. Both for $e_g=0$ and for $e_g \ge 1$, the
cyclic hammock chain graph allows more $q$-flows than the open hammock
graph.  This is a consequence of the possibility that $q$-flows on the
cyclic graph can make a global circuit around the chain. In
particular, even if $e_g > 0$, so there are no $q$-flows on the open
hammock graph, there are still $q$-flows on the cyclic hammock graph
provided that the second term in Eq. (\ref{flow_ham_cyclic}) is
nonzero, which is the case if $r$ is odd or if $r$ is even and $q \ge
3$. As with the open hammock chain, for a sufficiently large fixed
value of$q$, $F(G_{\{e\}_r,e_g,m;c},q)$ is a monotonically increasing
function of $r$.

For fixed $m$, as $q \to \infty$, the flow polynomials
behave asymptotically as 
\beq
F(G_{\{e\}_r,e_g,m;o},q) \sim \delta_{e_g,0}q^{m(r-1)} \quad
{\rm as} \ q \to \infty
\label{flow_ham_open_large_q}
\eeq
and
\beq
F(G_{\{e\}_r,e_g,m;c},q) \sim
\delta_{e_g,0}q^{m(r-1)} + q^{m(r-1)+1} \quad {\rm as} \ q \to \infty \ .
\label{flow_ham_cyclic_large_q}
\eeq

Another comparison is with flow polynomials for strip graphs of
various lattice types.  It is natural to consider the simplest strip
graph for this comparison, namely a one-dimensional line graph of
length $n$ vertices.  If this has open boundary conditions, there are
no $q$-flows, so we restrict to the case of cyclic boundary
conditions, for which one has the elementary result $F(C_n,q) = q-1$.
We compare this with $F(G_{\{e\}_r,e_g,m;c},q)$. If $r=1$, then the
cyclic hammock graph reduces to a circuit graph $C_n$ with
$n=m(e_1+e_g)$ vertices, and hence the same flow polynomial, namely
$q-1$. For most choices of $r$, $e_g$, $q$, and $m$, there are more
$q$-flows (typically exponentially more) on a cyclic hammock graph
with $m$ hammock subgraphs than on $C_n$.  For example, if $r=3$,
$e_g=0$, and $q=3$, then $F(G_{\{e\}_3,e_g=0,m;c},q=3)=2^m+2 \cdot
3^m$, while if $r=3$, $e_g=1$, and $q=3$, then
$F(G_{\{e\}_3,e_g=1,m;c},q=3) = 2 \cdot 3^m$, as contrasted with
$F(C_n,q=3)=2$.  However, in certain cases, there may be an equal
number of $q$-flows on cyclic hammock graphs and the circuit graph, or
even no $q$-flows on cyclic hammock graphs and a nonzero number of
$q$-flows on a circuit graph. For example, if $r \ge 2$ and $e_g=0$,
then, from Eq. (\ref{flow_ham_cyclic}) we have
$F(G_{\{e\}_r,e_g=0,m;c},q)=[(q-1)D_r]^m +
(q-1)[D_{r+1}]^m$. Continuing to hold $e_g=0$, if also $q=2$ and $r$
is even, then, because $D_{\rm odd}=0$ at $q=2$, the second term
vanishes, and $D_{r \ {\rm even}}=1$ at $q=2$, so $F(G_{\{e\}_{r
    \ {\rm even}},e_g=0,m;c},q=2)=1$, the same as for the circuit
graph with $q=2$.  If $e_g > 1$, then $F(G_{\{e\}_r,e_g,m;c},q) =
(q-1)[D_{r+1}]^m$, so if also $r$ is even and $q=2$, then
$F(G_{\{e\}_{r \ {\rm even}},e_g,m;c},q=2)=0$, and hence there are no
2-flows on these cyclic hammock graphs, whereas in contrast, there is
one 2-flow on $C_n$.  Finally, we observe that the zeros of flow
polynomials and reliability polynomials are of interest but are beyond
the scope of the current work; an analysis of these will ge 
given elsewhere. 

% ====================================================================

\section{Reliability Polynomials}
\label{reliability_section}

A communication network, such as the internet, can be represented by a
graph, with the vertices of the graph representing the nodes of the
network and the edges of the graph representing the communication
links between these nodes.  In analyzing the reliability of a network,
one is interested in the probability that there is a working
communications route between any node and any other node
\cite{colbourn,brown_colbourn_review}.  This quantity is the
(all-terminal) reliability function.  This is often studied using a
simplification in which one assumes that each node is operating with
probability $p_{node}$ and each link (bond, abbreviated $b$) is
operating with probability $p_b$. As probabilities, $p_{node}$ and
$p_b$ lie in the interval [0,1]. The dependence of the all-terminal
reliability function $R_{tot}(G,p_{node},p_b)$ on $p_{node}$ is an
overall factor of $(p_{node})^n$; i.e., $R_{tot}(G,p_{node},p_b) =
(p_{node})^n R(G,p_b)$.  The difficult part of the calculation of
$R_{tot}(G,p_{node},p_b)$ is thus the part that depends on the bonds,
$R(G,p_b)$.  The function $R(G,p_b)$ is given by
\beq
R(G,p_b) = \sum_{\tilde G \subseteq G} p_b^{e(\tilde G)} \, 
(1-p_b)^{e(G)-e(\tilde G)}
\label{rgen}
\eeq
where $\tilde G$ is a connected spanning subgraph of $G$.  Each term in this
sum is the probability that the communication links $\tilde E \in \tilde G$ are
functioning (equal to $p_b^{e(\tilde G)}$) times the probability that the other
links, $E - \tilde E$, are not functioning (equal to $(1-p_b)^{e(G)-e(\tilde
  G)}$).  Since $e(\tilde G)$ and $e(G)-e(\tilde G)$ are both nonnegative,
Eq. (\ref{rgen}) shows that $R(G,p_b)$ is a polynomial in $p_b$. From its
definition, $R(G,p_b)$ is clearly a monotonically increasing function of $p_b
\in [0,1]$ with the boundary values $R(G,0)=0$ and $R(G,1)=1$.  $R(G,p_b)$ is
given in terms of the Tutte polynomial, evaluated with $x=1$ (guaranteeing that
$\tilde G$ is a connected spanning subgraph of $G$) and $y=y_b$, where
\beq
y_b = \frac{1}{1-p_b} \quad i.e., \ v_b = y_b - 1 = \frac{p_b}{1-p_b} \ , 
\label{yb}
\eeq
by the relation
\beq
R(G,p_b)=p_b^{n(G)-1}(1-p_b)^{e(G)+1-n(G)} \, 
T(G,1,\frac{1}{1-p_b}) \ . 
\label{rtrel}
\eeq

From our calculation of the general Tutte polynomials
for $G_{\{e\}_r,e_g,m;o}$ and 
$G_{\{e\}_r,e_g,m;c}$, using the relation Eq. (\ref{rtrel}), we can 
compute the corresponding reliability polynomials 
$R(G_{\{e\}_r,e_g,m;o},p_b)$ and $R(G_{\{e\}_r,e_g,m;c},p_b)$. 
For this purpose, it is helpful to use Eqs. (\ref{tham_open_x1}) and 
(\ref{tham_cyclic_x1}).  For the open hammock chain graphs we obtain 
\beq
R(G_{\{e\}_r,e_g,m;o},p_b) = 
[(p_b^{(\sum_{j=1}^r e_j)+e_g-r} \, \mu_{R,0})]^m \ , 
\label{reliability_ham_open}
\eeq
where
\beq
\mu_{R,0} = 
\prod_{j=1}^r \Big \{ (1-p_b)e_j + p_b \Big \} - (1-p_b)^r\prod_{j=1}^r e_j 
 \ .  
\label{mu0_reliability}
\eeq
The quantity $\mu_{R,0}$ always has a factor of $p_b$, so for general
$r$, $R(G_{\{e\}_r,e_g,m;o},p_b)$ has an overall factor of 
$(p_b)^{m[(\sum_{j=1}^r e_j)+e_g-(r-1)]}$.  For example, if $r=2$, then 
\beq
R(G_{\{e\}_{r=2},e_g,m;o},p_b) = \Big [  p_b^{e_1+e_2+e_g-1} 
[ (e_1+e_2)(1-p_b) + p_b ] \Big ]^m 
\label{reliability_open_r2}
\eeq
and if $r=3$, then 
\beqs
&& R(G_{\{e\}_{r=3},e_g,m;o},p_b) = \Bigg [ p_b^{(\sum_{j=1}^3 e_j) + e_g -2} 
\times 
\cr\cr
&\times&
\Big \{ (e_1e_2+e_1e_3+e_2e_3)(1-p_b)^2 + (e_1+e_2+e_3)p_b(1-p_b) + p_b^2 
 \Big \} \Bigg ]^m \ . 
\label{mu0_r3}
\eeqs

For the cyclic hammock chain graphs, we find the general formula 
\beqs
&& R(G_{\{e\}_r,e_g,m;c},p_b) = p_b^{m[(\sum_{j=1}^r e_j)  + e_g - r]} \, 
(\mu_{R,0})^{m-1} \times 
\cr\cr
&\times& \bigg [ p_b^{-1}[ me_g(1-p_b)+p_b] \mu_{R,0} + 
m(1-p_b)^r \prod_{j=1}^r e_j \bigg ] \ . 
\label{reliability_ham_cyclic}
\eeqs

In the cases where $r=2$ and $r=3$, Eq. (\ref{reliability_ham_cyclic}) 
yields the following: 
\beqs
& & R(G_{\{e\}_{r=2},e_g,m;c},p_b) = p_b^{m(e_1+e_2+e_g-1)-1} \Big [
  (e_1+e_2)(1-p_b)+p_b \Big ]^{m-1} \times \cr\cr
& \times & \Bigg [\bigg \{me_g(1-p_b)+p_b \bigg \} 
  \bigg \{ (e_1+e_2)(1-p_b) + p_b \bigg \} + m e_1 e_2 (1-p_b)^2 \Bigg ]  \ .
 \cr\cr
&& 
\label{reliability_ham_cyclic_r2}
\eeqs
and
\beqs
&& R(G_{\{e\}_{r=3},e_g,m;c},p_b) = p_b^{m[(\sum_{j=1}^3 e_j) +e_g-2]-1} \times
\cr\cr
&\times& 
\Big [ (e_1e_2+e_1e_3+e_2e_3)(1-p_b)^2 + (e_1+e_2+e_3)p_b(1-p_b) + p_b^2 
\Big ]^{m-1} \times \cr\cr
&\times& \bigg [ \Big ( me_g(1-p_b)+p_b \Big ) \Big \{ (e_1e_2+e_1e_3+e_2e_3)(1-p_b)^2 
+ (e_1+e_2+e_3)p_b(1-p_b)+ p_b^2 \Big \} \cr\cr
&+& m e_1e_2e_3(1-p_b)^3  \bigg ]
\label{reliability_ham_cyclic_r3}
\eeqs
%

% ======================================================================

\section{Percolation Clusters}
\label{percolation_section}

In this section we use our results to calculate a quantity of interest in the
area of bond percolation.  We first briefly mention some necessary background
(reviews include \cite{stauffer_aharony,brp}). 
Consider a connected graph $G$ and assume that the vertices are definitely
present, but each edge (bond, $b$) is present only with a probability $p_b
\in [0,1]$.  In the usual statistical mechanical context, one usually considers
a limit in which the number of vertices $n \to \infty$. An important quantity
is the average number of connected components (= clusters) in $G$.  For a given
$G$, we denote this average cluster number per vertex as $\langle k
\rangle_G$. This is given by
\beqs
\langle k \rangle_G & = & \frac{(1/n)\sum_{G^\prime} k(G^\prime)
p_b^{e(G^\prime)}(1-p_b)^{e(G)-e(G^\prime)} }{
\sum_{G^\prime} p_b^{e(G^\prime)}(1-p_b)^{e(G)-e(G^\prime)} } \cr\cr
& & 
\cr
& = &
\frac{(1/n)\sum_{G^\prime} k(G^\prime)v_b^{e(G^\prime)}}
{\sum_{G^\prime} v_b^{e(G^\prime)}} \ ,
\label{keq}
\eeqs
where $G^\prime$ is a spanning subgraph of $G$, as above, and 
$v_b$ was defined in Eq. (\ref{yb}).  
Hence, in the $n \to \infty$ limit, the average
cluster number per vertex, $\langle k \rangle_{\{G\}}$, is given by
\beq
\langle k \rangle_{\{G\}} = \frac{\partial f(\{ G \},q,v)}{\partial q} 
\bigg |_{q=1, \  v=v_b} \ . 
\label{kdfdq}
\eeq
This is independent of the longitudinal boundary conditions on the chain.  Two
general properties holding for an arbitrary graph $G$ may be noted. First, if
$p_b=1$, then there is just one cluster, consisting of the entire graph, and
hence the division by $n$ yields 0, so
\beq
\langle k \rangle_G = 0 \quad {\rm if} \ p_b=1 \ . 
\label{kp1}
\eeq
This implies that $\langle k \rangle_G$ contains at least one factor of
$(1-p_b)$.  Second, if $p_b=0$, then there are $n$ clusters each
consisting of one of the $n$ vertices of $G$, so
\beq
\langle k \rangle_G = 1 \quad {\rm if} \ p_b=0 \ . 
\label{kp0}
\eeq
For a given $G$, $\langle k \rangle_G$ is a monotonically decreasing
function of $p_b$, decreasing from $\langle k \rangle_G=1$ at $p_b=0$
to $\langle k \rangle_G=0$ at $p_b=1$.

For the hammock chain graphs under consideration here, the average cluster
number $\langle k \rangle_{\{G_{\{e\}_r,e_g} \}}$ is determined by substituting
Eq. (\ref{fham}) in Eq. (\ref{kdfdq}). We calculate
\beqs
&& \langle k \rangle_{\{G_{\{e\}_r,e_g} \}} =
\frac{1}{(\sum_{j=1}^r e_j)+e_g-(r-1)} \times
\cr\cr
&\times& \Bigg [ -r + e_g(1-p_b) +
  \sum_{j=1}^r \Big \{ e_j(1-p_b)+p_b^{e_j} \Big \}
  + \prod_{j=1}^r (1-p_b^{e_j}) \Bigg ] 
\label{kav}
\eeqs
For $r=2$, Eq. (\ref{kav}) reduces to the expression given as
Eq. (10.3) in \cite{neca}, which reads, in our current notation, as
\beqs
\langle k \rangle_{\{G_{\{e\}_{r=2},e_g}\}} & = & 
\frac{1}{(e_1+e_2+e_g-1)} \, 
\bigg [ -1 + (e_1+e_2+e_g)(1-p_b)+p_b^{e_1+e_2} \bigg ] 
\cr\cr & = & \Bigg (\frac{1-p_b}{e_1+e_2+e_g-1} \Bigg ) 
\Bigg [e_1+e_2+e_g - \Big ( \sum_{j=0}^{e_1+e_2-1} p_b^j \Big ) \Bigg ] \ . 
\label{kave_r2}
\eeqs
The expression in the second line of Eq. (\ref{kave_r2}) explicitly shows 
the factor of $(1-p_b)$ in 
$\langle k \rangle_{\{G_{\{e\}_{r=2},e_g}\}}$. 
As an illustration of a result for higher $r$, if $r=3$, then
the general result (\ref{kav}) yields 
\beqs
\langle k \rangle_{ \{ G_{\{e\}_{r=3},e_g } \} } &=& 
\frac{1}{( (\sum_{j=1}^3 e_j) +e_g-2 )} \, 
\Bigg [ -2 + \Big \{ (\sum_{j=1}^3 e_j)+e_g\Big \}(1-p_b)  \cr\cr
&+& \Big \{ p_b^{e_1+e_2} + p_b^{e_1+e_3} + p_b^{e_2+e_3} \Big \} 
- p_b^{e_1+e_2+e_3}  \Bigg ] \ . 
\label{kave_r3}
\eeqs
Again, one can reexpress this with an expicit overall factor of $(1-p_b)$
by using the identity $p_b^t-1 = (p_b-1)\sum_{i=0}^{t-1}p_b^i$:
\beqs
\langle k \rangle_{ \{ G_{\{e\}_{r=3},e_g } \} } &=& 
\Bigg ( \frac{1-p_b}{(\sum_{j=1}^3 e_j) +e_g-2} \Bigg ) \, 
\Bigg [ \Big \{ (\sum_{j=1}^3 e_j) + e_g \Big \} \cr\cr
&-& \Big \{ \sum_{i=0}^{e_1+e_2-1} p_b^i + 
            \sum_{i=0}^{e_1+e_3-1} p_b^i + 
            \sum_{i=0}^{e_2+e_3-1} p_b^i \Big \} 
          + \sum_{i=0}^{e_1+e_2+e_3-1} p_b^i \Bigg ] \ . \cr\cr
&& 
\eeqs
Expressions for $\langle k \rangle_{ \{ G_{\{e\}_r,e_g } \} }$ for the $m \to
\infty$ limits of higher-$r$ hammock chain graphs can be calculated from our
general formula (\ref{kav}) in a similar manner.

% ===================================================================

\section{Some Graphical Quantities}
\label{graphical_quantities_section}

\subsection{General}
\label{graphical_general}

Special valuations of the Tutte polynomial of a graph for particular $x$ and
$y$ yield enumerations of various types of subgraphs of the given graph. Recall
that a tree graph is a connected graph with no circuits.  A spanning tree of a
graph $G$ is a spanning subgraph of $G$ that is also a tree.  A spanning forest
of a graph $G$ is a spanning subgraph of $G$ that may consist of more than one
connected component but contains no circuits. The special valuations of
interest include (i) $T(G,1,1)=N_{ST}(G)$, the number of spanning trees ($ST$)
of $G$; (ii) $T(G,2,1)=N_{SF}(G)$ the number of spanning forests ($SF$) of $G$;
(iii) $T(G,1,2)=N_{CSSG}(G)$, the number of connected spanning subgraphs
($CSSG$) of $G$; and (iv) $T(G,2,2)=N_{SSG}(G)=2^{e(G)}$, the number of
spanning subgraphs ($SSG$) of $G$.
For both the open and cyclic strips, the last of these quantities is directly
determined by Eq. (\ref{eham}) as 
\beq
N_{SSG}(G_{\{e\}_r,e_g,m;BC}) = 2^{e(G_{\{e\}_r,e_g,m;BC})} 
= 2^{m[(\sum_{j=1}^r e_j)+e_g]} \ . 
\label{ssg_ham}
\eeq
The evaluation of our general
results for $T(G_{\{e\}_r,e_g,m;o},x,y)$ and $T(G_{\{e\}_r,e_g,m;c},x,y)$ for
the requisite values of $x$ and $y$ yield the above-mentioned graphical
quantities (i)-(iii).  We discuss these next.

% ================================================================

\subsection{Spanning Trees} 
\label{st_section}

To evaluate the number of spanning trees of the hammock chain graphs, 
$N_{ST}(G_{\{e\}_r,e_g,m;BC})$, we require the evaluation of 
$T(G_{\{e\}_r,e_g,m;BC},1,1)$. Using the general formulas 
given in the text, we calculate 
\beq
N_{ST}(G_{\{e\}_r,e_g,m;o}) = (\lambda_{T,0;x=1,y=1})^m  \ , 
\label{st_ham_open}
\eeq
where
\beq
\lambda_{T,0;x=1,y=1} =  \sum_{r \ {\rm terms}} ( \prod_j e_j)_{r-1} \ , 
\label{tlam_x1y1}
\eeq
and the notation $(\prod_j e_j)_{r-1}$ means the product of the
edges in the set $\{e\}_r$ with one edge removed. Since there are $r$ ways of
removing one edge from this product, there are $r$ terms in the sum of 
$(r-1)$-fold products in Eq. (\ref{tlam_x1y1}). For the cyclic chain, we 
find 
\beqs
&& N_{ST}(G_{\{e\}_r,e_g,m;c}) = m (\lambda_{T,0;x=1,y=1})^{m-1} 
\Big [ e_g \lambda_{T,0;x=1,y=1} + \prod_{j=1}^r e_j \Big ] \ . \cr\cr
&&
\label{st_ham_cyclic}
\eeqs
For $r=2$, these general formulas yield 
\beq
N_{ST}(G_{\{e\}_{r=2},e_g,m;o}) = (e_1+e_2)^m
\label{st_ham_open_r2}
\eeq
and
\beq
N_{ST}(G_{\{e\}_{r=2},e_g,m;c}) = m(e_1+e_2)^{m-1}
\Big [ e_g(e_1+e_2) + e_1e_2 \Big ] \ , 
\label{st_ham_cyclic_r2}
\eeq
in agreement with Eqs. (11.2) and (11.3) in \cite{neca}. 
As an example of an explicit result for higher $r$, for 
$r=3$, the general results above yield 
\beq
N_{ST}(G_{\{e\}_{r=3},e_g,m;o}) = (e_1e_2 + e_1e_3 + e_2e_3)^m
\label{st_ham_open_r3}
\eeq
and
\beqs
&& N_{ST}(G_{\{e\}_{r=3},e_g,m;c}) = m(e_1e_2 + e_1e_3 + e_2e_3)^{m-1}\bigg [
e_g(e_1e_2 + e_1e_3 + e_2e_3) + e_1e_2e_3 \bigg ] \ . \cr\cr
&& 
\label{st_ham_cyclic_r3}
\eeqs

% ============================================================

\subsection{Spanning Forests}
\label{sf_section}

Substituting $x=2$ and $y=1$ in our general calculation of the Tutte
polynomials, we obtain $N_{SF}(G_{\{r\}_r,e_g,m;o})$ and
$N_{SF}(G_{\{r\}_r,e_g,m;c})$ for arbitrary $r$.  These
\beq
N_{SF}(G_{\{e\}_r,e_g,m;o})=(\lambda_{T,0;x=2,y=1})^m
\label{sf_ham_open_gen}
\eeq
and
\beq
N_{SF}(G_{\{e\}_r,e_g,m;c})=(\lambda_{T,0;x=2,y=1})^m - 
                            (\lambda_{T,1;x=2,y=1})^m \ ,  
\label{sf_ham_cyclic_gen}
\eeq
where these $\lambda_{T,0}$ and $\lambda_{T,1}$ evaluated at $x=2$ and
$y=1$ (and hence $q=0$) are
\beqs
&& \lambda_{T,0;x=2,y=1} = 2^{e_g}\Bigg [ 2^{(\sum e_j)_r} 
-\theta(r-3)\sum_{{r \choose 2} \ {\rm terms}} 2^{(\sum e_j)_{r-2}} 
\cr\cr
&+& 2 \theta(r-4) \sum_{{r \choose 3} \ {\rm terms}}2^{(\sum e_j)_{r-3}} 
+ ... + (-1)^{r-1}(r-1) \Bigg ]
\label{tlam0_x2y1}
\eeqs
and
\beqs
&& \lambda_{T,1;x=2,y=1} =
     \theta(r-2)\sum_{r \ {\rm terms}} 2^{(\sum e_j)_{r-1}} 
   -2\theta(r-3)\sum_{{r \choose 2} \ {\rm terms}} 2^{(\sum e_j)_{r-2}} 
\cr\cr
&+& 3\theta(r-4)\sum_{{r \choose 3} \ {\rm terms}} 2^{(\sum e_j)_{r-3}} 
+ ... + (-1)^{r+1}r \ . 
\label{tlam1_x2y1}
\eeqs
Here we have used the identity $D_n=(-1)^n(n-1)$ at $q=0$, 
from Eq. (\ref{dn_q0}). 

For the special case $r=2$ these general formulas reduce to
\beq
N_{SF}(G_{\{e\}_{r=2},e_g,m;o}) = \Big [2^{e_g}\Big ( 2^{e_1+e_2}-1 \Big ) 
\Big ]^m
\label{sf_ham_open_r2}
\eeq
and
\beq
N_{SF}(G_{\{e\}_{r=2},e_g,m;c}) = \Big [2^{e_g}\Big ( 2^{e_1+e_2}-1 \Big ) 
\Big ]^m 
- \Big [2^{e_1} + 2^{e_2} -2 \Big ]^m \ , 
\label{sf_ham_cyclic_r2}
\eeq
in agreement with Eqs. (11.4) and (11.5) in \cite{neca}. 

As an illustration of the application of our general results to 
a higher-$r$ case, we consider $r=3$.  Here we calculate
\beq
N_{SF}(G_{\{e\}_{r=3},e_g,m;o}) = 
\Big [ 2^{e_g} \Big \{ 2^{e_1+e_2+e_3} - (2^{e_1}+2^{e_2}+2^{e_3}) + 2 
\Big \} \Big ]^m 
\label{sf_ham_open_r3}
\eeq
and
\beqs
&& N_{SF}(G_{\{e\}_{r=3},e_g,m;c}) = 
\Big [ 2^{e_g} \Big \{ 2^{e_1+e_2+e_3} - (2^{e_1}+2^{e_2}+2^{e_3}) + 2 
\Big \} \Big ]^m \cr\cr
&-& \Big [ \Big ( 2^{e_1+e_2}+2^{e_1+e_3}+2^{e_2+e_3} \Big ) 
- 2\Big ( 2^{e_1}+2^{e_2}+2^{e_3} \Big ) + 3 \Big ]^m \ . 
\label{sf_ham_cyclic_r3}
\eeqs
The numbers $N_{SF}(G_{\{e\}_r,e_g,m;o})$ and 
            $N_{SF}(G_{\{e\}_r,e_g,m;c})$ for higher $r$ can be computed
in a similar manner from our general expressions for the Tutte polynomials. 

% ============================================================

\subsection{Connected Spanning Subgraphs}
\label{cssg_section}

By substituting $x=1$ and $y=2$ in our general results, one can calculate
$N_{CSSG}$ for these hammock chain graphs. Again, it is useful to utilize
Eqs. (\ref{tham_open_x1}) and (\ref{tham_cyclic_x1}) for this purpose.
Carrying out this procedure, we obtain, for the open hammock chain graphs, 
\beq
N_{CSSG}(G_{\{e\}_r,e_g,m;o}) = (\lambda_{T,0;x=1,y=2})^m  \ , 
\label{cssg_ham_open}
\eeq
where
\beq
\lambda_{T,0;x=1,y=2} = \prod_{j=1}^r (e_j+1) - \prod_{j=1}^r e_j 
\label{tlam0_x1y2}
\eeq
For the cyclic hammock chain graphs we calculate
\beqs
&& N_{CSSG}(G_{\{e\}_r,e_g,m;c}) = (\lambda_{T,0;x=1,y=2})^{m-1} 
 \bigg [ (me_g+1)\lambda_{T,0;x=1,y=2} + m\prod_{j=1}^r e_j \bigg ] \ .
\cr\cr
&&
\label{cssg_ham_cyclic}
\eeqs
For $r=2$, these general results yield 
\beq
N_{CSSG}(G_{\{e\}_{r=2},e_g,m;o}) = (e_1+e_2+1)^m 
\label{cssg_ham_open_r2}
\eeq
and
\beqs
&& N_{CSSG}(G_{\{e\}_{r=2},e_g,m;c}) = (e_1+e_2+1)^{m-1} 
\Big [ (me_g+1)(e_1+e_2+1)+me_1e_2 \Big ] \ , \cr\cr
&&
\label{cssg_ham_cyclic_r2}
\eeqs
in agreement with Eqs. (11.6) and (11.7) of \cite{neca}. 
As an example of a result for higher $r$, in the case $r=3$, 
the general results above yield 
\beqs
&& N_{CSSG}(G_{\{e\}_{r=3},e_g,m;o}) = (\lambda_{T,0;x=1,y=2,r=3})^m
\label{cssg_ham_open_r3}
\eeqs
and
\beqs
&& N_{CSSG}(G_{\{e\}_{r=3},e_g,m;c}) = (\lambda_{T,0;x=1,y=2,r=3})^{m-1}
\times 
\cr\cr
&\times& \Big [ (me_g+1) \lambda_{T,0;x=1,y=2,r=3}+me_1e_2e_3 \Big ] \ , 
\label{cssg_ham_cyclic_r3}
\eeqs
where 
\beq
\lambda_{T,0;x=1,y=2,r=3} = (e_1e_2+e_1e_3+e_2e_3)  + (e_1+e_2+e_3) + 1 \ . 
\label{sum_eiej_r3}
\eeq
%

% =================================================================

\subsection{Acyclic and Cyclic Orientations} 
\label{acyclic_section}

Assigning an orientation to each edge $e \in E$ of a graph $G=(V,E)$ 
yields a directed graph, denoted $G=(V,\vec E)$. Among the 
$2^{e(G)}$ edge orientations, an acyclic
orientation of $G$ is defined as an orientation that does not contain any
directed cycles. A directed cycle is a cycle in which, as one travels
along the cycle, all of the oriented edges have the same direction.  The number
of such acyclic orientations is denoted $a(G)$. This quantity depends only on
the structure of the basic graph $G$ itself and given by the evaluation
of the Tutte polynomial of $G$ with $x=2$ and $y=0$ \cite{stanley}:
\beq
a(G)= T(G,2,0) \ . 
\label{atrel}
\eeq
Equivalently, this is obtained by the evaluation of the chromatic polynomial at
$q=-1$: $a(G)= (-1)^{n(G)}P(G,-1)$.  For the oriented graph $G=(V,\vec E)$, a
totally cyclic orientation is one in which every oriented edge is a member of a
directed cycle.  The number of totally cyclic edge orientations, denoted
$b(G)$, again depends only on the structure of the basic graph $G$ is given by
\cite{vergnas}
\beq
b(G) = T(G,0,2) \ .
\label{btrel}
\eeq

From our general calculation of the Tutte polynomials for the open and cyclic
hammock chain graphs, we have, for general $r$,
\beq
a(G_{\{e\}_r,e_g,m;o}) = (\lambda_{T,0;x=2,y=0})^m
\label{a_ham_open_r}
\eeq
and
\beq
a(G_{\{e\}_r,e_g,m;c}) = (\lambda_{T,0;x=2,y=0})^m -2(\lambda_{T,1;x=2,y=0})^m
\ ,
\label{a_ham_cyclic_r}
\eeq
where
\beq
\lambda_{T,0;x=2,y=0} = 2^{e_g} \bigg [ 
2\prod_{j=1}^r (2^{e_j}-1) - \prod_{j=1}^r (2^{e_j}-2) \bigg ]  
\label{tlam0_x2y0}
\eeq
and
\beq
\lambda_{T,1;x=2,y=0} = \prod_{j=1}^r (2^{e_j}-1) - \prod_{j=1}^r (2^{e_j}-2)
\ . 
\label{tlam1_x2y0}
\eeq

For $r=2$, these general formulas reduce to 
\beq
a(G_{\{e\}_{r=2},e_g,m;o}) = \Big [ 2^{e_g} \Big ( 2^{e_1+e_2}-2 \Big ) \Big ]^m 
\label{a_ham_open_r2}
\eeq
and
\beq
a(G_{\{e\}_{r=2},e_g,m;c}) =
\Big [ 2^{e_g}\Big ( 2^{e_1+e_2}-2 \Big ) \Big ]^m 
- 2\Big [ 2^{e_1}+2^{e_2}-3 \Big ]^m \ , 
\label{a_ham_cyclic_r2}
\eeq
in agreement with Eqs. (11.9) and (11.10) in \cite{neca}.
As an explicit example of cases with higher $r$, for $r=3$, we have
\beq
a(G_{\{e\}_{r=3},e_g,m;o}) = (\lambda_{T,0;x=2,y=0,r=3})^m 
\eeq
and
\beq
a(G_{\{e\}_{r=3},e_g,m;c}) = (\lambda_{T,0;x=2,y=0,r=3})^m 
-2 (\lambda_{T,0;x=2,y=0,r=3})^m  \ , 
\eeq
where
\beq
\lambda_{T,0;r=3,x=2,y=0} = 2^{e_g} \Big [ 2^{e_1+e_2+e_3} 
-2 \Big ( 2^{e_1}+2^{e_2}+2^{e_3} \Big ) + 6 \Big ]
\label{tlam0_r3_x2y0}
\eeq
and
\beqs
&& \lambda_{T,1;r=3,x=2,y=0} = 
\Big ( 2^{e_1+e_2} + 2^{e_1+e_3} + 2^{e_2+e_3} \Big ) 
-3 \Big ( 2^{e_1}+2^{e_2}+2^{e_3} \Big ) + 7 \ . \cr\cr
&&
\label{tlam1_r3_x2y0}
\eeqs

From our general results for Tutte polynomials of hammock chain graphs, we 
find, for the number of totally cyclic orientations on these graphs, 
\beq
b(G_{\{e\}_r,e_g,m;o}) = \delta_{e_g,0}(2^r-2)^m 
\label{b_ham_open_r}
\eeq
and
\beq
b(G_{\{e\}_r,e_g,m;c}) = 2(2^r-1)^m - \delta_{e_g,0}(2^r-2)^m \ . 
\label{b_ham_cyclic_r}
\eeq
%

% ==================================================================

\section{Conclusions}

In conclusion, in this paper we have presented exact calculations of
the Potts/Tutte polynomials for hammock chain graphs
$G_{\{e_1,...,e_r\},e_g,m;BC}$ comprised of $m$ repeated hammock
(series-parallel) subgraphs $H_{e_1,...,e_r}$ connected with line
graphs of length $e_g$ edges, such that the chains have open or cyclic
boundary conditions (BC). We study the general case where the hammock
subgraphs $H_{e_1,...,e_r}$ have $r$ separate paths along ``ropes''
with respective lengths $e_1, ..., e_r$ edges connecting the two
end-vertices of each hammock subgraph.  We have discussed the special
cases of these results that yield chromatic polynomials, flow
polynomials, reliability polynomials, and various quantities of
graph-theoretic interest, including numbers of spanning trees,
spanning forests, connected spanning subgraphs, acyclic orientations,
and cyclic orientations. We have presented a detailed study of the
continuous accumulation set ${\cal B}_q$ of chromatic zeros in the
complex $q$ plane in the limit of infinite chain length, $m \to
\infty$, for a variety of choices of edge sets $(\{e\}_r,e_g)$. This
work involves a very interesting confluence of statistical mechanics,
graph theory, complex analysis, and algebraic geometry. 

% =================================================================

\section{Acknowledgments}

This research was partially supported by the U.S. National Science Foundation
grant NSF-PHY-22-10533. RS expresses his gratitude to Shan-Ho Tsai for the
valuable collaboration on the earlier related work in Refs. \cite{wa3,nec}.

% ==================================================================

\section{Ethics Statements on Conflict of Interest and Data Accessibility}

\bigskip

As required by current rules for article submission,
the authors include the following ethics statements:

\begin{enumerate}

\item Concerning any possible conflicts of interest, the authors have
  no financial or non-financial conflict of interest or competing
  interests that are relevant to this article. The authors have stated
  their grant support in the Acknowledgments.

\item Concerning data accessibility, this article is theoretical and
does not contain any experimental data. Calculational data supporting
the conclusions of the article are included herein.

\end{enumerate}

% =================================================================

\section{Appendix}

For reference we list some special values of $D_n$ here \cite{wa2}.
\beq
D_n = (-1)^n\Big ( 2^{n-1}-1 \Big ) \quad {\rm if} \ q=-1 
\label{dn_qm1}
\eeq
\beq
D_n = (-1)^n (n-1) \quad {\rm if} \ q=0
\label{dn_q0}
\eeq
\beq
D_n = (-1)^n \quad {\rm if} \ q=1
\label{dn_q1}
\eeq
and
\beq
D_n = \cases{ 1 &if $q=2$ and $n$ is even \cr
              0 &if $q=2$ and $n$ is odd \cr} \ . 
\label{dn_q2}
\eeq
A factorization property is that if $n$ is odd, say $n=2k+1$, then 
\beq
D_{2k+1} \ {\rm contains \ the \ factor} \ q-2. 
\label{dnodd_factor}
\eeq
One can characterize the Taylor expansion of $D_{2k+1}$ further. In 
addition to $D_3=q-2$, the Taylor series of $D_{2k+1}$ about $q=2$ is
\beq
D_{2k+1}=k(q-2) + k(k-1)(q-2)^2 + ... \quad {\rm as} \ q \to 2 \ , 
\label{dnoddq2}
\eeq
where the $+...$ indicate higher-order terms in the Taylor expansion of 
$D_{2k+1}$ about $q=2$ that are present if $k \ge 2$. 
$D_n$ satisfies a number of other identities such as 
\beq
D_n = (q-1)D_{n-1} + (-1)^n \ .
 \label{identity1}
\eeq
These identities can be proved from the definition of $D_n$ in Eq. (\ref{dk}). 

% ===============================================================

\end{document}